# Direct current thermo-mechanical testing: Principles, uncertainty hierarchy, and its role in advanced materials characterisation

Abdalrhaman Koko[1*], Sodiq Abiodun Kareem[2], Olajesu Favor Olanrewaju[3], Rachael Williams[1], Justus Uchenna Anaele[4], Yuanbo T. Tang[5], and Bryan Roebuck[1]

[1] Advanced Engineering Materials, National Physical Laboratory, United Kingdom

[2] School of Materials Engineering, Purdue University, USA

[3]Department of Materials Science and Engineering, Iowa State University, USA

[4] Department of Metallurgical and Materials Engineering, Federal University of Technology, Nigeria

[5] School of Metallurgy and Materials, University of Birmingham, United Kingdom

## Abstract

Direct current thermo-mechanical testing (DC-TMT), based on resistive Joule heating, enables rapid heating and cooling, steep thermal gradients and simultaneous mechanical loading, making it a powerful tool for probing deformation, phase transformations, oxidation-assisted damage and creep under conditions inaccessible to conventional furnace-based methods. Despite its growing use, DC-TMT lacks formal standardisation and is often misinterpreted as equivalent to bulk isothermal testing, overlooking intrinsic differences in thermal and mechanical fields. This review addresses that gap by consolidating four decades of research on specimen geometry, temperature measurement, strain characterisation and environmental control, and by classifying uncertainty sources as dominant, secondary and conditional. Evidence from modelling and experiment shows that temperature gradients, heating rate and gauge representativeness govern the reliability of inferred material behaviour. Applications across aluminium, steels, nickel-based superalloys, titanium alloys, hardmetals, zirconium alloys, shape memory alloys and additively manufactured systems are critically assessed. The review highlights domains where DC-TMT provides reproducible mechanistic insight and conditions where direct equivalence with bulk data is not warranted. Implications include the need for transparent reporting, multi-sensor temperature validation and integration with electro-thermal modelling to enable rigorous, mechanism-focused interpretation.

[*] Corresponding authors E-mail: abdo.koko@npl.co.uk

**Keywords:** Joule heating; thermo-mechanical testing; measurement uncertainty; Gleeble; ETMT; creep testing;

# 1. Introduction

The ability to characterise materials under coupled thermal and mechanical loading is central to understanding and modelling manufacturing processes, such as welding, casting, forging, additive manufacturing, and high-temperature service degradation. In these environments, materials experience steep temperature gradients, rapid heating and cooling rates, and evolving mechanical constraints [1–3]. Yet, most conventional high-temperature mechanical testing methods rely on furnace-based heating and macroscopic specimens, implicitly assuming thermal homogeneity and quasi-equilibrium conditions. This mismatch between laboratory testing paradigms and real thermo-mechanical histories introduces systematic uncertainty when extrapolating measured properties to industrial processes or in-service behaviour.

Direct current thermo-mechanical testing (DC-TMT), based on resistive (Joule[†]) heating of the specimen itself, emerged as a response to this mismatch. By combining rapid, localised heating with simultaneous mechanical loading, DC-TMT enables the imposition of non-uniform, transient thermo-mechanical states that are inaccessible using conventional furnaces or induction-based systems [4,5]. Early implementations in the 1970s, most notably the Gleeble platform [6], demonstrated the potential of resistive heating to simulate welding heat-affected zones and high-temperature deformation mechanisms with unprecedented temporal control [7,8]. Subsequent developments extended this concept to miniature electro-thermal mechanical testing (ETMT), enabling high-throughput testing, remnant-life assessment, and alloy development when material availability is severely constrained [9,10].

Despite its widespread adoption across aerospace [11,12], automotive [13–16], nuclear, and advanced manufacturing research [17–20] (Figure 1), DC-TMT occupies an ambiguous position within the broader landscape of materials characterisation. On one hand, it offers unparalleled flexibility in heating rate, thermal cycling, and coupled loading. On the other hand, it inherently introduces temperature gradients, evolving gauge lengths, and strong sensitivity to specimen geometry, electrical contact, and thermal boundary conditions [14,21]. As a result, DC-TMT measurements cannot be interpreted as direct analogues of bulk, isothermal test data without careful consideration of these intrinsic features [22,23].

[†] The term Joule heating will be used hereafter in this document to mean direct current Joule heating, whereas when alternating current (AC) Joule heating will be discussed as induction heating.

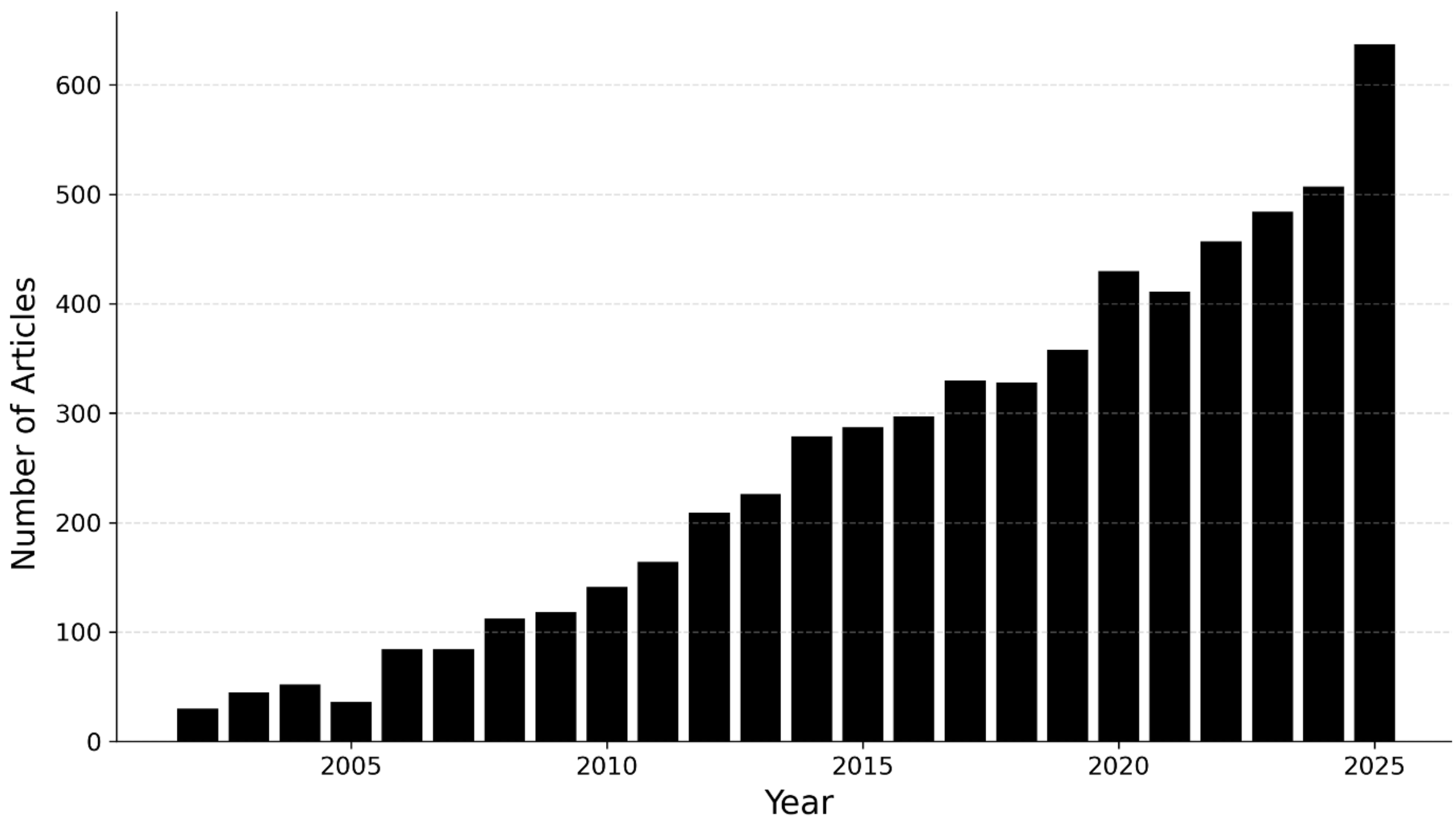

Figure 1: Publication trends for DC-based thermo-mechanical testing (including Gleeble and ETMT) from 2002 to 2025, based on a Scopus database search, showing rapid growth in recent years.

This ambiguity has led to two competing narratives in the literature. In one, DC-TMT is treated as a high-fidelity surrogate for conventional testing, with discrepancies attributed primarily to experimental error or calibration deficiencies [24]. In the other, DC-TMT is viewed as a fundamentally different experimental paradigm, probing localised, non-equilibrium material responses that are highly relevant to processing and service, but not directly interchangeable with standardised bulk measurements [25–27]. Resolving this tension requires moving beyond descriptions of hardware capability toward a principled understanding of uncertainty, representativeness, and inference in Joule-heated miniature testing.

Accordingly, this review does not aim to validate DC-TMT against conventional methods or propose it as a universal replacement for furnace-based testing. Instead, it critically examines the physical origins and hierarchy of uncertainty in DC-TMT, distinguishing between dominant, secondary, and conditional sources of error, and clarifying which material parameters can be meaningfully inferred under specific testing objectives. By synthesising developments in temperature control, strain measurement, environmental effects, and data interpretation, the review establishes when DC-TMT should be used, how its results should be interpreted, and where its limits are intrinsic rather than technological. This framework provides a foundation for exploiting DC-TMT as a powerful yet constrained tool for mechanism discovery, comparative studies, and model calibration in modern materials research.

## 2. Physical principles of DC thermo-mechanical testing

DC-TMT operates through direct resistive (Joule) heating of an electrically conductive specimen [28]. The instantaneous electrical power dissipated in the specimen is

$$P = IV = I^2R = \frac{V^2}{R}, \tag{1}$$

where $I$ is the applied current, $V$ is the voltage drop across the specimen, and $R$ is the electrical resistance of the heated section. Over a heating interval, the electrical energy supplied is

$$Q_{\text{in}} = \int_0^t I^2\, R(t)\, dt, \tag{2}$$

which reduces to $Q_{\text{in}} = I^2Rt$ only when both current and resistance remain constant over time, and this electrical input should be distinguished from the net thermal energy retained by the specimen, since heat is simultaneously lost to the specimen supports and surroundings during DC-TMT.

In DC-TMT, a direct current is passed through the specimen, and resistive dissipation produces rapid internal heating, often concentrated in the reduced gauge section. Temperature monitoring and control may be achieved using either contact methods, such as a spot-welded thermocouple, or non-contact methods, depending on the apparatus design, specimen geometry, and test conditions [29]. Likewise, heat extraction at the specimen ends depends on the grip configuration and may involve water cooling, forced-air cooling, passive heat sinking, or no active cooling. These boundary conditions commonly produce axial temperature gradients, with the central gauge region often corresponding to the highest-temperature part of the specimen.

A DC-TMT system (Figure 2) generally comprises a mechanical loading frame with a load cell and a precision actuator for force- or displacement-controlled testing. Together with the temperature measurement, these signals are used by the control unit to regulate both the mechanical and thermal histories, while recording force, displacement, temperature, current, and voltage drop [30]. The apparatus may additionally be installed within a vacuum chamber with gas-flow capability, allowing testing under vacuum or controlled atmospheres, as shown in Figure 2c. Figure 2 should therefore be regarded as an illustrative example of a common DC-TMT configuration rather than a universal arrangement.

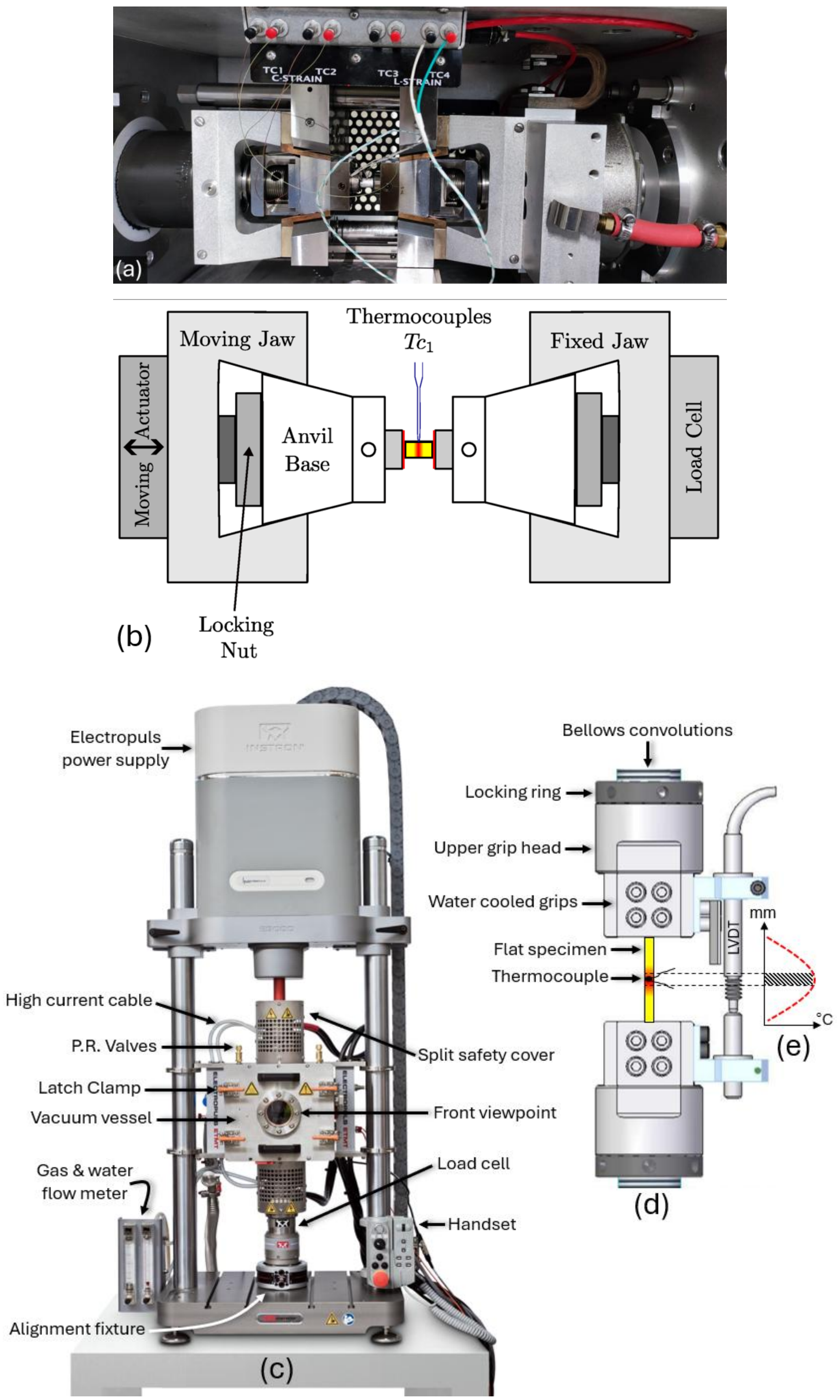

Figure 2: Representative DC-TMT configurations. (a) Aluminium 2017-T4 specimen mounted between ISO-T tungsten–carbide anvils in a Gleeble 3500 thermo-mechanical simulator, illustrating the electrical contact geometry during resistive heating. (b) Schematic of a compression configuration, showing jaw motion, load transmission, thermocouple placement, and cooling mechanisms in the Gleeble system. (c) Full view of an Instron ETMT system configured for high-temperature mechanical testing under vacuum or controlled gas environments. (d) Close-up schematic of a flat specimen clamped between water-cooled grips with a centrally attached thermocouple. (e) Schematic axial temperature profile along the specimen length, illustrating the formation of a central hot zone between cooled grips. Panels (a) and (b) are adapted from [31], licensed under CC BY 4.0.

# 3. Intrinsic constraints and sources of uncertainty in DC-TMT

In DC-TMT, discrepancies in reported material behaviour arise primarily not from statistical scatter, but from identifiable sources of experimental and interpretive uncertainty whose relevance depends on material class and testing objective. In this context, uncertainty refers to the bounded indeterminacy inherent to Joule-heated testing, rather than experimental error in the conventional metrological sense.

This section examines the principal sources of uncertainty in DC-TMT, the physical mechanisms by which they influence measured response, and the experimental conditions required for their effects to be meaningfully interpreted. Many of the experimental and interpretive practices discussed here have been established and refined through long-term metrology-led research, and are now widely adopted as best practice rather than corrective measures. Importantly, these uncertainty sources do not contribute equally to data interpretation. They are therefore discussed within an implicit hierarchy, synthesised explicitly in Section 5.

## 3.1. Specimen geometry and representativeness

While many of the uncertainties discussed in this section are present in bulk thermo-mechanical testing, their magnitude and impact are strongly amplified in miniaturised DC-TMT specimens, i.e., ETMT, due to reduced gauge volume, increased field gradients, and limited separation between measurement and boundary effects. Compared to standardised bulk Gleeble specimens, ETMT miniaturised specimens enable the characterisation of material behaviour using limited material volumes. Although such specimens are well suited for material ranking and comparative studies, they do not inherently produce universally transferable constitutive data, because several specimen-related uncertainty sources scale unfavourably at reduced length scales, including:

#### 3.1.1. Size effect

Miniaturised specimens often exhibit mechanical behaviour that deviates from that of their bulk counterparts due to scale-dependent phenomena. The term "size effect" is frequently used ambiguously to describe both geometric scaling effects associated with specimen dimensions and statistical effects related to microstructural representativeness, the latter of which are addressed separately under the representative volume element (RVE) effect in the following subsection.

For specimens thicker than approximately 0.25 mm, yield and ultimate tensile strengths are largely insensitive to size, whereas ductility parameters, particularly uniform and total elongation, are highly sensitive to gauge volume [27,32]. In ductile materials tested in tension, insufficient specimen thickness promotes premature necking and shear-dominated failure, whereas thicker specimens accommodate continued plastic deformation and fail predominantly by void nucleation and growth [33,34]. Specimens with different geometric scaling, often expressed through the ratio $L_0/\sqrt{A_0}$, deviate from conventional proportional specimens and exhibit differences in post-necking behaviour and total elongation [35], where $L_0$ is the initial gauge length and $A_0$ the initial cross-sectional area. Empirical relationships such as the Bertella–Oliver or Barba laws provide a means of normalising elongation data; however, their validity presupposes uniform deformation and does not eliminate the underlying geometric origin of size-dependent ductility [36,37].

As the specimen thickness decreases to the grain-level scale, grain boundaries, phase interfaces, and dislocation activity become disproportionately influential, leading to altered yield strength, ductility, and creep characteristics that complicate extrapolation to macro-scale performance. In particular, surface grains exhibit reduced dislocation storage due to dislocation escape at free surfaces, resulting in local softening and modified plastic deformation behaviour [32]. Moreover, even for specimens well above the grain scale, increasing thickness alters crack-front morphology from a tunnelled to a saddle-shaped profile [38], indicating that size effects persist beyond the microstructural scale. More details about size effects on the plastic behaviour of polycrystalline materials can be found in Ref [32]. Overall, at smaller length scales, specimen size also increases sensitivity to geometric and procedural deviations, such that manufacturing tolerances and handling-induced variability contribute disproportionately to scatter in measured tensile response [35].

As a rule of thumb, for a specimen with a nominal cross-section of approximately 1 mm, a thickness exceeding ~0.25 mm is a necessary condition for avoiding premature localisation and for interpreting measured ductility in a manner consistent with bulk behaviour. Consistent elastic modulus, yield strength, and ultimate tensile strength have been reported for additively manufactured miniaturised specimens of comparable thickness tested in accordance with ASTM E8 geometries (Figure 3a-c) [39]. However, for comparisons with bulk-sized specimens to be meaningful, the gauge cross-section must be statistically representative of the underlying microstructure. This requires the cross-section to contain a sufficient number of grains, typically on the order of at least ~20, and a thickness-to-grain size ratio $t/d_g$ exceeding ~10–12, below which measured properties increasingly reflect microstructural sampling rather than aggregated material response [39].

Representative specimen geometries and dimensional ranges commonly used in miniaturised DC-TMT studies are summarised in Table 1. These geometries do not eliminate size effects, particularly for ductility and tertiary creep behaviour, but they bound their influence sufficiently to enable comparative and trend-based interpretation. Below these limits, measured elongation and creep response increasingly reflect geometric and statistical artefacts rather than intrinsic material behaviour.

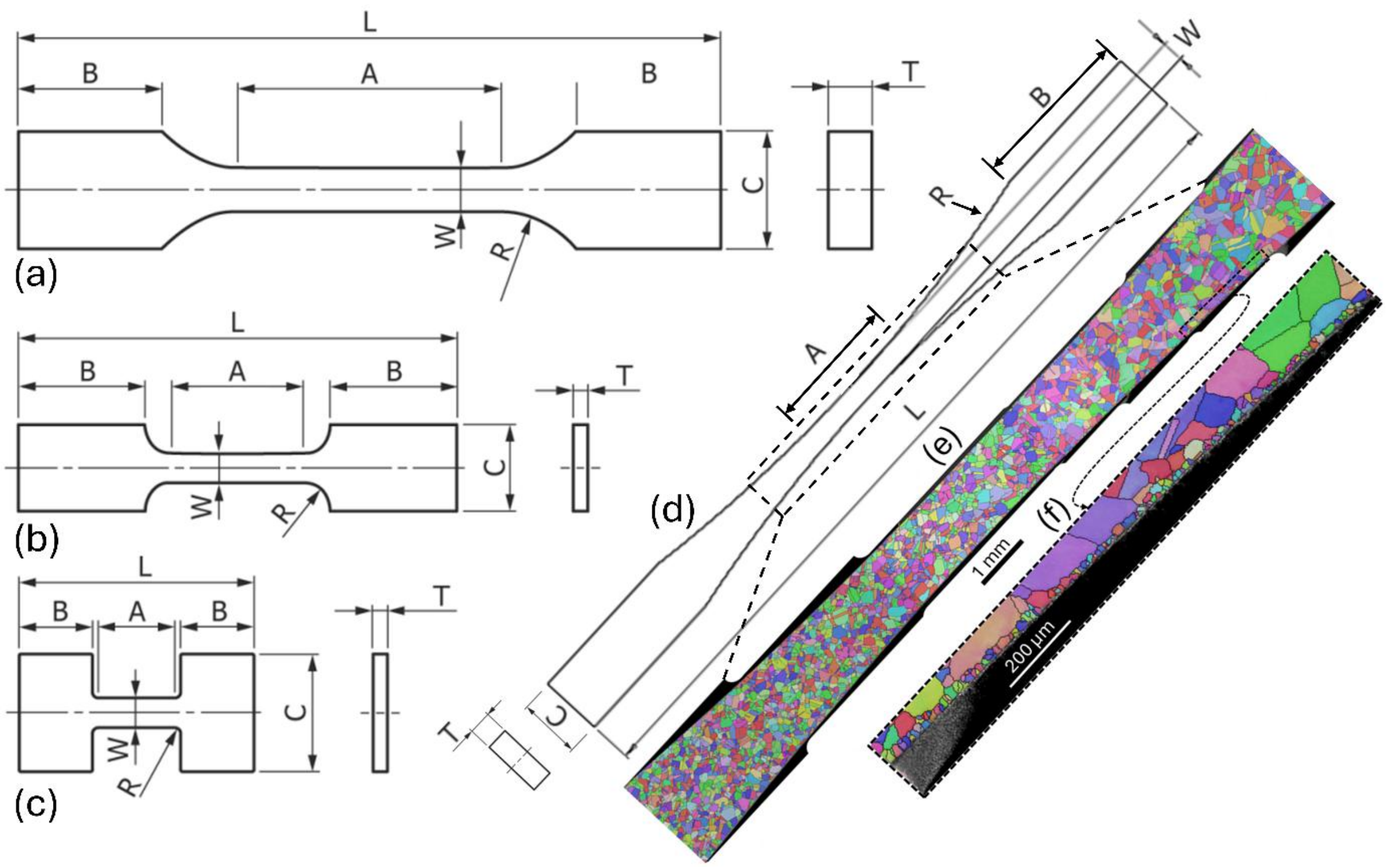

Figure 3: (a–c) ASTM E8 and (d) conventional waisted miniature ETMT specimen geometries illustrating commonly employed gauge and grip dimensions (see Table 1). (e) EBSD map of an IN718 miniature specimen and (f) localised recrystallisation near the specimen edge induced by machining, highlighting the sensitivity of miniaturised specimens to fabrication-related thermal and deformation effects.

Table 1: Representative dimensions of ASTM E8 and waisted miniature specimens used in DC-TMT studies.

| | Specimen (a), mm | Specimen (b), mm | Specimen (c), mm | Specimen (d), mm |
|---|---|---|---|---|
| Width (W) | 1.50 ± 0.02 | 1.00 ± 0.02 | 1.00 ± 0.02 | 1.10 ± 0.02 |
| Thickness (T) | 1.50 ± 0.02 | 0.50 ± 0.02 | 0.50 ± 0.02 | 1.00 ± 0.01 |
| Filet radius (R), min | 3.0 | 0.90 | 0.20 | 75.65 |
| Length of reduced section (A) | 10.0 ± 0.2 | 4.5 ± 0.2 | 2.6 ± 0.1 | 16.0 ± 0.2 |
| Width of grip section (C) | 4 | 3 | 4 | 3 |
| Length of grip section (B), min | 4.9 | 4.3 | 2.5 | 12 |
| Length (L), min | 25 | 15 | 8 | 40 |

### 3.1.2. Representative volume element effect

In essence, an RVE refers to the minimum volume of material that statistically captures the full heterogeneity of the microstructure, such as grain size, phase distribution and texture, so that the measured mechanical response is representative of the bulk material [40,41]. When the specimen size, particularly the gauge volume, becomes comparable to or smaller than the RVE, the mechanical response is increasingly dominated by a limited number of grains or phases, leading to non-representative stress–strain behaviour [42,43].

Additional uncertainty arises when the specimen volume does not statistically sample material inhomogeneity within the gauge section due to, for example, variations in grain size, phase morphology, and inclusion distribution [44]. This is especially problematic in anisotropic or textured materials like additively manufactured alloys [45,46], resulting in increased scatter and reduced reproducibility in properties like creep rate, yield strength, and ductility [47–49]. As a result, creep parameters derived from specimens lacking a representative gauge volume may reflect localised deformation mechanisms rather than intrinsic bulk behaviour, particularly at high homologous temperatures, e.g., grain boundary sliding may dominate over bulk diffusion, skewing lifetime predictions [50–53].

More generally, the extraction of activation parameters assumes that the measured response samples a statistically representative microstructural volume and that the governing deformation mechanism remains well defined over the measurement interval. In practice, activation parameters for thermally activated dislocation-mediated deformation are not measured directly but are inferred from families of creep or constant-strain-rate tests performed under controlled temperature and stress conditions. At the slip-system level, the shear strain rate may be written using the Orowan relation and an Arrhenius-type mobility law as [54]

$$\dot{\gamma} = \rho_m b v_0 \exp\left(-\frac{\Delta G(\tau)}{kT}\right), \qquad 3$$

where $\rho_m$ is the mobile dislocation density, $b$ is the Burgers vector, $v_0$ is a reference velocity, $k$ is Boltzmann's constant, $T$ is the absolute temperature, $\tau$ is the resolved shear stress, and $\Delta G$ is the stress-dependent activation barrier. The activation volume is defined as

$$V^* = -\left(\frac{\partial \Delta G}{\partial \tau}\right)_T = kT\left(\frac{\partial \ln \dot{\gamma}}{\partial \tau}\right)_T. \quad 4$$

For macroscopic experiments, this is commonly estimated from the stress dependence of the measured strain rate at constant temperature, giving an apparent activation volume

$$V_{\mathrm{app}}^* = MkT\left(\frac{\partial \ln \dot{\varepsilon}}{\partial \sigma}\right)_T. \quad 5$$

where $\sigma$ and $\dot{\varepsilon}$ are the applied stress and axial strain rate, respectively, and $M$ is a factor relating macroscopic and resolved shear quantities. Over a limited stress range, the barrier may be linearised as $\Delta G(\tau) \approx \Delta G_0 - \tau V^*$, yielding

$$\dot{\gamma} \approx \rho_m b v_0 \exp\left(-\frac{\Delta G_0 - \tau V^*}{kT}\right). \quad 6$$

Likewise, an apparent effective activation barrier may then be estimated from temperature-dependent strain-rate data at constant stress using

$$\Delta G_{\mathrm{eff}}(\sigma) \approx -k\left(\frac{\partial \ln \dot{\varepsilon}}{\partial (1/T)}\right)_\tau. \quad 7$$

These experimentally derived quantities should, however, be interpreted with caution. Their extraction assumes that the gauge section samples a statistically representative microstructural volume; that deformation remains spatially uniform; that a single rate-controlling mechanism dominates over the analysed temperature and stress interval; and that the underlying microstructural state, including mobile dislocation density, obstacle population, grain structure, and phase constitution, does not evolve significantly during the measurement. It also assumes that the kinetic prefactor varies only weakly with temperature, that the stress dependence of the barrier can be reasonably linearised over the fitting range, and that the measured temperature and stress are homogeneous and accurately known. In DC-TMT, these assumptions may be compromised by steep thermal gradients, control lag, Joule-heating non-uniformity, localisation, phase transformation, recovery, recrystallisation, or enhanced contributions from grain-boundary-mediated deformation at high homologous temperature. Under such conditions, the extracted $V_{\mathrm{app}}^*$ and $\Delta G_{\mathrm{eff}}$ are more appropriately regarded as apparent, test-configuration-dependent parameters rather than intrinsic bulk material constants.

For example, variance in grain size distributions directly contributes to uncertainty in measured activation parameters, including apparent activation energy [55]. Additionally, creep is frequently governed by grain-boundary-mediated mechanisms, such as diffusion and sliding; the grain size-to-specimen size ratio becomes a critical parameter. In miniaturised specimens, grain boundary contributions may be exaggerated [27], particularly in additively manufactured materials with columnar grains aligned along the build direction [56].

Consequently, when the gauge volume approaches the material's RVE, microstructural characterisation of the tested gauge section becomes mandatory. Grain size distributions, phase fractions, and texture should be quantified and reported in conjunction with mechanical data. Without this information, parameters derived from DC-TMT – such as activation energy, activation volume, or inferred creep mechanisms – cannot be assumed to represent bulk behaviour, even if stress–strain curves appear reproducible [57,58].

#### 3.1.3. Specimen shape

The specimen shape, whether flat or waisted (Figure 3d) directly controls deformation localisation and thermal field distribution in DC-TMT, and therefore conditions the interpretability of measured mechanical response. Flat specimens possess a uniform cross-section that promotes diffuse necking across the gauge length, particularly in ductile materials. The resulting non-localised deformation complicates stress–strain interpretation, while sharp corners associated with square miniature cross-sections act as preferential crack initiation sites, accelerating crack initiation and propagation [40,59], which can be avoided by using cylindrical specimens instead. By contrast, waisted, dogbone (Figure 3a), and bow-tie (Figure 3c) specimens introduce a reduced central cross-section that enforces deformation localisation and constrains failure to the intended gauge region. This constraint becomes necessary when testing alloys exhibiting anomalous temperature-dependent strengthening, such as many superalloys, for which insufficient localisation can shift deformation outside the intended measurement zone [60,61].

Alternative specimen geometries have been proposed to impose specific mechanical or thermal constraints, including wedge designs intended to suppress misalignment, winged geometries aimed at flattening axial temperature gradients and improving thermal homogeneity [62], and notched profiles used to enforce stress localisation [22]. Generally, Specimen shape also modifies the imposed temperature field, with waisted geometries producing steeper axial temperature gradients than flat specimens, an effect discussed in detail in Section 3.3.

#### 3.1.4. Fabrication challenges

The production of defect-free, miniaturised specimens with consistent geometry is inherently demanding, requiring high-precision microfabrication techniques while avoiding the introduction of edge defects, such as burrs, microcracks and microstructural effects, or plastic deformation. Variability in machining accuracy, surface finish, and dimensional tolerances can introduce significant test-to-test scatter. Unlike standard-sized specimens, where a deviation of ±0.1 mm is negligible, in miniaturised specimens, this margin can represent a substantial portion of the cross-sectional area [63]. Additionally, small geometric inconsistencies, such as edge roundness or notch radius, can significantly impact outcomes, particularly in fatigue testing, where micro-notches can serve as crack initiation points [64].

Miniaturised specimens amplify the impact of surface phenomena such as oxidation, contamination, and roughness, because the damaged layer can alter the local stress distribution and mechanical response, most notably by promoting premature necking and non-representative deformation behaviour [4,39,65]. Furthermore, residual stresses and thermal effects during waisting or localised grinding can induce recrystallisation (Figure 3f) or phase transformation in the heat-affected zone [66,67].

Electrical discharge machining (EDM), while widely used for its dimensional accuracy, often leaves behind a recast layer, up to 50 µm thick, containing microvoids and brittle regions, which can distort ductility measurements if not adequately removed [35]. EDM can also introduce hydrogen into the surface layer, particularly in steels and other hydrogen-sensitive alloys [68,69]. This hydrogen uptake can lead to embrittlement, altering fracture behaviour and reducing ductility, especially under tensile or creep loading [70].

Fabrication-induced thermal and chemical artefacts constitute a primary source of uncertainty in miniaturised testing and must be controlled for the measured response to remain interpretable. Limiting heat input during specimen manufacture, eliminating hydrogen uptake associated with EDM, and removing recast surface layers are necessary conditions for avoiding fabrication-dominated ductility and fracture behaviour. Verification of post-fabrication microstructural stability, for example, through electron backscatter diffraction (Figure 3f) is likewise required to ensure that the tested gauge volume reflects the intended material state rather than manufacturing-induced transformations, particularly when deformation occurs at temperatures insufficient to fully anneal prior damage [24].

#### 3.1.5. Alignment and gripping

Accurate mechanical characterisation of specimens demands ensuring concentricity of the gripping axis and meticulous specimen alignment within the test apparatus. Even minor misalignments can introduce unintended bending or torsional stresses due to the low flexural rigidity, significantly distorting axial load measurements and compromising the validity of stress–strain data [71,72]. This is especially critical in miniaturised specimens, where geometric sensitivity is amplified [64]. Alignment must therefore be verified under load using stiff, non-compliant reference specimens, such that off-axis moments arise in the test frame rather than in the specimen itself. Only once this condition is satisfied can subsequent deformation be attributed to the material response within the gauge section.

Gripping concentricity can be ensured by first using a rigid calibration specimen with negligible compliance and applying a high axial preload approaching the machine's allowable limit. Under this condition, any misalignment manifests as off-axis forces and moments borne by the grips and load frame rather than by specimen deformation, enabling iterative adjustment of the fixture until the axial force is maximised with minimal bending. The test specimen is then installed with minimal initial contact pressure, centred within the grips, and aligned such that the thermocouple is positioned at the midpoint of the gauge length. Grip tightening is performed progressively using a criss-cross pattern to ensure uniform clamping. For slender specimens, a low-load seating step may be applied before final tightening to promote self-centring. Nevertheless, a small residual bending component remains unavoidable due to system compliance asymmetries.

Grip design must ensure secure load transfer without introducing stress concentrations, surface damage, or geometric distortion, as improper clamping alters necking behaviour and can precipitate premature failure. In miniaturised flat specimens, gripping-induced stress heterogeneity is amplified by geometric sensitivity, such that even small misalignments or compliance mismatches produce unequal stress distributions across the gauge section. In brittle or low-ductility materials, this promotes premature fracture, whereas in ductile materials, local yielding initiates preferentially on the higher-stress side, distorting the measured stress–strain behaviour [35,71,73].

Pinned gripping configurations further introduce uncertainty through hole deformation and evolving contact conditions, which directly contaminate elongation measurements in small-scale specimens, even when a 3:1 area ratio between the grip section and gauge section is maintained [33]. Additionally, it is challenging to create precise holes in specimens, and these holes may distort during testing, which can lead to uncertainties in the elongation data [35,71,73]. Shoulder-supported grips (Figure 4), or fixtures incorporating compliant interfaces, reduce local stress concentrations and are required where pin-induced deformation would otherwise dominate the measured response. To minimise deformation in the grip section, shoulder-supported grips maintain a 3:1 grip-to-gauge section area ratio, eliminating the need for holes. Overall, in situ imaging techniques, such as digital image correlation, provide an independent means of verifying alignment and detecting off-axis deformation that is not observable from force–displacement data alone [71]. More details about the effect of gripping on the specimen’s thermal profile are discussed in 3.3.3.

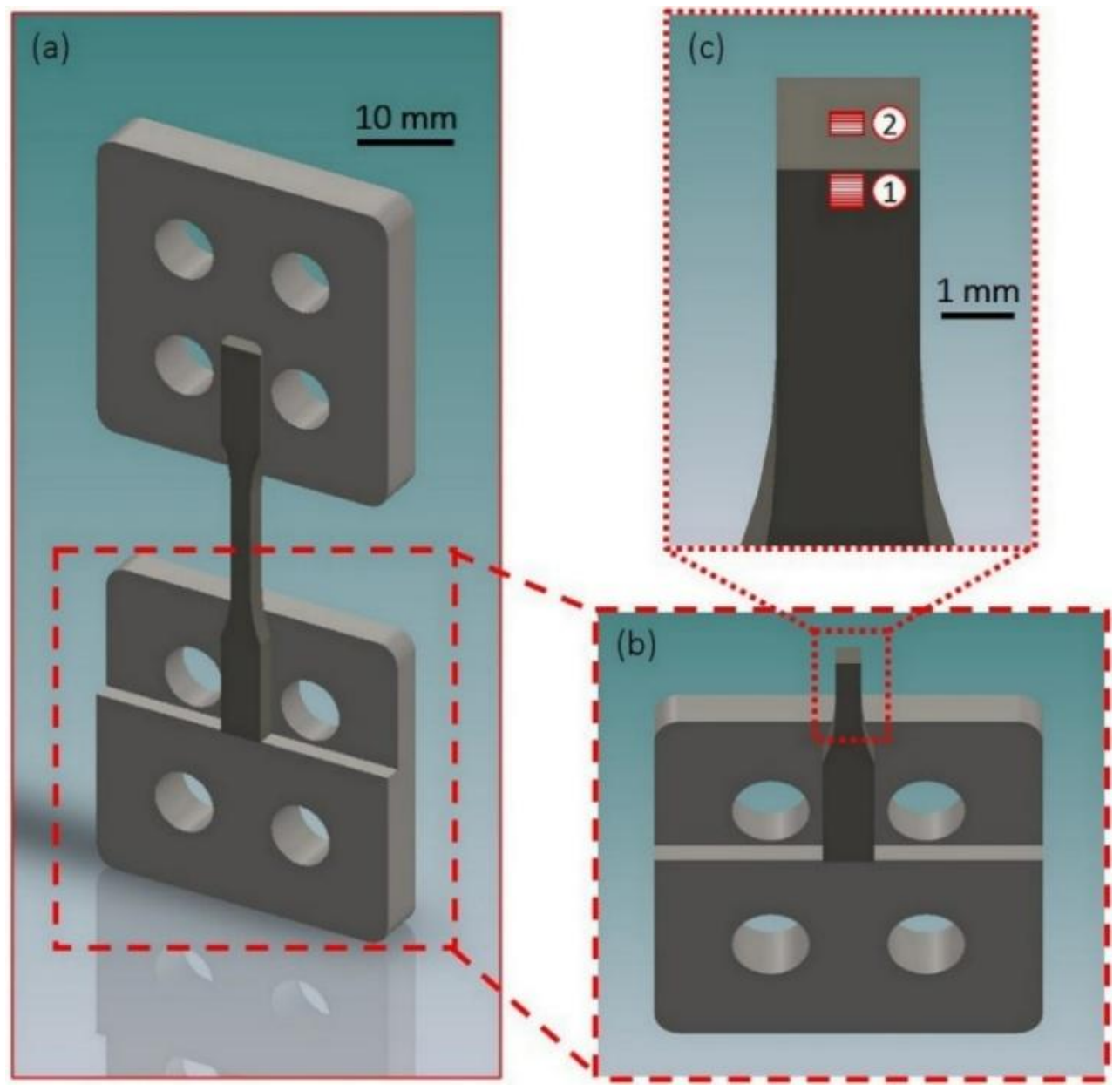

Figure 4: (a) ETMT specimen and shoulder-supported grip assembly during testing. (b) Cross-sectional view through the specimen centre illustrating grip–specimen contact. (c) Enlarged view indicating regions analysed by HR-EBSD and cECCI, highlighting microstructural sensitivity near grip-constrained zones. Adapted from [74].

## 3.2. Temperature control and measurement fidelity

Accurate temperature measurement and control are necessary conditions for interpreting thermo-mechanical response in DC-TMT, because temperature enters both constitutive behaviour and damage evolution in a strongly non-linear manner. Temperature in DC-TMT may be measured using either contact methods, most commonly spot-welded thermocouples, or non-contact methods such as pyrometry, infrared imaging, and phosphor thermometry. The choice depends on specimen geometry, surface condition, optical access, heating rate, and whether the objective is single-point closed-loop control or spatial mapping of the thermal field.

In the DC-TMT literature surveyed here, bare-wire thermocouples fused into a bead and attached near the nominal centre of the specimen gauge (Figure 5a-d) remain the most commonly reported feedback sensor for closed-loop temperature control. Their importance in DC-TMT is not merely instrumental because the specimen itself acts as the heating element, a welded thermocouple interacts directly with the Joule-heated system and can introduce DC-specific uncertainty through parasitic electrical pickup, local thermal shunting at the weld, interaction with the specimen's electrical circuit, and perturbation of the local temperature field. These effects become more severe in miniature specimens, under high current densities, and in the presence of steep thermal gradients.

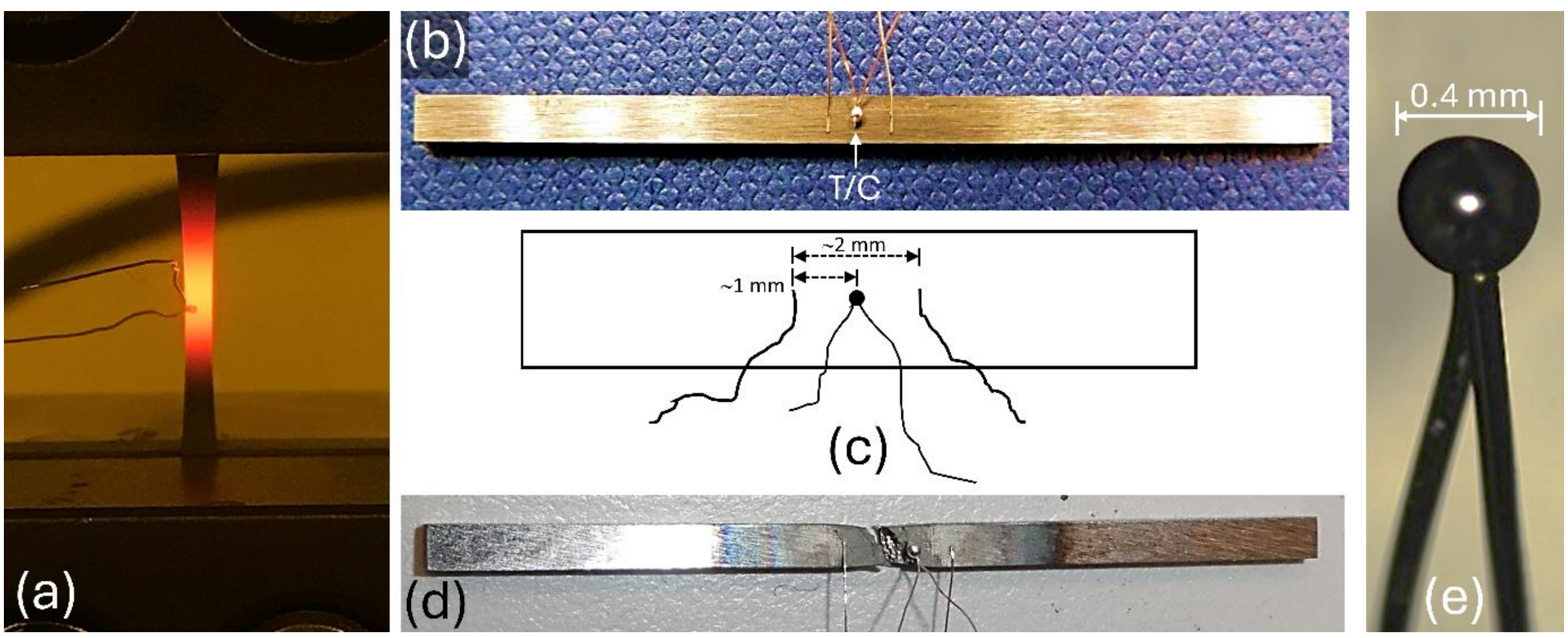

Figure 5: Example of one common implementation of a thermocouple-based temperature measurement arrangement in DC-TMT. (a) Thermocouple used for temperature measurement and control during high-temperature testing. (b-d) Attachment of a thermocouple near the centre of a miniaturised specimen with adjacent potential/resistivity leads, shown before and after testing. (e) Example of a fused thermocouple bead; a round, symmetric bead is generally indicative of clean fusion and more reproducible attachment.

Thermocouple-related uncertainty, therefore, arises from a combination of thermal conduction along the wires, junction geometry, contact resistance at the attachment point, and electrical parasitics associated with current flow through the specimen. Unless these effects are constrained, the measured thermocouple signal does not uniquely represent the local specimen temperature. Reducing this uncertainty requires stable thermal and electrical coupling at the junction, reproducible attachment geometry, appropriate lead routing, and calibration under representative current, temperature, and specimen-geometry conditions. The principal considerations are discussed below.

### 3.2.1. Thermocouple type

Under DC-TMT conditions, thermocouple selection is governed less by nominal temperature range alone than by susceptibility to electrical interference, chemical drift, and survivability during repeated thermal cycling [75,76]. In practice, base-metal thermocouples such as Type

K (chromel–alumel) are widely used at low to intermediate temperatures, whereas noble-metal thermocouples such as Type R (87%Pt/13%Rh–Pt) or S are preferred at higher temperatures or where signal stability under Joule heating is critical [75,77].

Under direct current heating, thermocouple signals can be distorted by parasitic voltages superimposed on the intended thermoelectric output. This effect is particularly problematic for high-output thermocouple types and becomes more pronounced at high current density and temperature. Experimental comparisons have shown that Type K and Type R thermocouples can agree reasonably within approximately 10 °C up to about 1000 °C, but larger deviations emerge at higher temperatures under DC heating, with Type K underestimating temperature by 36 – 51 °C in one study [27]. In addition, Type K thermocouples are more susceptible to chemical drift under aggressive atmospheres, including oxidation-related degradation of the Ni–Cr leg [78], which can further bias the measured temperature [79]. In the superalloy case reported in [27], temperature errors of this magnitude propagated directly into microstructural interpretation, altering predicted γ′ precipitate fractions by approximately 0.03 at 980 °C and 0.13 at 1100 °C.

Parasitic electrical contributions may be reduced by temporally decoupling current flow from temperature measurement, for example, by briefly interrupting the DC and extrapolating the initial cooling transient back to the moment of current cut-off [26]. Alternatively, dual-thermocouple configurations employing sensors with different voltage sensitivities can provide partial separation of thermal and electrical contributions to the measured signal while maintaining closed-loop control [27]. These approaches reduce electrical interference, but they do not eliminate coupling between Joule heating and temperature measurement; the residual uncertainty remains material-, geometry-, and temperature-dependent.

### 3.2.2. Lead configuration

The thermocouple lead arrangement in DC-TMT influences both signal integrity and the local thermal field. Wire diameter governs the trade-off between response time, mechanical robustness, and measurement perturbation[80]. Very fine wires (≈0.05–0.13 mm) provide low thermal mass and rapid response but are less durable during attachment and repeated thermal cycling. Larger wires (>0.5 mm) are mechanically more robust but increase thermal conduction and local heat sinking [80]. No single wire diameter is universally optimal;

thermocouple size should be selected relative to specimen dimensions, heating rate, and required junction robustness, with the smallest wire that provides stable attachment and acceptable signal quality generally preferred. For millimetre-scale specimens, wire diameters around 0.1 mm may provide a practical compromise for Type R thermocouples, whereas smaller specimens typically require proportionally finer wires to minimise thermal shunting and local perturbation.

Lead integrity is a necessary condition for temperature interpretability. Temperature gradients along the leads, for example, between the hot junction and cooler regions near the grips or connectors, can generate additional thermoelectric offsets that are superimposed on the intended signal (Figure 6b) [81]. In dynamic experiments, the finite thermal response time of the thermocouple junction introduces additional temporal lag, leading to apparent discrepancies relative to faster measurement techniques and obscuring transient thermal behaviour [75]. Additionally, defects, inhomogeneities, unintended electrical junctions, oxidation, or contamination along the wire length can further modify the local Seebeck response and produce time-dependent drift [82]. Also, repeated bending, cold work, or thermal cycling may likewise alter thermoelectric behaviour non-uniformly along the lead path, so that different parts of the same wire no longer respond identically under an imposed thermal gradient.

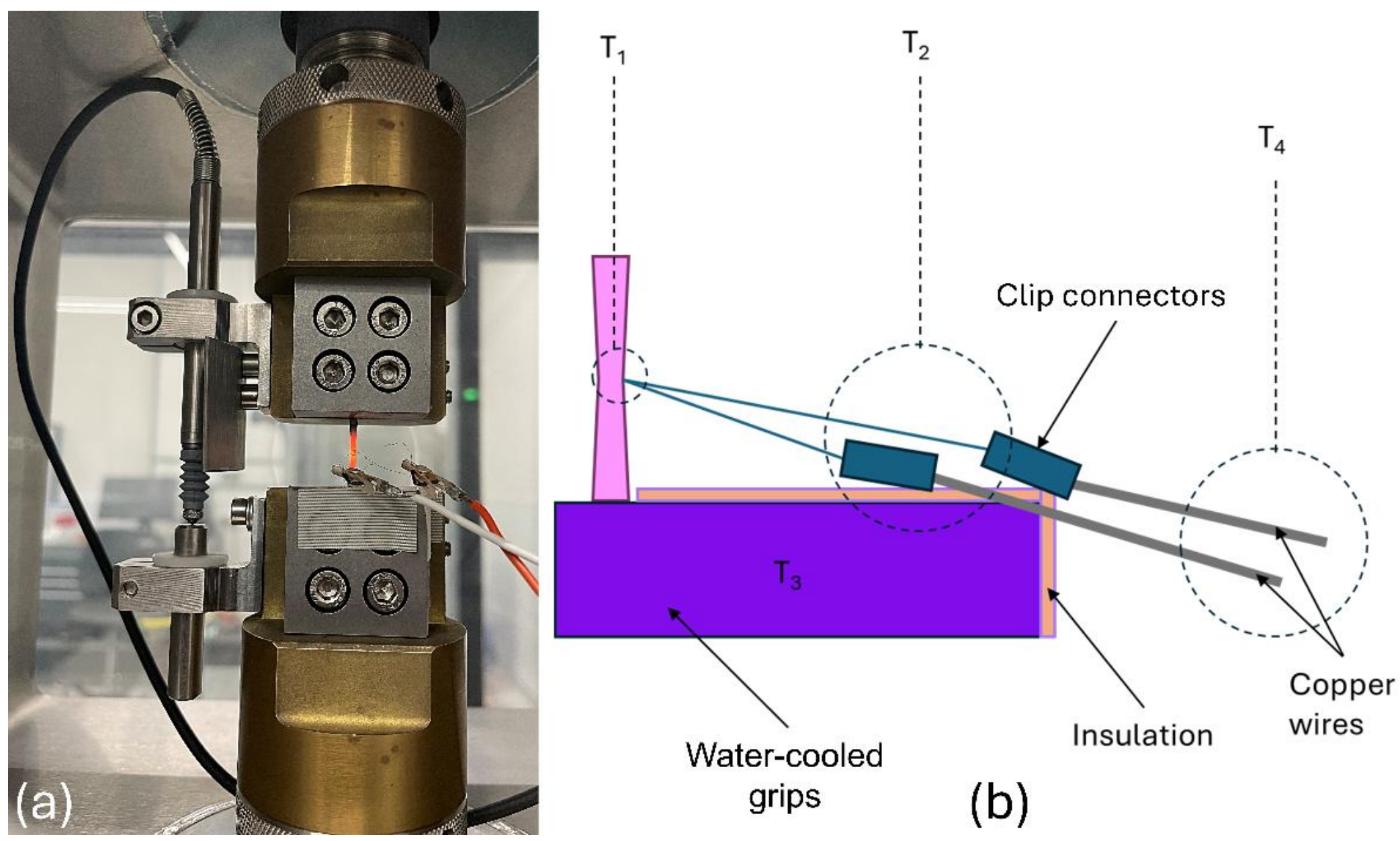

Figure 6: Example of a thermocouple wiring arrangement used to illustrate unintended thermoelectric junctions and parasitic voltage pathways in DC-TMT. (a) Photograph of one laboratory implementation

showing thermocouple lead routing near the specimen and grips. (b) Schematic indicating the hot junction at the specimen ($T_1$), an intermediate junction along the lead path ($T_2$), the grip/body temperature ($T_3$), and the cold or measurement junction ($T_4$), highlighting locations where additional thermoelectric offsets and thermal shunting may arise.

Additional spurious voltages may arise at connections between thermocouple wires, extension leads, connectors, or measurement electronics, while long or poorly routed leads may also pick up electromagnetic interference under Joule-heated conditions [83]. Chemical or galvanic potentials at dissimilar-metal interfaces or corroded contacts can introduce further offsets [84]. Accordingly, lead routing, electrical isolation, polarity verification, minimisation of unnecessary junctions, and avoidance of excessive bending or contamination are not secondary wiring details, but part of the temperature-measurement uncertainty budget in DC-TMT. These measures reduce spurious voltage generation, although they cannot fully eliminate drift during long-duration or very high-temperature testing.

### 3.2.3. Junction geometry and attachment

The geometry of the thermocouple bead directly affects thermal response and measurement fidelity. Bead size and shape determine the effective thermal mass of the junction (Figure 5e); larger beads increase thermal inertia and respond more slowly to rapid temperature changes, introducing temporal lag relative to the true specimen surface temperature [85]. In transient experiments, this can cause the thermocouple to read below the actual temperature [86]. For fused thermocouples, particularly noble-metal systems such as Pt/Pt–Rh, the ratio between bead diameter and wire diameter governs the balance between junction robustness and thermal response, and excessive bead growth disproportionately increases response time and measurement lag.

Reliable temperature inference further requires stable thermal and electrical coupling between the thermocouple junction and the specimen surface. Poor fusion, internal voids, or oxide inclusions within the bead can modify the effective Seebeck response relative to the parent wires and introduce systematic temperature error [87]. Specimen surfaces should therefore be prepared to ensure clean, reproducible attachment, for example, by light surface abrasion followed by solvent cleaning before spot welding. Achieving a clean weld is especially important in highly conductive materials and in very small specimens, where current spreading and heat dissipation during attachment make junction quality strongly geometry dependent. Accordingly, welding parameters should be optimised on sacrificial specimens representative of the actual test geometry.

Interfacial thermal resistance between the bead and the specimen can also cause the junction temperature to deviate from the true local surface temperature, particularly under high heating rates or steep thermal gradients. In miniature specimens, where the gauge volume is small, local microstructural modification or damage introduced during thermocouple attachment may directly influence the mechanical response, making it difficult to separate material behaviour from temperature-measurement artefacts if such effects are not controlled. Because the thermocouple is often attached near the hottest central region, which may also coincide with the fracture-prone zone (as in Figure 5d), post-test assessment may be required to confirm that the weld did not influence localisation or failure.

The thermocouple bead is also susceptible to chemical and mechanical degradation during service. Oxidation, corrosion, or grain growth within the welded junction can gradually alter thermoelectric properties, while repeated thermal cycling may induce embrittlement or cracking that further destabilises the signal [88]. Clean fusion, controlled bead geometry, and stable thermal contact are therefore necessary to reduce temperature-measurement bias, but they do not eliminate drift entirely, especially during long-duration testing or exposure in air.

#### 3.2.4. Signal conditioning and conversion

Thermocouple temperature measurement in DC-TMT requires conversion of the Seebeck-generated electromotive force into temperature using standard reference relations (e.g., ITS-90 [89]) or tabulated calibration data [75]. In practice, however, the dominant limitation is rarely the mathematical linearisation itself. Rather, under Joule-heated conditions, the measured signal may be contaminated by parasitic voltages, temperature gradients along the measurement circuit, alloy-dependent drift, and uncertainty in cold-junction compensation, all of which can introduce systematic temperature offsets [75,90].

For this reason, thermocouples in DC-TMT should be treated as system-dependent sensors rather than absolute thermometers. Reliable use therefore requires calibration and validation under representative specimen geometry, thermal history, and electrical configuration, with explicit recognition that the inferred temperature is a bounded estimate rather than an exact state variable. In high-gradient or high-temperature tests, signal conditioning and conversion should be supported by independent verification, ideally using an additional contact sensor or a non-contact temperature-measurement method.

### 3.2.5. Calibration, drift, and traceability

Thermocouple calibration defines the lower bound of temperature uncertainty in DC-TMT. Under controlled conditions, achievable accuracy is typically on the order of ±5 °C, corresponding to approximately 0.2–0.5% and 0.5–1% of the reading for Type R and Type K thermocouples, respectively [91]. In DC-TMT, however, the practical validity of this calibration must be assessed under the actual electrical and thermal test conditions, because parasitic voltages, thermal gradients, and drift may dominate the uncertainty budget. One useful check is a comparison between thermocouples with dissimilar EMF–temperature sensitivities operating under nominally identical thermal conditions. Since Type K thermocouples generate substantially higher thermoelectric voltages than Type R at the same temperature, systematic divergence between their readings that increases with temperature may indicate parasitic electrical contributions, although compositional degradation of the Type K thermocouple must also be considered [27].

Additional validation can be obtained by reference to materials exhibiting well-defined intrinsic transitions. Such "structural thermocouples" exploit abrupt changes in electrical or magnetic response at known temperatures, thereby providing fixed internal reference points for the full measurement chain rather than for the thermocouple alone (see Section 3.5.2). For example, pure Ni experiences the Curie transition at 358.28 ± 0.04 °C (Figure 7a) [92,93], pure Ti experiences a phase transform from α (HCP) to β (BCC) at 882.5 °C (Figure 7b) [94,95], and eutectoid steel experiences a phase transformation at approximately 727–742 °C, depending on composition (Figure 7c) [96].

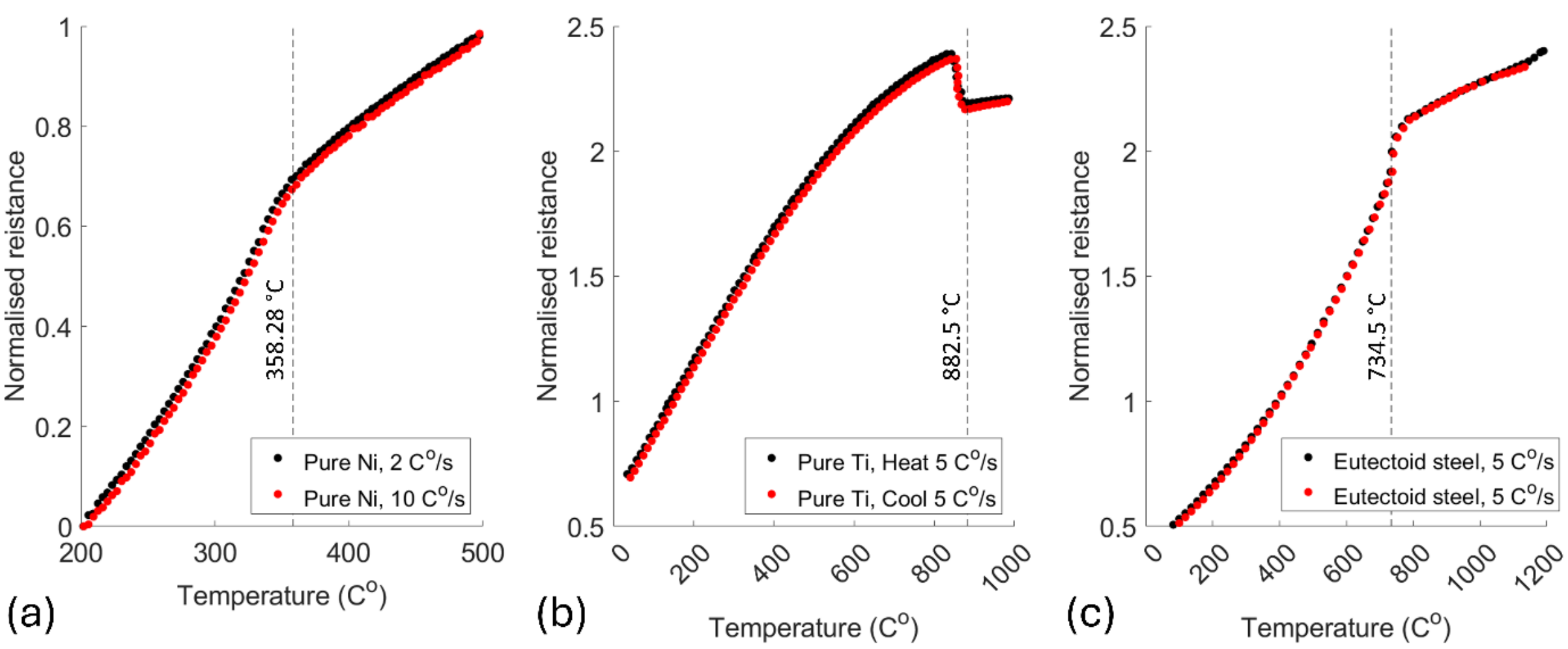

Figure 7: Normalised electrical resistance as a function of temperature for (a) pure Ni, (b) pure Ti, and (c) eutectoid steel, illustrating intrinsic magnetic or phase transitions that can serve as internal reference points for validation of temperature measurement in DC-TMT.

Independent temperature measurement techniques should also be used, where possible, to cross-validate thermocouple readings and assess spatial temperature distributions. In this respect, non-contact methods such as pyrometry, infrared imaging, and phosphor thermometry are especially valuable in DC-TMT because they provide information that a single welded thermocouple cannot, namely, independent verification of absolute temperature and spatial mapping of the thermal field [97]. Each method, however, introduces its own uncertainty sources. Radiative methods depend on emissivity, which is not generally constant but may vary with temperature, wavelength, oxidation state, surface roughness, and viewing angle [98,99]. Consequently, single-colour pyrometry and infrared thermography can suffer large systematic errors if emissivity is assumed incorrectly or evolves during the test. This is particularly relevant in DC-TMT, where oxidation, surface-condition changes, and thermal gradients may develop during heating.

Moreover, two-colour pyrometry can reduce sensitivity to uniform emissivity changes by comparing radiation intensities at two wavelengths, but it does not remove emissivity-related uncertainty completely, since its validity depends on assumptions such as near-grey-body behaviour or a stable emissivity ratio over the selected wavelength pair, as well as a common unobstructed field of view for both channels [100,101]. Thermal imaging additionally remains sensitive to reflected background radiation, line-of-sight effects, and spatial variations in surface condition [98,99]. Phosphor thermometry is much less sensitive to emissivity and background thermal radiation, but requires stable optical access, a suitable phosphor coating, and careful calibration of decay time or intensity ratios [100,102].

Consequently, disagreement between contact and non-contact measurements cannot be attributed unambiguously to thermocouple drift, Joule-heating artefacts, or true thermal non-uniformity unless emissivity evolution, optical geometry, and calibration are all controlled. For this reason, temperature traceability in DC-TMT is strongest when contact and non-contact methods are used in a complementary manner, as thermocouples provide practical closed-loop feedback, while independent techniques verify absolute temperature and spatial uniformity under the actual specimen geometry and electrical configuration.

## 3.3. Spatial temperature gradients and non-isothermal deformation

In DC-TMT, resistive (Joule) heating inherently produces non-uniform axial temperature distributions, with the maximum temperature located near the specimen centre and decreasing toward the water-cooled grips [98,102–104], as illustrated in Figure 8. These temperature gradients directly influence oxidation kinetics, damage accumulation, and crack propagation under thermo-mechanical fatigue conditions [105]. The axial temperature profile typically exhibits a flattened central region of relatively uniform temperature, often referred to as the "effective" gauge length (Figure 8c), bounded by regions of increasing gradient toward the grips. The extent and steepness of this profile depend on the specimen geometry, absolute temperature, and heat loss mechanisms. At elevated temperatures, radiative heat loss increasingly dominates over axial conduction. Because radiative emission scales with the fourth power of absolute temperature ($\propto T^4$, Stefan–Boltzmann law), heat dissipation is enhanced near the hottest region, resulting in a broader and flatter temperature distribution within the central gauge section. In other words, as radiative losses outweigh conductive cooling, the temperature gradient becomes less steep, producing a larger, more uniform hot zone.

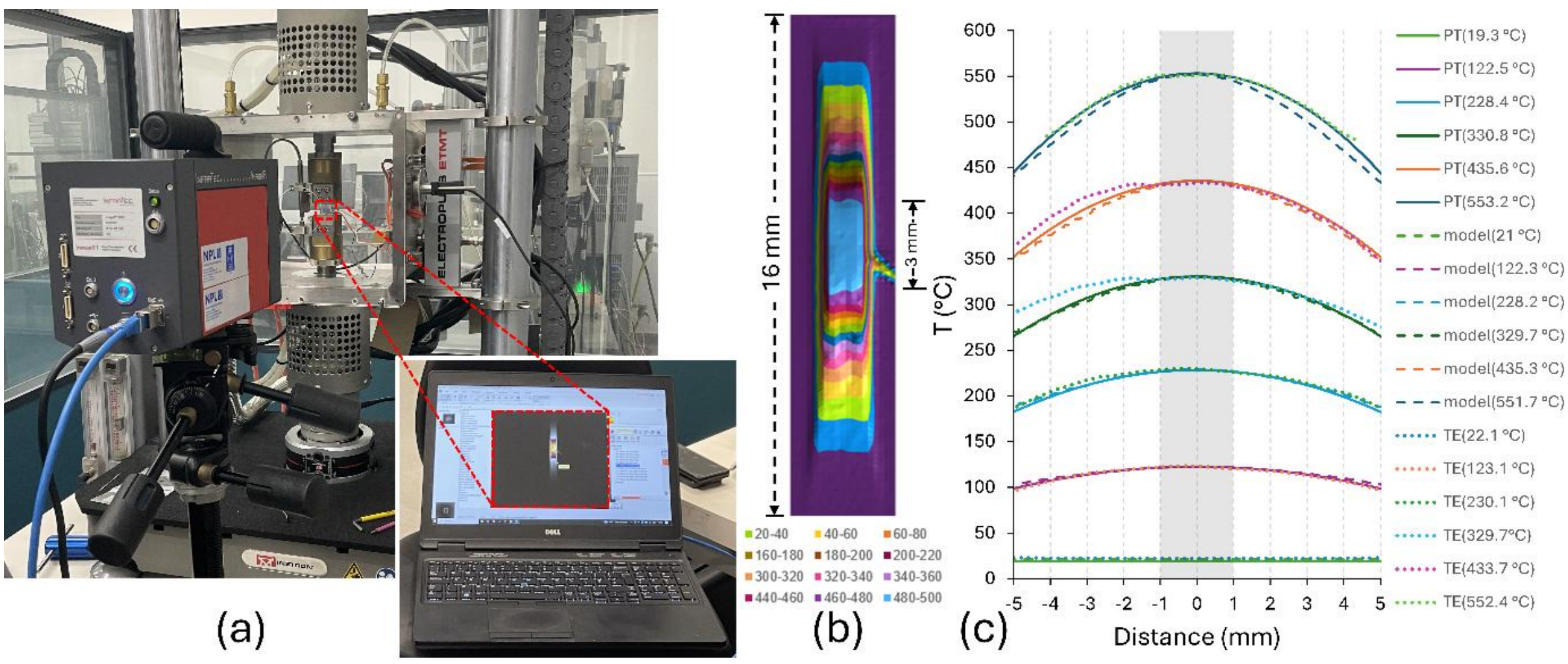

Figure 8: (a) High temperature testing of IN718 waisted specimen at 500 °C, while using thermal imaging. (b) The temperature gradient on a waisted specimen as measured using thermal imaging. (c) Comparison of the fitted vertical temperature profiles for the IN718 waisted specimen characterised using phosphor thermometry (PT), COMSOL model, and thermal imaging (TE) [98]. For more details about the model, see Ref [98]. The shaded area highlights the 2 mm central region, which has a relatively homogeneous temperature distribution.

Nevertheless, even under the assumption of a homogeneous microstructure, the resulting temperature distribution is sensitive to several additional factors, including:

### 3.3.1. Thermo-electrical property coupling

In DC-TMT, the specimen itself acts as the heating element; consequently, its intrinsic material properties directly govern the resulting temperature field (see Appendix A). Two distinct but coupled processes must be considered: first, the spatial distribution of heat generation by Joule dissipation, which is governed primarily by electrical resistivity and current distribution; and second, the subsequent redistribution of that heat, which is governed mainly by thermal conductivity and heat capacity.

Electrical resistivity is the primary parameter controlling Joule heat generation. For a given current distribution, regions of higher resistivity dissipate more electrical power per unit volume. As a result, high-resistivity alloys, such as many stainless steels and Ni-based alloys, can generate substantial internal heating at relatively modest currents, whereas low-resistivity metals such as Cu and Al require higher currents to reach the same temperature. In the latter case, the larger imposed current can increase sensitivity to local resistance variations arising from geometry, surface condition, electrical contact, or microstructural heterogeneity. Because resistivity is strongly temperature-dependent, the heating rate and heat-generation profile evolve dynamically during testing [106,107].

Thermal conductivity then governs how efficiently the generated heat is redistributed along the specimen gauge length. High-conductivity materials, such as Cu and Al, spread heat rapidly and therefore tend to develop flatter axial temperature gradients between the gauge centre and the grips. In contrast, materials with lower thermal conductivity, such as Ti alloys and many steels, redistribute heat less efficiently and therefore develop steeper axial temperature gradients, such that a single-point temperature measurement may not be representative of the thermal state across the deforming gauge volume. This directly affects the homogeneity of the stress-strain field and the reliability of measured mechanical properties.

Specific heat capacity further influences the temporal evolution of the temperature field. Materials with high specific heat require greater energy input to achieve a given temperature rise, slowing thermal transients and damping overshoot, but potentially introducing control lag. Conversely, low-heat-capacity materials respond more rapidly to applied power, increasing susceptibility to thermal oscillations and control instability. Phase transformations and magnetic transitions can abruptly modify resistivity and heat capacity, leading to transient redistribution of Joule heating. For example, steels exhibit pronounced resistivity changes during austenite-ferrite transformations [96], producing discontinuities in the temperature profile, while Curie transitions in Ni- or Fe-based alloys alter both electrical and thermal response, further perturbing the temperature field [108].

Because resistivity governs heat generation, while thermal conductivity and heat capacity govern heat redistribution and transient response, temperature-control parameters and effective gauge length must be tuned for each material class rather than transferred directly between alloys. Materials exhibiting phase or magnetic transitions require explicit verification of temperature stability through these transitions, as abrupt property changes can transiently distort the thermal field. These measures reduce temperature-field uncertainty but cannot eliminate redistribution arising from intrinsic property evolution.

### 3.3.2. Dimensional control of thermal gradients

The specimen cross-section directly affects both electrical resistance and thermal conduction, and therefore strongly influences the temperature field established during DC-TMT (Figure 9). Because electrical resistance scales inversely with cross-sectional area, thinner specimens experience greater Joule heating for a given applied current and generally heat more rapidly. Even when electrical loading is instead compared at similar nominal current density, differences in specimen geometry and boundary conditions still alter the resulting thermal field. At the same time, their reduced cross-section limits heat conduction away from the gauge section, which tends to sharpen axial temperature gradients. In contrast, thicker specimens exhibit lower electrical resistance, require higher currents to achieve equivalent temperatures, and conduct heat more effectively along the specimen length, which tends to broaden the temperature distribution.

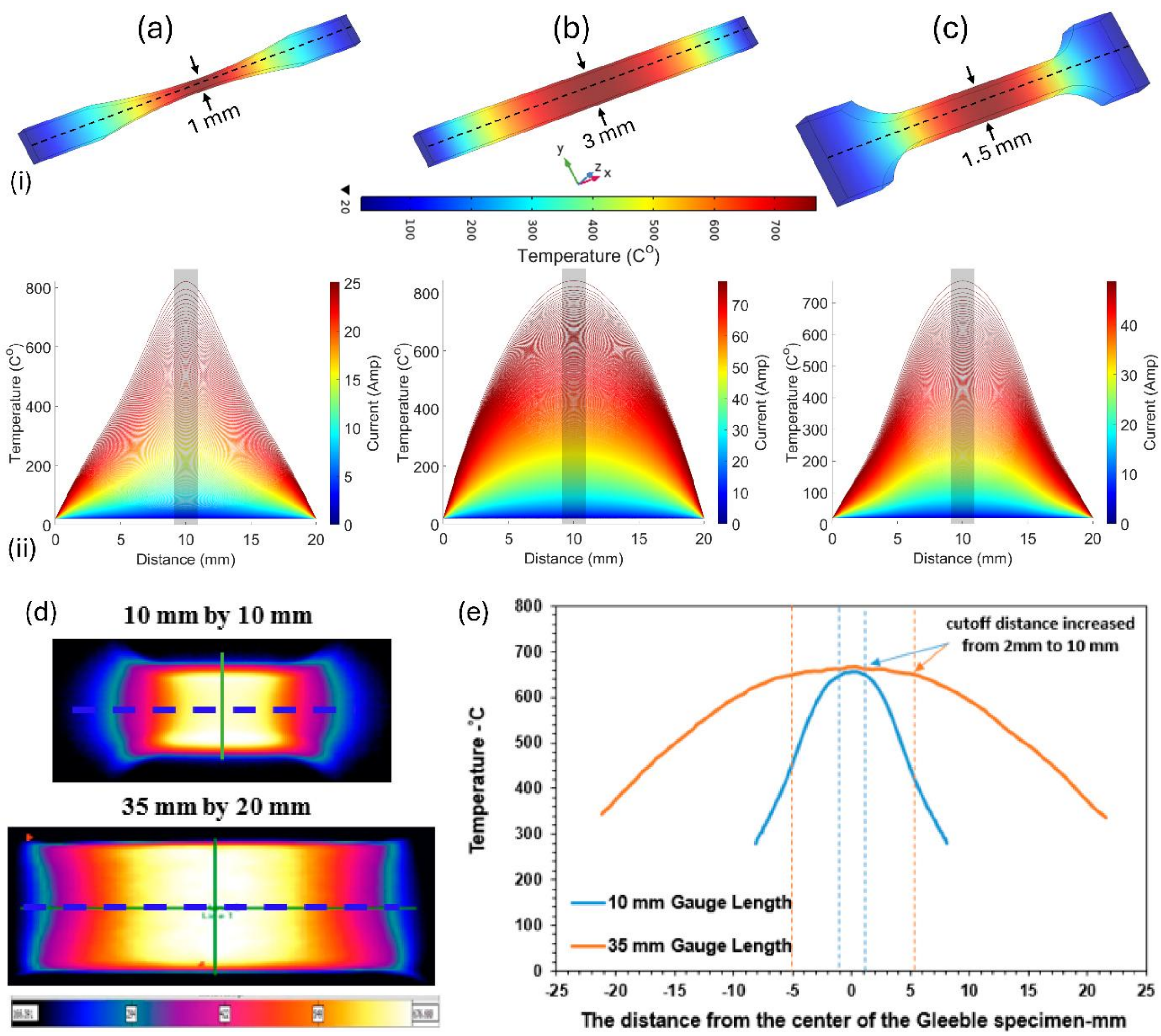

Figure 9: (a-c) Modelled temperature distributions and corresponding axial centreline temperature profiles for 20 × 3 $mm^2$ IN718 specimens with waisted, flat, and dogbone geometries, respectively, showing how specimen shape alters the extent of the central region with relatively uniform temperature (shaded). (d) Thermal images for two coupon geometries, 10 mm × 10 mm and 35 mm × 20 mm, during DC-TMT. (e) Corresponding axial temperature profiles, illustrating that the longer gauge length produces a broader high-temperature plateau. Panel (d-e) adapted from [109], licensed under CC BY 4.0.

Gauge length then controls how strongly the cooled specimen ends influence the heated gauge section. In practical terms, longer gauge lengths generally develop a broader central region of relatively uniform high temperature, because the centre is farther from the grips and less strongly affected by end cooling. Shorter gauge lengths remain more strongly constrained by the grips and therefore exhibit a narrower effective hot zone, even when the centre reaches the same nominal peak temperature. Thus, the main effect of increasing gauge length is not merely to increase total electrical resistance, but to enlarge the region over which the specimen remains close to the target temperature. This behaviour is evident in Figure 9d-e, where the longer specimen exhibits a substantially wider high-temperature plateau along the gauge length.

Consequently, specimen aspect ratio governs the balance between Joule heating and grip-mediated cooling, and different specimen geometries can be used deliberately to reshape the axial temperature profile [22,62,110,111]. In materials exhibiting anomalous yielding, deformation may initiate outside the intended gauge region, suppressing measurable strain accumulation in the centre and producing heterogeneous deformation fields [53]. This is particularly pronounced in alloys that strengthen with increasing temperature, such as many single-crystal nickel-based superalloys in the 600-800 °C range [61]. In such cases, specimen geometry strongly conditions both the temperature field and the spatial distribution of deformation, as illustrated by the contrasting profiles for waisted and flat geometries in Figure 9.

These geometric dependencies imply that, even for identical material properties and comparable nominal electrical loading, for example, when expressed in terms of applied current density, the resulting temperature field remains strongly specimen-specific. This is because current density governs local Joule heat generation, whereas the final temperature distribution is also controlled by specimen length, cross-sectional shape, surface-area-to-volume ratio, and heat extraction through the grips (See Appendix A). Thus, temperature homogeneity, thermal gradients, and control stability in DC-TMT cannot be generalised across specimen dimensions, and temperature measurements must be interpreted in the context of the specific length, width, thickness, and boundary conditions employed. Specimen dimensions are therefore not a secondary design choice, but a primary determinant of the effective gauge temperature and strain localisation.

#### 3.3.3. Grip heat-sinking and boundary conditions

Specimen gripping directly affects the electrical, thermal, and mechanical boundary conditions in DC-TMT, such that grip-related asymmetries can induce axial temperature offsets and distort the imposed temperature profile. Different grip architectures impose distinct mechanical and electrical boundary conditions depending on specimen geometry.

Axisymmetric specimens introduce different load-transfer and current-distribution constraints than flat or prismatic specimens, leading to geometry-dependent sensitivity to gripping-induced artefacts.

Because electrical current flows through both the specimen and the grips, the electrical resistivity of the grips determines how the applied voltage drop is partitioned between these components. Low-resistivity grips dissipate minimal heat and concentrate Joule heating within the specimen gauge, whereas more resistive grips dissipate a non-negligible fraction of the input power, reducing heating efficiency and complicating temperature control within the gauge section. Grip thermal conductivity further shapes the axial temperature profile. Highly conductive grips (e.g., copper alloys) act as strong heat sinks, enhancing axial temperature gradients by extracting heat from the specimen ends, whereas lower-conductivity grips (e.g., steels) reduce end cooling but increase the risk of grip–specimen interface overheating [110]. Modifications to grip electrical and thermal boundary conditions have been shown to substantially alter temperature and strain uniformity in thermo-mechanical testing, demonstrating that measured constitutive response may reflect grip-mediated artefacts rather than intrinsic material behaviour if such effects are not constrained [31,110].

Active cooling of the grips establishes the cold boundary condition for DC-TMT and strongly influences the flatness, symmetry, and stability of the axial temperature profile. Unequal cooling between grips shifts the peak temperature position, while temporal fluctuations in cooling-water flow alter grip temperature, producing thermal expansion mismatches between the specimen and the load frame. These effects manifest as apparent load relaxation, displacement drift, or control instabilities, even when the central specimen temperature appears to be stable. Consequently, cooling-system variability introduces time-dependent uncertainty into long-duration measurements such as stress relaxation and creep, which propagates into both temperature control and potential-drop-based measurements. Thus, independent temperature measurement techniques should be employed to measure the spatial temperature distributions.

## 3.4. Control-loop dynamics and transient thermal response

A PID controller (Proportional–Integral–Derivative) is used in DC-TMT to regulate specimen temperature by adjusting the applied electrical current during Joule heating. In this configuration, closed-loop control dynamically couples thermocouple response, specimen thermal inertia, and resistivity evolution, such that the control system itself becomes a source of temperature uncertainty rather than a passive stabilisation mechanism [112]. Consequently, the effectiveness of PID temperature control depends on material properties, specimen geometry, thermocouple behaviour, and electrical contact conditions at the grips, all of which evolve during DC-TMT experiments.

In practice, PID tuning directly governs how closely the specimen temperature matches the intended test condition, and thus how reproducible and accurate the measured mechanical properties are. The proportional term provides an immediate response to deviations, the integral term eliminates steady-state error by accumulating offsets over time, and the derivative term anticipates future changes by responding to the rate of temperature variation [112].

Control-loop dynamics introduce temperature uncertainty through overshoot, oscillation, finite-time tracking error, and response lag. Excessive proportional gain can promote oscillatory behaviour (Figure 10b), while insufficient integral action slows the removal of constant tracking error, so that a noticeable residual temperature offset may persist over the finite duration of a heating ramp or short hold segment (Figure 10c). In ideal PID control, however, any nonzero integral gain removes asymptotic steady-state error for a constant setpoint. Derivative action, although capable of damping transients, can amplify measurement noise when thermocouple signals are noisy or delayed. These effects are particularly consequential in DC-TMT because material behaviour, such as yield stress, creep rate, and phase transformation kinetics, is highly temperature-sensitive. Derivative action, although capable of damping transients, amplifies measurement noise when thermocouple signals are noisy or delayed.

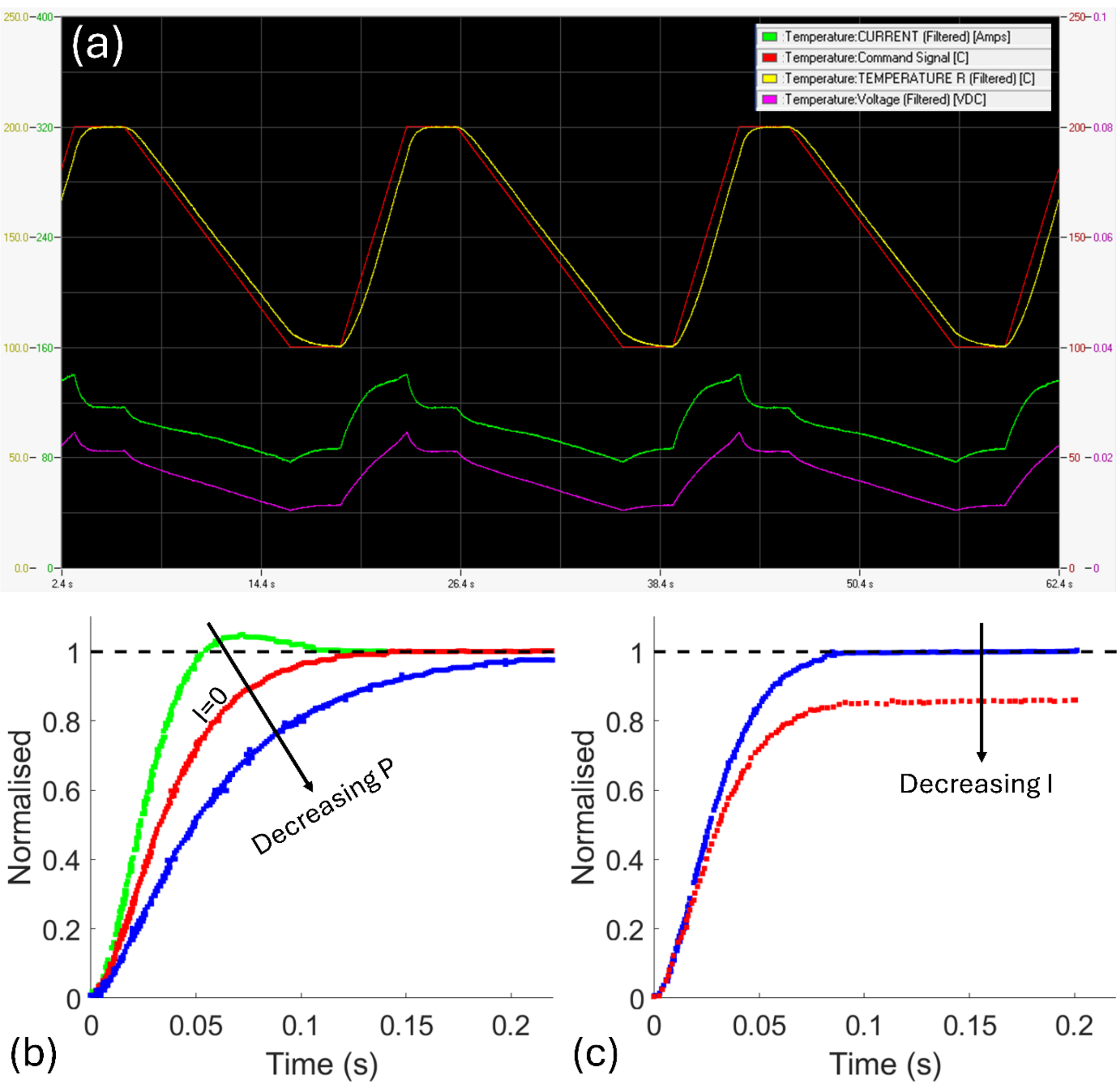

Figure 10: Representative temperature responses illustrating control-loop-induced deviations during PID-regulated Joule heating. (a) Target and measured temperature during trapezium cyclic heating used to assess control stability. (b–c) Sensitivity of temperature response to proportional and integral parameters, highlighting overshoot, lag, and oscillatory behaviour.

These effects are particularly consequential in DC-TMT because material behaviour, such as yield stress, creep rate, and phase transformation kinetics, is highly temperature-sensitive. Even small fluctuations (e.g., ±10 °C) around a nominal set point at high homologous temperatures can produce order-of-magnitude differences in creep rates or alter deformation mechanisms [113,114]. Likewise, a delayed control response during rapid heating or cooling shifts the effective thermal history, thereby modifying the transformation onset and recovery kinetics. Hence, as shown in Figure 10a, a proper PID tuning should be performed, for both temperature and load, on a sacrificial/dummy specimen of the same alloy before the test, rather than relying on auto-tune functions, as these often fail to account for system-specific thermal inertia and can lead to instability.

## 3.5. Strain measurement and localisation

Uncertainty sources associated with load application and displacement control in mechanical testing are well documented elsewhere and are not revisited here [115–117]. Instead, this section focuses specifically on strain measurement in DC-TMT, where specimen, thermal gradients, and combined electro-thermal loading introduce additional uncertainty mechanisms beyond those encountered in conventional bulk testing.

### 3.5.1. Contact and virtual extensometry

Strain measurement, especially in miniaturised specimens, is intrinsically challenging due to the small gauge lengths and low absolute displacements that approach the resolution limits of conventional mechanical extensometry. Contact-based techniques, including clip-on extensometers and capacitance gauges, introduce additional compliance and mass that are non-negligible relative to the specimen stiffness, potentially perturbing the deformation field and contaminating the measured strain response. As a result, strain is often inferred indirectly from actuator displacement or LVDT measurements. However, such global measurements inherently conflate specimen deformation with elastic and inelastic compliance of the test frame, grips, and load train. Even when compliance corrections are applied, residual uncertainty remains because system stiffness may evolve with temperature, load, and time, particularly under DC-TMT conditions [35,39].

Non-contact or virtual (point-based) extensometry techniques, including optical line-scan methods and laser interferometry, decouple strain measurement from the mechanical load mechanism and provide localised strain information within the gauge section. Nevertheless, these approaches introduce their own uncertainty sources, such as sensitivity to rigid-body motion and vibration, discrepancies between tracked surface features and bulk deformation, and degradation or loss of surface markings in highly ductile materials [34,42]. Consequently, no strain measurement technique is inherently free of artefacts in miniaturised DC-TMT, and measured strain must be interpreted in the context of the specific limitations of the chosen measurement approach.

### 3.5.2. Potential-drop/resistance method

In DC-TMT, Joule heating enables the simultaneous use of electrical resistance as a strain-sensitive signal by measuring the potential difference between two fixed points along the specimen gauge length. Because electrical resistance depends on specimen length, cross-sectional area, and intrinsic resistivity, any change in geometry or material state during deformation modifies the measured potential drop. Consequently, resistance-based strain measurement intrinsically couples mechanical deformation, thermal state, and microstructural evolution [93,118,119].

Electrical resistance is given by

$$R = \rho \frac{L}{A}, \tag{8}$$

where $\rho$ is the intrinsic resistivity. Any change in specimen length ($L$), cross-sectional area ($A$), or resistivity modifies the measured resistance. The method, therefore, relies on the assumption that changes in resistance can be attributed primarily to irreversible geometric changes associated with plastic deformation.

Under elastic loading, changes in $L/A$ are small and reversible, and resistance returns to its initial value upon unloading. Plastic deformation, by contrast, produces irreversible changes in specimen geometry and therefore a permanent resistance offset. For this reason, resistance-based methods are commonly interpreted as measures of plastic strain. Assuming isochoric plastic deformation ($L_0 A_0 = L_t A_t$), plastic strain can be estimated as [24,25,120]:

$$\varepsilon_p = \frac{1}{2}\ln\left(\frac{R_t}{R_0}\right) - \frac{1}{2}\ln\left(\frac{\rho_t}{\rho_0}\right) \approx \frac{1}{2}\ln\left(\frac{R_t}{R_0}\right), \tag{9}$$

where $R_0$ and $R_t$ represent the specimen resistance before and during the test, and the approximation assumes $\rho_t/\rho_0 \approx 1$ and plastic incompressibility. For more information on initial resistivity measurements at room temperature, please refer to Appendix B.

Equation 9 is consistent with the assumption that plastic strain arises from shear-dominated deformation in miniature specimens, where the hydrostatic component (volume change) is negligible. This justifies neglecting elastic effects in the derivation, even though there are small elastic contributions. Thus, it is typically used to estimate the plastic strain at a resolution of about ± 0.05%, with the upper and lower bounds of the plastic strains approximated using equation 10:

$$\varepsilon_p \approx \ln \sqrt{\frac{R_t \pm \Delta R_t}{R_0 \pm \Delta R_0}} \quad 10$$

Comparison tests using Nimonic 75 at 600 °C showed that strain measurements obtained from resistance changes in ETMT agree well with conventional extensometer data (Figure 11a), validating the method for plastic strain measurement, with the curves exhibiting serrated steps [5], attributed to progressive yielding events [121,122]. Further tests on IN718 at 950–1200 °C under stresses of 5–100 MPa demonstrated good repeatability, with strain calculated from resistance changes. When plotted against the Zener–Hollomon parameter, as shown in Figure 11c, the miniature ETMT data matched closely with established reference data from Lewandowski and Overfelt [123]. Recent work on Ni-base alloys also demonstrated the method's reliability [124]. Additionally, a tensile test on 304 stainless steel at 600 °C demonstrated close agreement between strain measurements obtained from resistance changes and DIC (Figure 11b) over the same effective cross-section. However, such an agreement does not eliminate the underlying assumptions inherent to the method, because fundamentally, the assumption that the resistivity ratio $\rho_t/\rho_0$ remains constant is invalid in high-temperature DC-TMT.

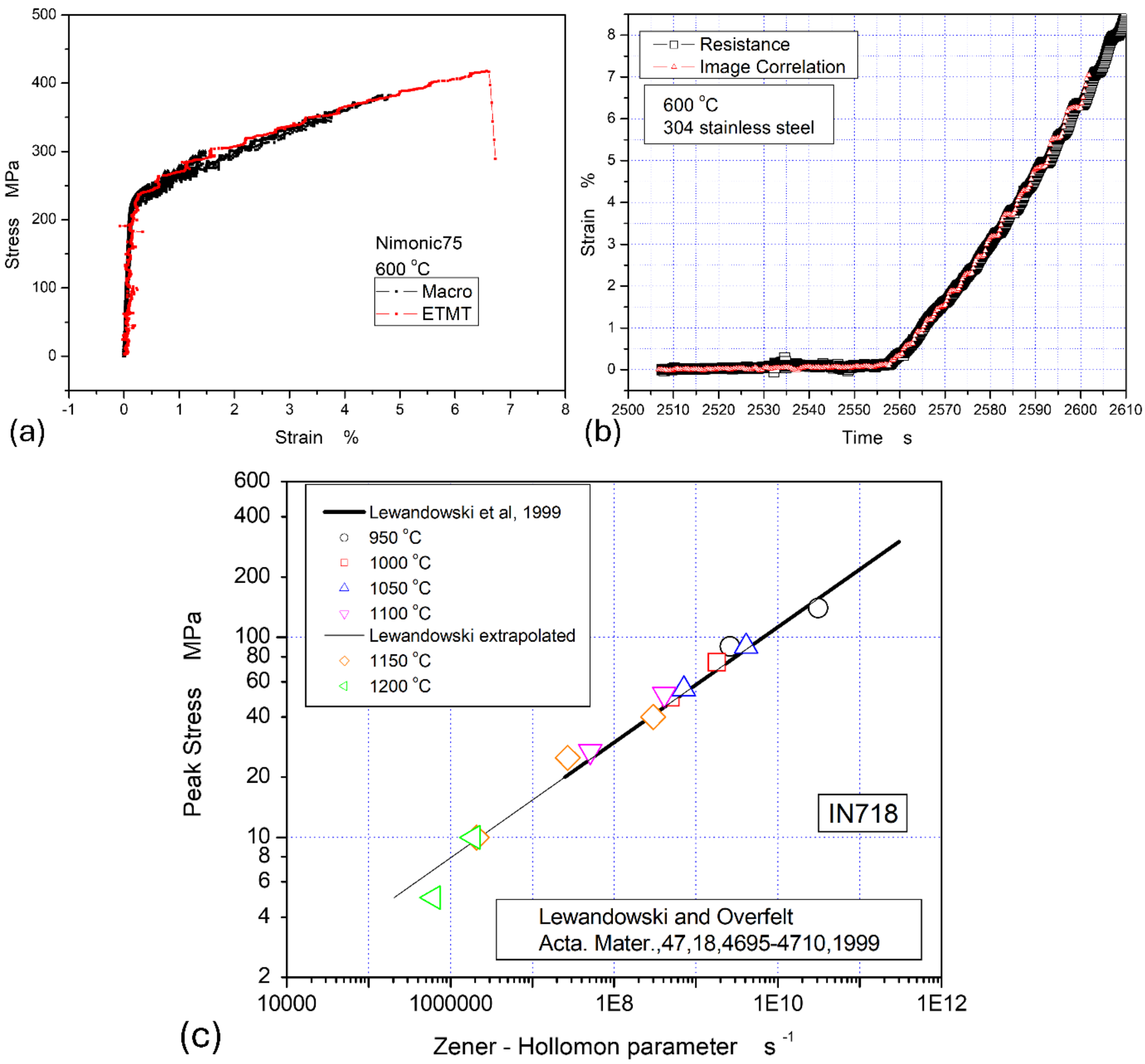

Figure 11: (a) Comparison of potential drop measurement and LVDT optioned during tensile testing of Nimonic 75 at 600 °C. (b) Tensile test on 304 stainless steel at 600 °C with the strain characterised using potential drop measurement and digital image correlation. (c) Comparison of DC-TMT Zener-Hollomon parameters obtained on a Ni-base alloy with reference data on the same alloy. These comparisons demonstrate internal consistency but do not remove uncertainty arising from thermal, microstructural, and phase-dependent resistivity evolution.

Electrical resistivity depends strongly on temperature, dislocation density, phase fraction, and microstructural state. Recovery, recrystallisation, and phase transformation processes modify resistivity independently of geometric strain, introducing ambiguity into resistance-based strain estimates. Consequently, resistance changes cannot be uniquely attributed to plastic deformation without additional constraints. This ambiguity is further amplified in multi-phase or transformation-prone materials, where deformation alters the phase fractions, resulting in distinct resistivities. In such cases, resistance evolution reflects both geometric strain and evolving phase topology, further complicating interpretation if these effects are not explicitly taken into account. For example, in Co-W-C alloys at temperatures below 500 °C, the microstructure can be a mixture of HCP and FCC phases that depends on composition and whether the heating increases or decreases due to hysteresis in phase transformation kinetics. Deformation alters the FCC/HCP ratio in the mixed-phase region, affecting the changes in resistance [123]. If this factor is not considered in the analysis, it adds uncertainty to the conclusions.

Resistance-based strain measurement further assumes well-defined electrical contact geometry and spatial homogeneity within the effective gauge region. Uncertainty arises from inaccuracies in potential-lead placement, lead-to-lead spacing relative to the thermal gradient, and inadvertent coupling with thermocouple wiring. Because the spatial extent of the quasi-isothermal region depends on specimen geometry and peak temperature (Section 3.3), resistance measurements may integrate strain over a temperature-varying volume, introducing systematic bias. For example, as shown in Figure 12, for a IN718 waisted specimen (20 × 3 × 1 mm$^3$, $L \times C \times W$), a potential leads that is 1.5 mm from the centre will estimate the strains across a temperature variation of 16 °C at peak temperature of 700 °C, compared to 5 °C for a 20 × 3 × 3 mm$^3$ straight/flat specimen and 9 °C for 20 × 3 × 1.5 mm$^3$ dogbone specimen, as inferred from a COMSOL modelling that was calibrated via phosphor thermometry and thermal imaging (Figure 8c).

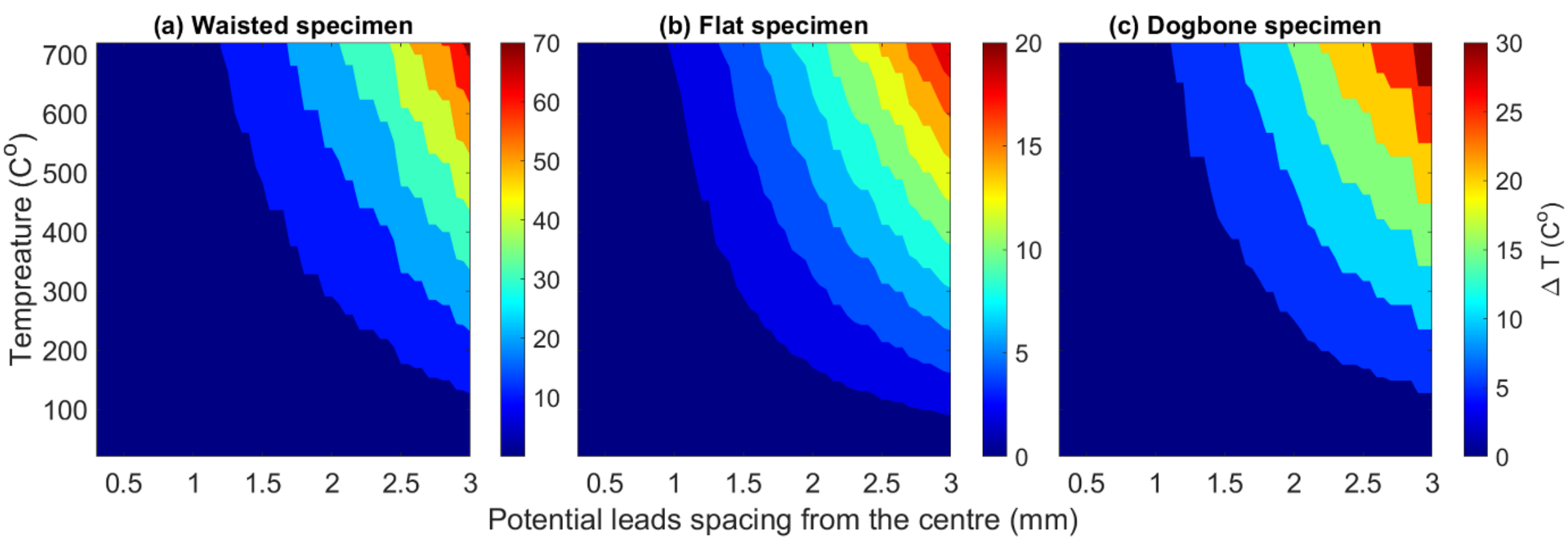

Figure 12: Temperature variation (ΔT) between potential measurement leads positioned at different distances from the centre for three IN718 specimen geometries under varying peak (central)

temperatures: (a) waisted specimen (20 × 3 × 1 mm$^3$, $L \times C \times W$), (b) straight/flat specimen (20 × 3 × 3 mm$^3$), and (c) dogbone specimen (20 × 3 × 1.5 mm$^3$).

The assumption that Seebeck effects are negligible is valid only when the potential leads lie entirely within the quasi-isothermal region of the specimen. When the potential lead spacing is small relative to the flat central temperature plateau, the temperature difference between the two leads can be kept below approximately 5 °C, which limits the thermoelectric error to well below 1% of the measured signal. Under such conditions, the Seebeck or parasitic voltages contribution is small compared with the geometric contribution to resistance and may be treated as negligible. However, when potential leads extend into regions with measurable temperature gradients, or when the gradient steepens due to material properties or geometry, the Seebeck voltage becomes non-negligible as thermoelectric (Seebeck/parasitic) voltages develop both within the specimen and within the voltage-measurement leads. These voltages superimpose on the ohmic potential drop used for strain estimation. Because typical thermopower values for engineering alloys lie in the range of 1–10 μV/K [125], a temperature difference of only 50–100 °C between voltage leads can introduce parasitic voltages on the order of 0.05–0.1 millivolt. These values are comparable in magnitude to the resistance changes associated with small strains in miniature specimens at high temperature, so Seebeck/parasitic voltages contributions can bias the inferred strain if not explicitly controlled.

Additionally, at low and moderate temperatures (< 200 °C), resistance-based strain estimation becomes increasingly noisy (Figure 13a), due to reduced thermal sensitivity of resistivity, limiting its practical resolution [77]. Additional uncertainty arises once local necking initiates, as the assumption of uniform deformation between potential leads breaks down.

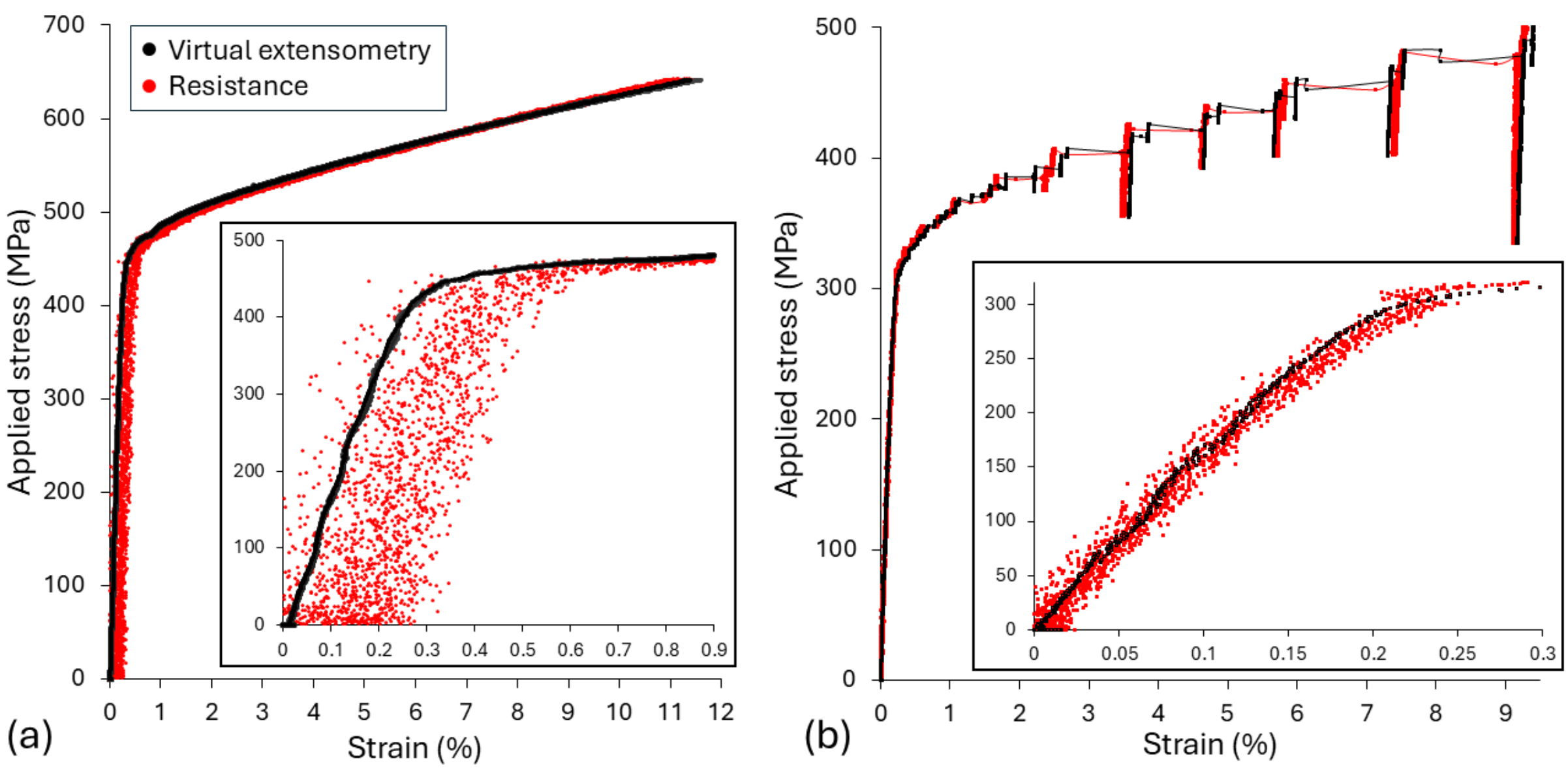

Figure 13: Comparison between virtual extensometer and resistance measurements obtained during a tensile test of Haynes230 at (a) 40 °C and (b) 500 °C.

Thus, the resistance-based strain measurement method is quantitatively valid only under a restricted set of conditions. The method assumes that (i) plastic deformation is isochoric, so changes in electrical resistance arise primarily from geometric changes in length and cross-section, (ii) the intrinsic resistivity of the material remains approximately constant during deformation, and (iii) the region between the potential leads is both mechanically homogeneous and quasi-isothermal. These assumptions hold in single-phase alloys over moderate temperature ranges, where microstructural evolution is limited, and the temperature gradient across the potential leads is negligible. Under such conditions, resistance-derived plastic strains have been shown to agree with independent extensometry and DIC measurements (Figure 13). Outside these regimes, particularly at high temperatures, in multiphase or transformation-prone alloys, after the onset of necking or other localisation phenomena, or during long-duration tests where resistivity evolves with microstructure and temperature, resistance changes cannot be uniquely attributed to geometric strain. In these cases, the method should be interpreted qualitatively, as a comparative or trend-based indicator rather than a source of absolute constitutive data.

### 3.5.3. Full-field methods

Full-field strain measurement techniques address a fundamental limitation of point-based extensometry in DC-TMT: the coexistence of spatial temperature gradients and non-uniform deformation within the gauge section. Because, in general, strain cannot be uniquely associated with a single temperature outside the central quasi-isothermal region, deformation often localises preferentially in regions of lower stress, particularly when anomalous yielding occurs, or specimen geometry is poorly matched to the thermal profile [77]. Full-field optical and interferometric techniques, including moiré interferometry, digital speckle pattern interferometry, digital image correlation (DIC), photoelastic stress analysis, and thermoelastic stress analysis, provide spatially resolved displacement and strain fields that expose deformation heterogeneity otherwise obscured by averaged measurements [126–129]. Among these approaches, DIC has seen the widest adoption in DC-TMT due to its flexibility, spatial resolution, and ability to track localised strain accumulation, necking, and crack initiation under combined thermal and mechanical loading [77,130,131].

The dominant uncertainty in high-temperature DIC arises from the degradation or evolution of the surface speckle pattern during testing. Above approximately 500 °C, the stability of surface contrast becomes a dominant limitation, as conventional speckle patterns degrade due to oxidation, peeling (Figure 14c), or chemical interaction with the substrate, resulting in a loss of correlation fidelity [130,132,133]. This necessitates the use of specialised high-temperature coatings [134], ceramic sprays [135], or laser-etched patterns [136] that can withstand both heat and thermal cycling. However, the specimen surface kinetics should be considered when deciding on the thermally stable spray-painting to minimise interactions between the coating and substrate. These mitigation strategies extend the usable temperature range of DIC but do not prevent gradual loss of correlation quality during prolonged exposure or under large thermal gradients.

In addition, when testing in air, oxidation will cause the speckle to flake off during the test (Figure 14c). A vacuum chamber eliminates the oxidation problem, so the speckle pattern remains stable. However, introducing a glass window in front of the optics can create its own issues, including reflections, refraction, and image distortion. These effects are especially problematic at high temperatures because of the intense thermal radiation from the glowing specimen, which can bounce off the window and reduce contrast in the recorded images. To minimise this, experimental setups often use anti-reflection-coated quartz or sapphire windows, angled slightly relative to the optical axis to deflect reflections away from the camera.

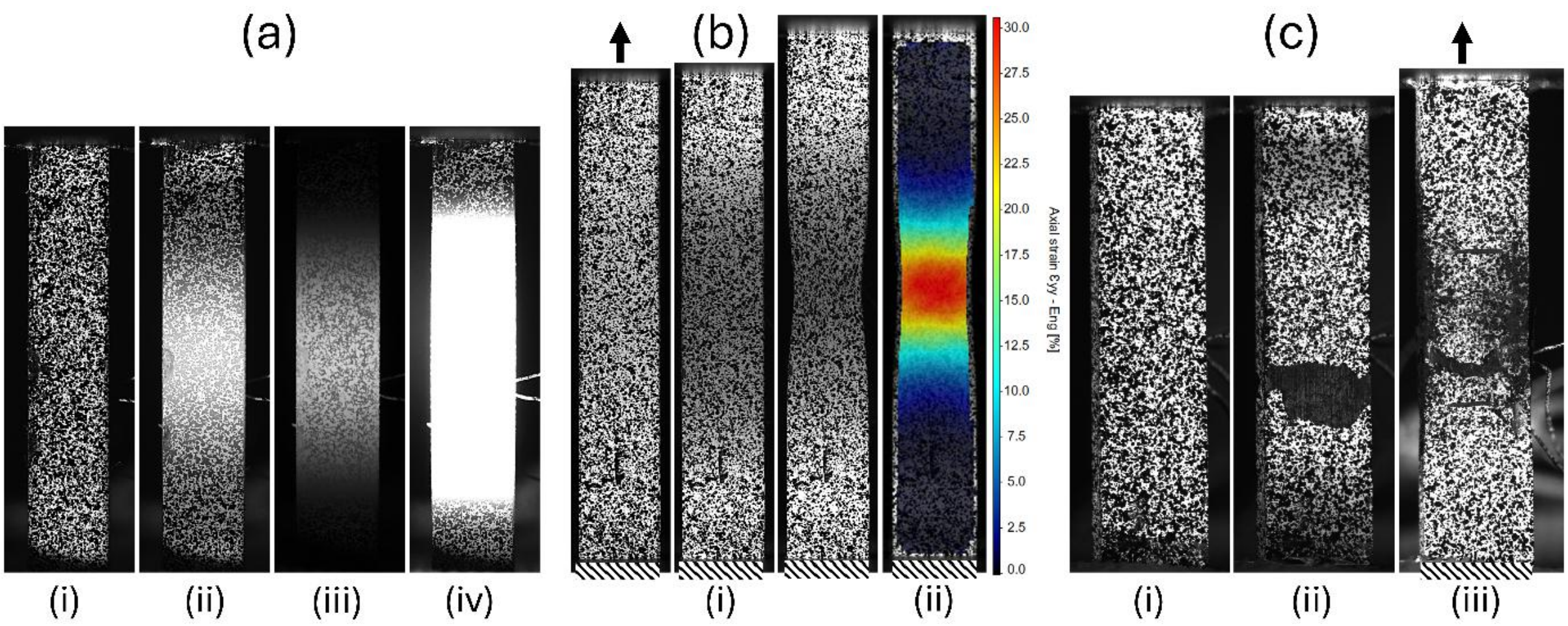

Figure 14: High-temperature DIC imaging in air of speckled flat miniaturised ETMT specimens. (a) 2 mm cross-section pure Ti specimen imaged with a conventional room temperature DIC setup at (i) room temperature, (ii) 790 ºC with lighting on and (iii) off, and (iv) 1000 ºC. (b.i) Imaging of a 2 mm cross-section pure Ti specimen with blue LEDs illumination and high-quality optical filters with polarising lenses during tensile testing at 1000 °C. (b.ii) DIC Strain field. (c) Imaging of a 3 mm cross-section carbon steel specimen at (i) room temperature, (ii) 1000 ºC without load, and (iii) 700 ºC with load.

Optical artefacts further complicate full-field measurements in DC-TMT. Intense thermal radiation from the heated specimen reduces image contrast and introduces wavelength-dependent noise, while refractive index fluctuations in the surrounding hot air generate apparent displacements unrelated to mechanical deformation [131,133,137]. These effects become increasingly severe with rising temperature and magnification, and cannot be fully eliminated by post-processing alone. As shown in Figure 14b, high-quality optical filters, careful shielding of the specimen, and synchronised illumination with narrow wavelength bands, such as using blue LEDs with polarising lenses, are essential to reduce this problem [138,139]. While narrow-band illumination and polarisation substantially reduce radiative noise, they do not eliminate distortion caused by air turbulence or specimen motion, particularly at very high temperatures and remain limited by oxidation-driven surface evolution, refractive index fluctuations, and camera dynamic range.

The small dimensions of miniaturised ETMT specimens necessitate high optical magnification, which amplifies sensitivity to vibration, rigid-body motion, and out-of-plane displacement. In conventional two-dimensional DIC, even sub-micrometre out-of-plane motion or camera drift can generate apparent in-plane strains exceeding 0.5%, a non-negligible error for low-ductility materials [140]. While stereo DIC can reduce this ambiguity by resolving three-dimensional motion, it introduces additional calibration, alignment, and synchronisation uncertainty that propagates into the strain field [141].

Other general DIC measurement and analysis-related error sources and good practices to reduce them can be found in Refs. [142–145], and caution needs to be exercised as these error sources are exacerbated in small-scale mechanical testing, particularly in the elastic regime [146,147]. Despite these challenges, DIC remains the most flexible and detailed non-contact strain measurement technique. To constrain these uncertainty sources, established good-practice measures include optimising speckle size and contrast, maintaining stable and spectrally filtered illumination, selecting appropriate spatial resolution and frame rate, using stereo imaging where out-of-plane motion is non-negligible, defining gauge regions based on prior thermal characterisation, and performing rigorous calibration before and after testing.

Although ASTM E2208 [148] can be used to evaluate non-contact optical methods for strain measurement, the accuracy evaluation of non-contact measuring techniques is not as standardised as that of contact methods. The goal of ongoing research into DIC by the International Digital Image Correlation Society (iDICs) is to increase the precision of measurements by reducing errors caused by software (interpolation and search algorithms), hardware (lens distortion, drift, picture capture frequency, and depth of field), and environmental factors (temperature, humidity, and illumination), as finding the origins of errors aids in evaluating their effects and raising the precision of strain measurements [149–151]. Under controlled conditions, reported 2D DIC strain uncertainty can approach ~0.5%. In addition, by recommending ideal measurement conditions and calculating measurement uncertainty, assessments using finite element modelling (FEM) help correct strain measurement [35,152,153].

As with resistance-based strain measurement, full-field optical methods are fundamentally constrained by the specimen temperature distribution. Unless spatially resolved temperature measurements are performed concurrently, strain fields inferred outside the central quasi-isothermal region represent deformation integrated over a temperature gradient, complicating the interpretation of constitutive behaviour [154]. Coupling optical strain measurement with independent temperature mapping mitigates this limitation, but it does not eliminate the underlying thermal–mechanical coupling.

## 3.6. Vacuum and environmental testing

Many engineering materials, including steels and zirconium alloys, undergo rapid oxidation at elevated temperatures, introducing environment-dependent uncertainty into both deformation and failure behaviour. Therefore, it is vital to be able to test in a vacuum or controlled environments. Back-filling with high-purity argon can reduce oxidation provided that residual oxygen levels are sufficiently low and system leak rates are controlled. For example, a reducing atmosphere such as argon with 5% hydrogen (Ar–5% $H_2$) is often used to examine material systems that would otherwise oxidise very rapidly at high temperatures [155].

For the vacuum option, it is also helpful to purge the container several times with an inert gas before the test or to refill the container completely with an inert gas. However, testing in a vacuum demands more rigorous system sealing and stability, and induces a phantom stress that needs to be corrected once a stable vacuum is reached. This phantom stress arises from pressure differentials acting on the specimen–grip assembly and can be comparable to the applied mechanical load in miniaturised specimens if not corrected.

Testing in controlled gaseous environments enables the isolation of environment-assisted deformation and damage mechanisms. For example, one could perform fatigue tests on steels in high-temperature air vs in vacuum to isolate the role of oxidation in crack growth. Similarly, testing Ni-based alloys in different atmospheres (air vs $CO_2$ or $SO_2$) can simulate the environments of turbine engines. However, environmental composition can directly affect thermocouple stability and potential-drop measurement through oxidation, carburisation, or hydrogen uptake of exposed leads, especially for type-K T/C. Additionally, gas circulation systems introduce forced convection and mechanical vibration, which can induce current fluctuations and rigid-body motion of the specimen, thereby increasing noise in both temperature and strain measurements [24].

## 3.7. DC-specific material responses

In DC-TMT, DC Joule heating is both the enabling mechanism and a critical factor that shapes the outcome of experiments. By passing an electrical current directly through the specimen, DC-TMT enables rapid internal heating without the need for external furnaces, allowing thermo-mechanical testing under conditions that approximate service-relevant thermal loading. However, the passage of high electrical current densities through the specimen introduces additional physical phenomena that may influence microstructural evolution and deformation. Experimental observations reported by Conrad and others show that, at current densities of order $10^3$–$10^5$ A/cm$^2$, apparent acceleration of recrystallisation, crystallisation, or modification of grain growth kinetics can occur [156]; the extent to which these effects arise from purely thermal gradients, electromigration, electron–dislocation interactions, or other current-assisted mechanisms remains the subject of ongoing debate. Several mechanisms have been proposed to explain current-associated effects in DC-TMT, including:

### 3.7.1. Electromigration and current-driven mass transport

The oxidation behaviour of the DC-TMT and ASTM specimens differs significantly; under equivalent circumstances, the DC-TMT specimens exhibit thicker oxidation layers than the ASTM ones [27]. The high current density during Joule heating has been proposed as a contributing factor to accelerated oxidation kinetics through electromigration, a phenomenon in which the movement of electrons induces mass transportation. However, disentangling current-driven mass transport from temperature-gradient and size effects remains challenging [157]. In phenomenological treatments, electromigration-assisted degradation is sometimes represented by an empirical relation of the form:

$$\dot{\xi}_{\mathrm{EM}} \propto j^n \exp\left(-\frac{E_a}{kT}\right), \qquad 1 \lesssim n \lesssim 2, \tag{11}$$

where $j$ is the current density, $E_a$ is an apparent activation energy, $k$ is the Boltzmann constant, and $T$ is the absolute temperature. Here, $n$ is an empirical exponent, often taken between 1 and 2, depending on the dominant mechanism. This expression should not be interpreted as a fundamental law for electromigration flux. In physically based descriptions, the electromigration flux is approximately linear in current density and scales with the diffusivity, while stress gradients, temperature gradients, and microstructural constraints also contribute to the net mass transport [158,159].

For example, at the high temperatures and current densities used in DC-TMT, nickel-based superalloys may exhibit enhanced diffusion processes that have been attributed in some studies to electromigration effects [160], leading to microstructural alterations such as changes in precipitation kinetics, recrystallisation, and the development of intermetallic compounds. At the same time, a practical advantage of DC-TMT is that rapid Joule heating and cooling reduce the time spent at elevated temperature during thermal transients, which can lessen oxidation accumulated during the heating and cooling stages relative to slower furnace-based tests. The net oxidation response in DC-TMT therefore reflects competing effects: reduced thermal exposure time during ramps, but potentially enhanced local oxidation damage associated with high current density, specimen geometry, thermal gradients, and surface condition.

The observed material behaviour, which includes embrittlement from oxidation, surface failure [161,162] and quicker creep rates [26], is consistent with oxidation-assisted surface damage that may, in some cases, be exacerbated by the presence of high electrical current during DC-TMT. For instance, DC-TMT specimens failed at various strain levels due to oxidation-induced surface cracking at 980 °C and 200 MPa, yet exhibited creep parameters comparable to ASTM specimens. Consequently, differences between ASTM and DC-TMT results cannot be attributed uniquely to current effects without also accounting for thermal history, dwell time, geometry-dependent temperature gradients, and oxidation kinetics. Because oxidation-assisted grain-boundary damage becomes more severe as the effective grain-boundary area increases, DC-TMT may be particularly susceptible to oxidation artefacts in fine-grained materials during long-duration exposure. More thorough comparative studies are therefore required to separate intrinsic current-driven effects in oxidation from differences arising from heating rate, time at temperature, and specimen geometry.

Moreover, when the test was done in inert environments to inhibit oxidation, a unique aspect of Joule heating was the potential for electrically enhanced diffusion and precipitation kinetics. Above a threshold current density (on the order of $10^3$ A/cm$^2$ for many alloys), researchers have reported accelerated solid-state reactions, such as precipitation and phase growth, as the electric current induces electromigration of point defects (e.g., vacancies), effectively increasing mass transport and diffusion rates [25]. One striking example is a Ni–Ti intermetallic diffusion couple: under an applied DC, the growth of the intermetallic layer was observed to be ~43 times faster than under identical thermal conditions with no current [163]. The applied current dramatically lowered the effective activation energy for phase growth (e.g., the activation energy for $Ni_3Ti$ formation dropped from ~292 kJ/mol to ~86 kJ/mol under a certain current density) [163]. Experimental studies on Al- and Ti-based alloys have shown that such vacancy redistribution can either accelerate processes such as recovery and recrystallisation or retard precipitation kinetics, depending on the imposed stress state and thermal history [156].

Analogous coupled effects have been reported in microelectronic interconnect systems, where Joule heating and electromigration can reinforce one another through a positive feedback loop: local mass-transport-induced thinning raises resistance, which increases Joule heating and further accelerates electromigration. Such behaviour has been observed in redistribution layers of 2.5D/3D integrated circuits and is now recognised as an important reliability concern in high-current-density electronic packaging [164,165]. However, these studies were conducted on micron-scale Cu interconnects with very different geometries, thermal boundary conditions, and oxidation environments from those of DC-TMT specimens. They therefore support the plausibility of coupled electro-thermal degradation, but do not by themselves establish the governing mechanism in bulk DC-TMT alloys.

These observations suggest that, under sufficiently high current densities, electrically assisted diffusion may alter precipitation, vacancy populations, and oxidation kinetics. However, extrapolating such effects to DC-TMT conditions requires careful control of temperature gradients and current density. Nonetheless, in DC-TMT, apparent electromigration effects are intrinsically coupled to non-uniform Joule heating, oxidation kinetics and size effect, and temperature measurement uncertainty. Without independent verification of spatial temperature fields, heating rates, and oxidation rates, it remains challenging to separate intrinsic current-driven mass transport from indirect thermal effects. As a result, electromigration should be treated as a potential, but not definitive, contributor to observed behaviour unless corroborated by controlled comparative experiments.

### 3.7.2. Electropulsing and non-thermal current effects

Modulation of the applied current waveform, including pulsed or intermittent DC profiles, has been reported to amplify current-associated diffusion and transformation effects in some material systems. Electropulsing studies conducted outside the DC-TMT context demonstrate accelerated nucleation during recrystallisation, crystallisation, and phase transformations, often accompanied by suppressed grain growth and refined microstructures [156,166]. Electropulsing also affects cracks [167] and precipitation evolution [168].

For instance, an electrothermal treatment of Ti–6Al–4V formed a full basket-weave α+β microstructure in only ~6 s at ~806 °C, whereas a traditional heat treatment required ~15 minutes at ~980 °C to achieve a similar transformation [163]. In low-carbon steels, combining electrical pulses with heating has produced ultrafine-grained austenite that nearly doubled the strength without losing ductility, compared to heat treatment alone [163]. These enhancements arise because the electric current can reduce the thermodynamic barrier for nucleation and enhance atom mobility via electromigration. The current passing through a metal can locally lower nucleation energy barriers and generate inhomogeneous heating at the microscale (especially if the new phase has higher electrical conductivity). The net effect promotes more rapid phase transformation kinetics, effectively providing athermal assistance in addition to the pure thermal effect of Joule heat.

It was also observed that pulsed-DC effect deformation mechanisms during mechanical testing, i.e., electrically-assisted deformation (see Section 3.7.4). In in situ DIC-monitored tensile tests on Ti–6Al–4V with a single current pulse, Bao et al. [169] and Kadir et al. [170] found an immediate drop in stress and the sudden formation of new localised high-strain bands upon current application. These intersecting shear bands eventually coalesced into a narrow "flow zone" of intense deformation, indicating that electric pulses can promote strain localisation in the specimen. The net effect was a change in fracture mode (e.g., a larger fracture angle and sawtooth-like fracture surface under pulsing) and an overall higher local plastic strain compared to tests at the same temperature without current.

This suggests that under certain conditions, DC or pulsed-DC loading can redistribute or concentrate strain, possibly through localised Joule heating or differential microstructural softening. However, these studies typically involve peak current densities and pulse durations that differ substantially from those achievable in DC-TMT, and often decouple electrical effects from bulk Joule heating. Consequently, while waveform modulation in DC-TMT may enhance diffusion- or nucleation-related phenomena, extrapolation of electropulsing results to DC-TMT testing conditions must be treated with caution, as thermal gradients and control-loop dynamics remain dominant.

### 3.7.3. Magnetic-field coupling and magneto-mechanical artefacts

For a typical DC-TMT specimen, Joule heating induces a circumferential magnetic field with a low magnitude (~20 mT, see Appendix C), but it can introduce secondary field-dependent effects in magnetically active materials. For example, when testing ferromagnetic alloys below the Curie temperature, small field–microstructure interactions can appear [171,172], and when using multiferroics or magnetostrictive alloys, e.g., Galfenol Fe–Ga [173,174] and Terfenol-D Tb–Dy–Fe alloys [175], the induced magnetic field can weakly couple to phase stability, transformation kinetics, or magnetostrictive strain.

Azeem et al. [176] noted that in $Ni_{54}Mn_{25}Ga_{21}$, the hysteresis in both phases' lattice parameters is unrelated to the transformation itself. In the first thermal cycle, lattice parameters converge at 345 °C, while in later cycles, they converge at 300 °C, influenced by the resistance heating method. This raises questions about whether these alloys retain their ferromagnetic properties, given that the Curie temperature is around 110 °C. While lattice parameters remain constant at a given temperature, transformation temperatures change by about 8 °C per cycle for the austenite finish temperature. However, above the Curie temperature, direct magnetic ordering effects should vanish, implying that observed shifts are more likely linked to differences in heating rate, current distribution, or thermal gradients rather than sustained ferromagnetic coupling.

Molotskii and others have suggested that the current’s magnetic field can induce an electromagnetic force on dislocation loops, helping to release them from obstacles, i.e., magnetoplasticity [177–179]. However, such magnetic forces become significant only at very high current densities or in pulses; under typical DC-TMT conditions (currents on the order of 10–100 A/mm$^2$), the magnetic pressure is relatively small.

These examples illustrate that Joule heating can modify apparent transformation behaviour through mechanisms that are difficult to disentangle, and that magnetic contributions, where present, are likely indirect and highly material-specific. For example, comparisons between phase transformation data obtained under Joule heating and those obtained using conventional techniques, such as differential scanning calorimetry (DSC) [180], should be interpreted with caution. Differences in apparent solvus or transformation temperatures may arise from microsegregation, specimen size, heating rate, and temperature gradients, in addition to any potential current-associated effects. Without independent validation of spatial temperature fields and transformation kinetics, attributing such discrepancies uniquely to electrical or magnetic influences is not justified. This is because for the majority of structural alloys tested in DC-TMT at high temperatures, magnetisation effects associated with Joule-induced magnetic fields are expected to be negligible compared with thermal gradients. Magnetic coupling becomes relevant only in a narrow class of magnetically active materials and at temperatures near or below magnetic transition points. Consequently, magnetisation should be regarded as a conditional, second-order uncertainty source in DC-TMT experiments.

### 3.7.4. Electrically-assisted deformation

Electrically-assisted deformation, sometimes addressed as "electroplasticity", refers to changes in ductility or dislocation mobility under applied current. It is often discussed in terms of softening, altered activation barriers for dislocation motion, or electron–dislocation interactions. However, in essence, electrically-assisted deformation is not an isolated phenomenon but rather some combination of (1) electromigration or DC diffusion-mediated processes that provide another mechanism for plasticity changes via accelerated dynamic recovery, recrystallisation [181–183], or the so-called "electron wind" effect, i.e., moving electrons impart momentum to dislocations [184,185]; (2) localised high-strain bands due to electropulsing [169]; or (3) magnetism-induced dislocations unpinning [177–179].

Crucially, researchers strive to distinguish these electroplastic mechanisms from ordinary Joule heating. Joule heating inevitably accompanies any current flow and raises the specimen temperature, which, by itself, reduces the stress and accelerates softening [186]. To ensure an observed effect is truly electroplastic, experiments employ strategies such as isothermal conditions, i.e., actively cooling the specimen or using short current pulses that do not significantly heat the bulk. There appears to be a threshold current density required to observe non-thermal effects; literature estimates range around 10–100 A/mm$^2$, below which any current-induced softening is negligible compared to thermal expectations [187,188]. At or above such thresholds (corresponding to the regimes studied by Troitskii [184], Conrad [156,189], Sprecher [190], etc.), small but real deviations from thermally predicted behaviour emerge.

In many cases, careful modelling of the thermal profile confirms that the majority of the apparent softening during DC-TMT can be attributed to resistive heating, e.g., the well-known thermal softening captured by a Johnson–Cook model [187]. What remains as an unexplained increment – if any – is ascribed to athermal mechanisms, such as those described above. Recent studies with rigorous controls find this athermal increment to be essentially zero within experimental uncertainty [187,191]. Even miniature ETMT generally agreed with standard furnace-heated tests, but at the highest temperature (≈1100 °C), the ETMT measured a noticeably different creep rate than conventional tests (Figure 15) [27]. The discrepancy at 1100 °C was later rationalised by considering potential experimental factors (thermocouple accuracy, thermal gradients in the miniature specimen, etc.), emphasising that softening can sometimes arise from differences in temperature control or specimen size rather than a true athermal material response.

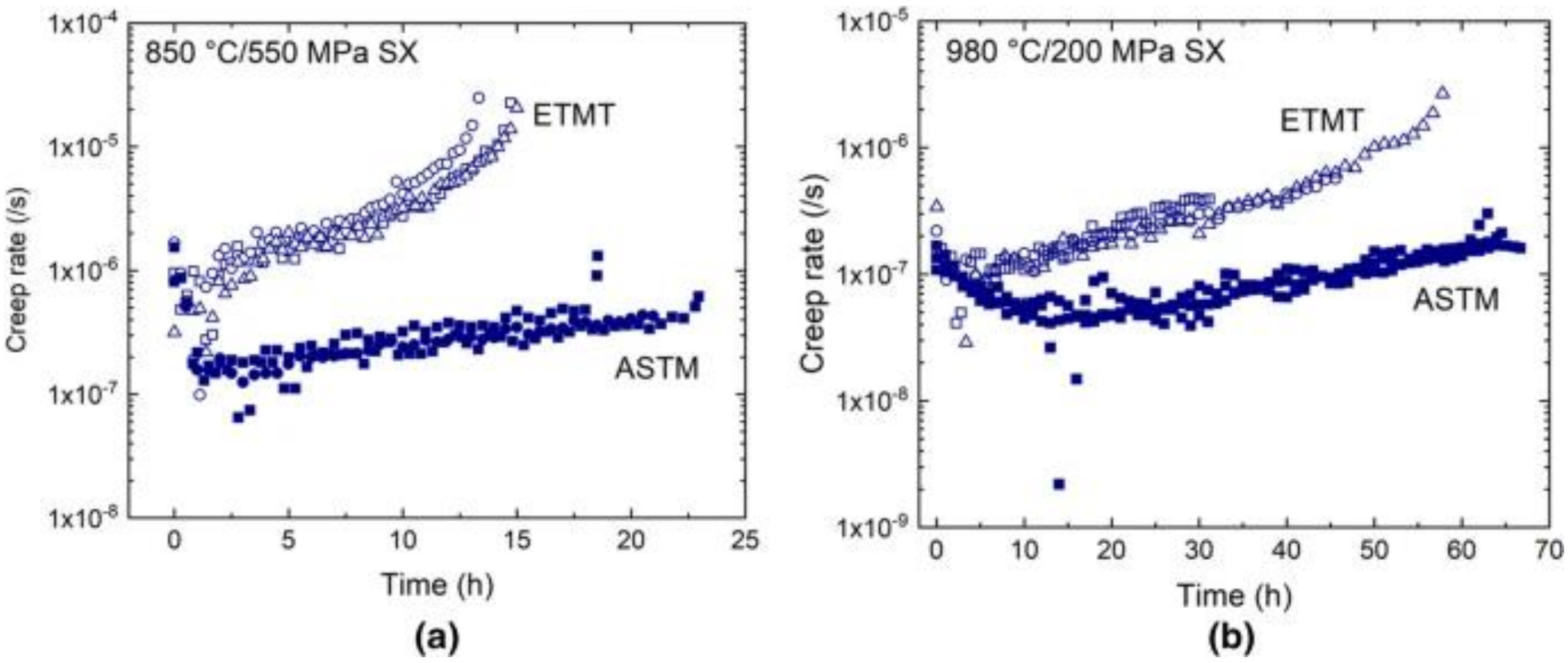

Figure 15: Comparison of instantaneous creep rate measured using miniaturised ETMT specimens and bulk ASTM specimens for single-crystal Mar-M-247 at (a) 850 °C/550 MPa and (b) 980 °C/200 MPa. While initial creep rates are comparable, the miniaturised ETMT specimens exhibit accelerated tertiary creep and order-of-magnitude deviations at longer times, associated with enhanced cracking and oxidation. The comparison illustrates the sensitivity of creep interpretation to specimen size and damage evolution mechanisms. Adapted from [27], licensed under CC BY 4.0.

Indeed, multiple concurrent phenomena (thermal, electrical, magnetic, and mechanical) are at play during electrically-assisted deformation, and their contributions can be difficult to decouple. No single theory yet fully captures all observations across different materials, so authors often cross-verify that any claimed electrically-assisted deformation is not an artefact of uneven heating, contact resistance, measurement lag, or other experimental biases.

## 3.8. Effect of heating rate

In DC-TMT, heating rate is a major determinant of transformation kinetics and final microstructure, but its effect is strongly material- and mechanism-dependent. In many alloy systems, rapid Joule heating reduces the time available for diffusional redistribution, shifts transformation start and finish temperatures to higher values, and promotes finer microstructures by limiting growth relative to nucleation. In other systems, however, especially where current-assisted diffusion or electromigration-related effects become important, the net response may not follow this simplified trend. Heating-rate effects in DC-TMT should therefore be interpreted as system-specific kinetic responses rather than as a universal consequence of rapid electrical heating.

For example, in steels, an ultrafast heating rate (~100–1000 °C/s) shifts the ferrite-to-austenite start and finish to higher temperatures, resulting in a narrower temperature range for transformation and refining the prior austenite grain size by approximately 50%, which in turn yields smaller ferrite grains upon cooling [192]. Additionally, ultrafast heating may not fully dissolve carbides or pearlite regions before austenitisation is complete, resulting in less carbon being taken into solution [192]. This occurs because diffusion of carbon and other solutes cannot keep pace, requiring a higher thermal driving force for the phase change. Similarly, in Ti–6Al–4V, increasing the heating rate also shifts the α→β transformation to higher temperatures. Recent in situ measurements during heating showed that the α-dissolution temperature range moves markedly upward with increasing heating rate, from about 600 °C at 0.25 °C $s^{-1}$ to about 960 °C at 200 °C $s^{-1}$, while also becoming narrower at the highest rates. This should not be interpreted as a complete transformation to single-phase β at 600 °C. Rather, full transformation is generally only approached near the β-transus, which for Ti–6Al–4V is around 995 °C [192,193].

Thus, heating rate constitutes a first-order source of uncertainty for phase transformation temperatures, microstructural state, and derived mechanical properties in DC-TMT. However, discrepancies between DC-TMT and conventional furnace-based measurements should not be attributed to heating rate alone. Depending on the material system, they may reflect a combination of reduced diffusion time, altered thermal gradients, non-equilibrium transformation pathways, and, in some cases, additional current-assisted mass-transport effects. Accordingly, comparison between DC-TMT and conventional data requires explicit consideration of both thermal history and material-specific sensitivity to applied current.

## 3.9. Inference and data interpretation

### 3.9.1. Assumptions, corrections, and identifiability

A distinctive and often dominant source of uncertainty in DC-TMT arises not only from instrumentation but from the assumptions embedded in data interpretation. Because DC-TMT often employs indirect measurement methods, such as strain inferred from resistance changes or strain rates derived from short test segments, the assumptions applied in data reduction strongly influence the results. For example, plastic strain is derived under the assumptions of constant volume and constant resistivity. In reality, resistivity evolves with dislocation density, recovery, recrystallisation, and oxidation. These microstructural effects can alter resistance independently of geometry, yet are often interpreted as mechanical strain, introducing an intrinsic interpretive uncertainty that cannot be resolved from the mechanical signal alone.

This problem is amplified in creep testing of superalloys, where the accuracy of strain–time curves depends critically on the thermal field within the specimen. Even small deviations from an ideal parabolic temperature profile across the gauge length can shift the local strain rate by several orders of magnitude, as the creep rate in superalloys depends exponentially on temperature. Even minor mischaracterisation of the axial thermal field propagates non-linearly into derived strain-rate parameters. In such cases, resistance–time data may capture not only creep strains but also artefacts from thermal gradients, drift, or oxidation. Misinterpreting these artefacts as genuine primary or secondary creep features can lead to erroneous conclusions about creep mechanisms. Furthermore, when stress–creep data are transformed into Zener–Hollomon plots, the chosen activation energy and assumed uniform temperature field have a significant influence on the derived constitutive parameters, thereby compounding interpretive uncertainty and obscuring the distinction between material response and experimental artefact.

Miniature specimen results must not be assumed to represent bulk behaviour without explicit justification, as that can be misleading, since surface-to-volume effects, gauge length, and specimen size alter deformation modes. We also observed from the literature that interpreting data from small-scale mechanical tests is inconsistent due to differing post-processing methods, data correction protocols, and yield definitions. Such variability in post-processing introduces investigator-dependent bias that can exceed the experimental scatter itself. Adjustments such as modulus scaling, necking corrections, or empirical curve shifting are applied variably across studies. For example, van Es et al. [64] applied a modulus-based shift to realign their stress-strain data, assuming ideal elasticity; a method that implicitly assumes homogeneous elasticity and disregards localised deformation inherent to miniature specimens.

Definitions of yield strength – via methods that implicitly assume homogeneous elasticity and disregard localised deformation inherent to miniature specimens – also vary, ranging from the standard 0.2% offset to alternative definitions based on maximum load or inflexion points, which creates inconsistency in reported values. Zheng et al. [35] noted that researchers often adopt custom approaches to interpreting stress-strain data, without universally accepted small-scale test standards. Additionally, differences in acquisition frequency (ranging from 1 Hz to 1000 Hz) affect the resolution of transient events, such as yield onset or crack formation. Mark et al. [146] cautioned that low data rates can under-specify critical deformation events, leading to systematic misinterpretation of deformation mechanisms and onset criteria. These inconsistencies underscore the need for harmonised data processing and reporting protocols to enhance reproducibility.

In DC-TMT, data interpretation constitutes an irreducible source of uncertainty arising from indirect measurement, non-uniform thermal fields, and investigator-dependent post-processing choices. While careful analysis can mitigate these effects, no universal data-reduction procedure exists that directly renders miniature Joule-heated test data equivalent to bulk measurements. Consequently, reported mechanical parameters must be interpreted within the context of the underlying assumptions used in their derivation.

### 3.9.2. Coupled electro-thermal–mechanical modelling

Finite element (FE) modelling, when combined with DC-TMT, has become a widely adopted and often necessary framework for understanding the coupled thermal, electrical, and mechanical behaviour of metallic materials under processing conditions representative of casting, welding, and forming. This hybrid approach bridges the gap between physical simulation and process-scale prediction, allowing the quantification of temperature gradients, phase transformations, and local strain fields that arise during resistive heating and deformation as the FE model resolves the actual thermal–electrical–mechanical fields the specimen experiences. In these experimentally calibrated (or data-driven) finite element models, the experiment provides time-resolved load, displacement, and thermocouple data from the test. The FE model, driven by the measured current or voltage boundary conditions, simulates the same cycle. A cost function, typically the normalised residual between measured and simulated force–temperature histories, is minimised by iteratively adjusting the constitutive parameters. However, temperature and strain are never uniform in specimens, so there is a need to correct or calibrate constitutive inferences accordingly. When researchers skip this coupling, their "constitutive laws" are frequently contaminated by hidden gradients and contact artefacts. This contamination represents an interpretive uncertainty rather than a numerical error, as it cannot be removed solely through post-processing.

Such modelling frameworks support constrained inference of local behaviour but do not constitute a unique or absolute reconstruction of intrinsic material properties. Table 2 illustrates that, even in well-instrumented DC-TMT experiments, uncertainty in inverse FE identification is dominated not by numerical error but by boundary condition assumptions, thermal measurement fidelity, and model regularisation choices. In many cases, these contributions exceed the intrinsic material scatter, emphasising that apparent variability in reported constitutive parameters often reflects analysis uncertainty rather than material behaviour.

Table 2: Principal sources of uncertainty and bias when using experimentally calibrated electro-thermo-mechanical finite element models to infer constitutive parameters from DC-TMT data.

| Stage | Primary source of uncertainty | Typical magnitude or range | Effect on derived parameters |
|---|---|---|---|
| Electrical & thermal input | Variability in electrical contact resistance and anvil pressure, as well as temperature-dependent resistivity. | ±10–20% in current density; ±15 °C in peak temperature. | Alters actual heat flux and temperature gradients; propagates into erroneous strain, stress, and activation energy [31,194]. |
| Thermal measurement | IR/TE emissivity error, oxidation, and thermocouple lag. | ±10–20 °C (IR/TE), ±5–10 °C (thermocouple). | Shifts computed activation energy ($\Delta Q \approx$ ±10–25%) [31,195]. |
| Mechanical field measurement | DIC speckle degradation, thermal shimmer, and calibration drift. | ±2–5% strain uncertainty. | Distorts the local strain-rate field, biasing the fitted strain-rate sensitivity $n$ [196]. |
| FE boundary conditions | Unknown or assumed heat-transfer and friction coefficients. | ±20% variation in coefficient values. | Affects stress localisation and recovery behaviour [31,195]. |

| Stage | Primary source of uncertainty | Typical magnitude or range | Effect on derived parameters |
|---|---|---|---|
| Material property inputs (resistivity, conductivity) | Poorly characterised temperature dependence beyond calibration range. | ±10–15% uncertainty. | Alters the electro-thermal coupling and stored energy terms in FE [194,197]. |
| Numerical discretisation | Mesh density, time step stability, and solver damping. | ±2–5% deviation in predicted stress or temperature. | Introduces smoothing or oscillations that mislead the inverse fit [31]. |
| Inverse-model optimisation | Non-unique minima; algorithm sensitivity to initial guess. | Parameter standard deviation ±5–10%. | Leads to apparent scatter in fitted Q, n, and A constants [195,196]. |
| Regularisation & weighting | Over-smoothing of local heterogeneity in multi-objective inversions. | Bias up to ±10% in fitted constants. | Masks genuine strain localisation or recrystallisation kinetics [195]. |
| Thermal–mechanical coupling omission | Neglect of deformation heating or latent heat during DRX. | ±10–20 °C underestimation of local temperature. | Underestimates DRX activation energy and kinetic rate [197]. |
| Reporting & data reduction | Using nominal gauge temperature or engineering stress–strain curves. | Systematic error: $Q$ underestimated by 10–30%; n overestimated by 5–15%. | Apparent material "scatter" arises from analysis error, not intrinsic variability [194,196]. |

Thus, as shown in Figure 16, a credible model begins with the electro-thermal problem, rather than the mechanics. Pantalé et al. [31] built an axisymmetric thermal–electrical model in Abaqus and drove it with a user amplitude (UAMP) that mimics the Gleeble PID loop. With measured anvil temperatures as boundary conditions and realistic contact layers, their predictions matched multi-T/C data within approximately two degrees, revealing that graphite interlayers suppress end losses and gradients during heating and cooling. That is the appropriate baseline.

Simplified assumptions that neglect these couplings systematically underestimate axial gradients and distort the inferred constitutive response, as demonstrated by Yu et al. [194]. When we assume the gauge is isothermal and isostrain, we recover a neat stress–strain curve that is incorrect. Yu et al.'s thermo-mechanically coupled FE model quantified the true temperature and strain fields during the DC heating compression test, then iteratively refined the constitutive law so that the forward FE response reproduced the experiment [194]. Their study demonstrates that inverse identification without resolving coupled fields yields internally consistent but physically incorrect constitutive parameters. Similarly, Rahul et al. [198] linked thermal compression data of $AlCoCrFeNi_{2.1}$ eutectic high-entropy alloys to FE simulations of material strain, enabling the construction of processing maps and identifying stability regimes during forging where strain localisation and thermal gradients coincide.

Moreover, if the model idealises the jaws as "20 °C sinks" and assigns arbitrary heat-transfer coefficients, it will overcool the gauge and underpredict softening; thus, similar to what Pantalé et al. [31] have done, the anvil thermal state should be measured, and correct electrical and thermal properties should be used to reproduce the cooling tail and the small but consequential axial gradients. These errors propagate directly into strain-stress, activation energy, and strain-rate sensitivity, particularly at high homologous temperatures.

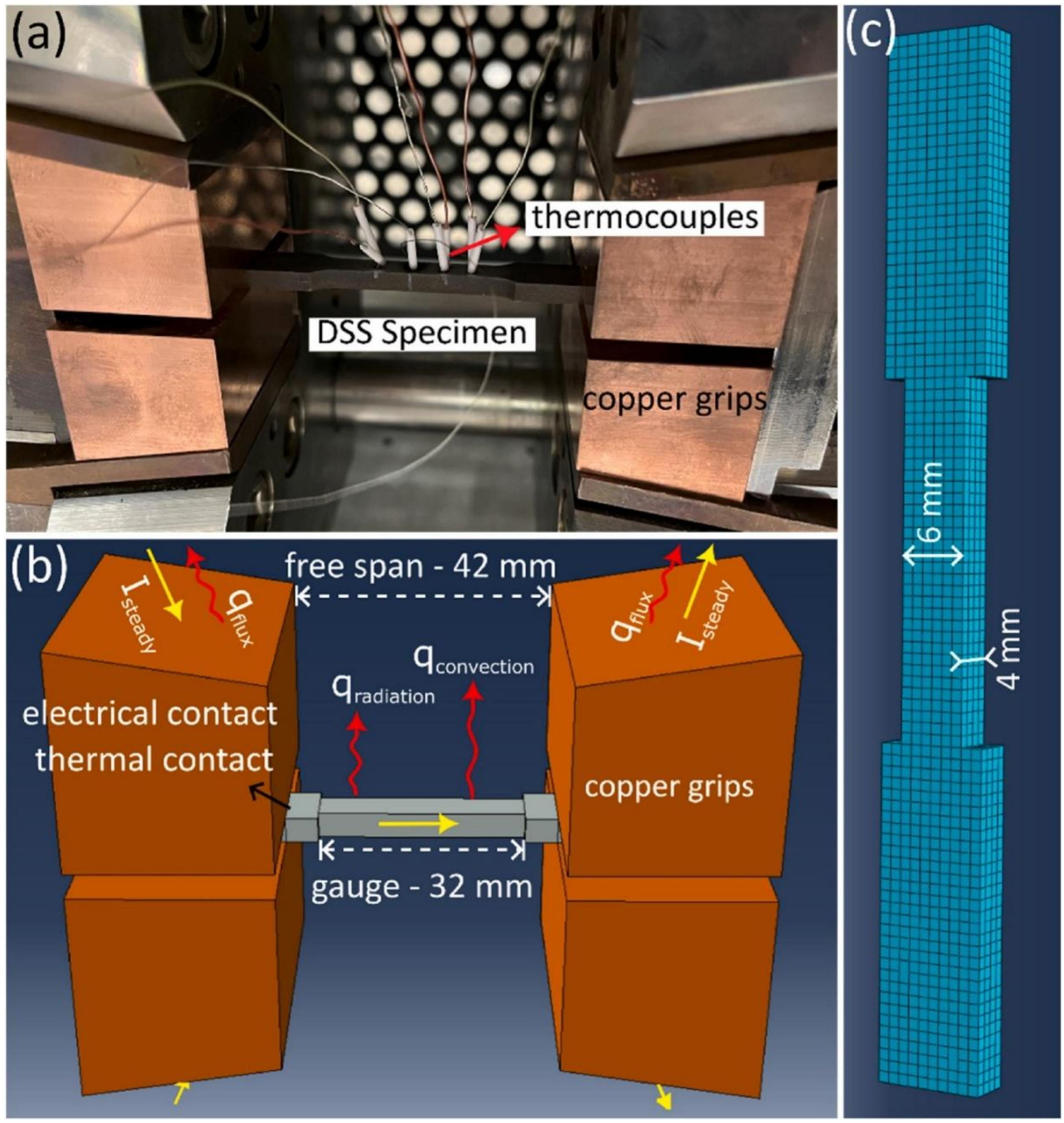

Figure 16: Experimental and modelling set-up used for Gleeble thermal simulations. (a) Gleeble chamber showing the duplex stainless steel specimen clamped between water-cooled copper grips with thermocouples welded at prescribed locations along the gauge. (b) Schematic of the coupled electrical-thermal finite element model replicating the experimental geometry and boundary conditions. (c) Mesh layout of the tensile coupon used for the simulations. Adapted from [199], licensed under CC BY 4.0.

The integration of microstructural evolution models within FE frameworks has also advanced markedly. Jin et al. [197] implemented incremental dynamic, meta-dynamic, and static recrystallisation models into a thermo-mechanical FE solver to simulate multi-pass deformation of AZ31 alloys, with kinetics parameters derived from thermal strain-stress curves. The piecewise-linear formulation reproduced local grain size distributions and verified experimental recrystallisation fractions, showing how direct current heating data can calibrate predictive models of grain evolution during hot working. While powerful, such integrated models introduce additional uncertainty through the identification of kinetic parameters and model form, which must be distinguished from experimental uncertainty.

On the other hand, during the DC-heated tensile test, the specimens develop non-uniform fields long before the required strain range is reached. Li et al. [196] addressed this problem by using shear specimens with DIC, then they coupled an electro-thermo-mechanical FE model with an optimisation-based identification to match the DIC field and obtain a viscoplastic hardening law up to failure. They also showed that shear tests are less sensitive to transformation noise during continuous cooling. These approaches highlight that alternative specimen geometries can reduce uncertainty arising from axial thermal gradients but introduce new challenges in stress interpretation.

Moreover, Chen et al. [200] coupled finite element thermal–mechanical models with hot tearing criteria for DC-cast AA6111 aluminium. Their model, validated against semi-solid constitutive laws, revealed the interplay between pore fraction, strain rate, and thermal gradient during the initiation of hot tearing. By integrating semi-solid stress–strain data from the tests, the model accurately predicted critical regions of cracking as functions of casting speed and secondary cooling rate. Comparable approaches have been used in spot-weld studies, where FE models integrate electrical contact resistance, local heating, and HAZ mechanical response to predict fatigue and crack initiation in low-carbon and advanced high-strength steels [201].

## 3.10. Residual and instrument-specific uncertainties

One of the most persistent and difficult-to-quantify sources of uncertainty in DC-TMT arises from calibration procedures and operator-dependent bias [202]. DC-TMT systems are typically complex to operate, requiring a series of pre-test checks that include load cell calibration, verification of thermocouple polarity, and tuning of the control system. Any deviation or oversight during these steps can introduce systematic error into the test data. Because these errors are systematic rather than stochastic, they do not average out with repeated testing and can propagate consistently across an entire dataset.

The thermal history of the specimen before testing is another critical factor. Heating steps undertaken for system calibration, such as tuning the PID controller or checking thermocouple polarity, can inadvertently alter the microstructure of sensitive alloys. Such pre-test thermal exposure is rarely documented and is typically indistinguishable from the formal test history in reported data. In superalloys, for example, prior supersolvus overheating introduces uncertainties in tensile and creep data by degrading ductility and shifting the balance of strengthening phases [203–207]. Similar sensitivity to unintended thermal exposure exists in other precipitation-strengthened and transformation-sensitive alloys, including Ti- and Al-based systems.

Statistical uncertainty constitutes a further, often underappreciated limitation of DC-TMT studies, arising from both limited specimen counts and intrinsic size effects associated with miniature testing. This limited sampling reduces statistical confidence and makes it challenging to distinguish between inherent material behaviour and experimental noise. The problem is compounded by the statistical size effect, whereby smaller specimens are less likely to contain critical flaws. As shown by Tomaszewski [40], this can skew fatigue results, as low-carbon steels exhibit little sensitivity, whereas stainless steels show pronounced differences in fatigue life with changes in specimen size. Despite these challenges, many published DC-TMT studies omit essential statistical measures such as standard deviations, error bars, or confidence intervals, rendering quantitative comparison and uncertainty propagation impossible. The absence of such metrics complicates cross-study comparisons and prevents meaningful synthesis of data for meta-analysis or design applications. As a result, apparent agreement or disagreement between studies may reflect statistical artefacts rather than genuine differences in material behaviour.

The uncertainty sources outlined in Section 3 are rarely dominant individually, but they interact with primary experimental and interpretive uncertainties to bias results in systematic ways. Their cumulative effect becomes particularly significant when DC-TMT data are used for model calibration, database construction, or cross-laboratory comparison, highlighting the need for transparent reporting of calibration procedures, thermal history, and statistical confidence.

# 4. Hierarchy of uncertainty and domains of validity

The uncertainty sources identified in Section 3 do not contribute equally to the interpretability of DC-TMT data. Their relative importance depends on the test objective, the material system, and the imposed temperature–stress regime, but not all sources are interchangeable or compensable. This section, therefore, establishes a hierarchy of uncertainty sources, distinguishing between dominant, secondary, and conditional contributions based on their impact on inferred material parameters, rather than their ease of mitigation. Several uncertainty sources are intrinsic to Joule-heated miniature testing and cannot be eliminated through calibration, modelling, or post-processing. Misclassification or neglect of dominant uncertainty sources invalidates the quantitative interpretation of DC-TMT data, irrespective of downstream experimental refinement.

Thus, the proposed hierarchy below highlights that uncertainty mitigation in DC-TMT should prioritise the control and characterisation of temperature fields, heating rate, and specimen geometry before refining secondary factors, such as strain measurement or control-loop tuning. Misclassification or neglect of dominant uncertainty sources invalidates the quantitative interpretation of DC-TMT data, regardless of subsequent calibration or data processing efforts.

## 4.1. Dominant uncertainties controlling inference

Dominant uncertainty sources are those that directly define the local thermo-mechanical state experienced by the material and whose mischaracterisation renders extracted material parameters non-physical, non-transferable, or mechanistically ambiguous. These uncertainties cannot be compensated for by improved instrumentation, control, or data processing once violated, but need to be eliminated or minimised first.

- Temperature field and thermal gradients: The spatial and temporal temperature field within the specimen is the dominant source of uncertainty in DC-TMT. Joule heating inherently produces axial and radial temperature gradients, such that any assumption of isothermal conditions – beyond the *effective gauge length* – introduces systematic bias into strain rate, phase stability, and constitutive parameter identification. If the axial temperature distribution over the effective gauge length is not independently characterised, DC-TMT data cannot be interpreted as a single-temperature constitutive response. Errors in peak temperature, gradient shape, or effective gauge definition propagate non-linearly into derived quantities such as activation energy, stress exponent, and creep mechanism attribution. Even when the central temperature is tightly controlled, unquantified thermal gradients can dominate the apparent material response.

Errors in peak temperature, effective gauge length definition, or gradient shape propagate non-linearly into derived quantities, such as activation energy, stress exponent, and attribution of creep mechanisms. Even when the central temperature is well controlled, unquantified axial gradients can dominate the apparent material response. Consequently, temperature-field uncertainty remains first-order even in well-instrumented experiments.

- Heating rate and thermal history: Closely coupled to the temperature field is the imposed heating rate and prior thermal history. In DC-TMT, rapid Joule heating defines a distinct kinetic regime rather than approximating equilibrium thermal conditions. Heating rate, therefore, does not merely introduce uncertainty but alters transformation temperatures, diffusion-controlled processes, and microstructural evolution relative to furnace-based experiments. A direct comparison of DC-TMT data with conventional thermal treatments, without accounting for heating-rate effects, is physically ill-posed. This uncertainty source is dominant whenever phase transformations, precipitation, recovery, or recrystallisation influence the measured response, particularly in steels, titanium alloys, and precipitation-strengthened superalloys.
- Specimen geometry and effective gauge definition: Inappropriate specimen geometry or misdefinition of the effective gauge volume renders extracted mechanical parameters physically meaningless in DC-TMT. Miniaturised specimens amplify sensitivity to temperature gradients, misalignment, and strain localisation, while non-standard geometries complicate stress and strain interpretation. Errors in effective gauge dimensions directly contaminate calculated strain, strain rate, and stress. For materials exhibiting anomalous yielding or strong temperature-dependent strength, unsuitable geometry can prevent deformation from localising within the intended

thermal–mechanical region. Once deformation occurs outside the defined gauge volume, no post-hoc correction can recover physically interpretable material parameters.

## 4.2. Secondary uncertainties and interaction effects

Secondary uncertainty sources influence the measured response but cannot dominate interpretation unless dominant uncertainties associated with temperature, heating rate, and gauge definition are already constrained.

- Strain measurement and data reduction: Strain measurement methods, including potential-drop techniques, crosshead displacement correction, and optical approaches, constitute secondary uncertainty sources in DC-TMT. These techniques rely on assumptions regarding deformation homogeneity, resistivity stability, and deformation mode, which introduce bias when these assumptions are violated. Data-reduction choices such as yield definition, modulus correction, and filtering further contribute to investigator-dependent variability. However, refinement of strain measurement or data processing cannot compensate for ill-defined temperature fields or ambiguity in effective gauge volume. Such uncertainties become significant only when compounded with first-order thermal and geometric effects.
- PID control and system dynamics: Control-system dynamics, including PID tuning and thermal lag, represent a secondary source of uncertainty in DC-TMT. Poor tuning can introduce temperature oscillations or overshoot that affect transient deformation behaviour and phase transformation timing. However, stable control of the thermocouple signal does not imply control of the specimen thermo-mechanical state. When properly tuned and validated, control-loop uncertainties are typically smaller than uncertainties arising from spatial temperature gradients and heating-rate effects.
- Finite-element-assisted interpretation: Finite-element modelling can assist in managing uncertainty associated with coupled thermal, electrical, and mechanical fields, but it also introduces model-form, boundary-condition, and inverse-identification uncertainties. When insufficiently constrained, finite-element correction can increase apparent internal consistency while reducing physical validity. Such modelling should therefore be treated as an uncertainty-management tool rather than a source of ground truth, and its outputs remain subordinate to the quality of the underlying experimental definition.

## 4.3. Conditional and material-specific uncertainty sources

- Several uncertainty sources discussed in Section 3 are conditional and become relevant only for specific material classes or testing conditions. These include current-assisted diffusion and electromigration, electropulsing phenomena, magnetisation-related coupling in magnetically active materials, and environment-dependent interactions such as oxidation or hydrogen uptake. For most structural alloys tested at high temperature, these effects are secondary to thermal gradients and heating-rate effects. Attribution of observed behaviour to current-assisted mechanisms is unjustified unless thermal explanations have been independently excluded. However, in magnetostrictive alloys, multiferroics, oxidation-sensitive systems, or long-duration creep tests, such mechanisms may become non-negligible and should be evaluated explicitly.
- Irreducible uncertainty and limits of equivalence: Some uncertainty sources in DC-TMT are irreducible within the current experimental paradigm. These include the co-existence of temperature gradients with mechanical loading, reliance on indirect strain measurement, and the kinetic divergence between rapid Joule heating and conventional thermal treatments. No amount of calibration can render DC-TMT data thermodynamically equivalent to bulk furnace-based measurements. Discrepancies between DC-TMT and conventional data should therefore not be automatically interpreted as experimental artefacts; in many cases, they reflect genuinely different thermo-mechanical histories and deformation mechanisms.

# 5. What DC-TMT reveals: capabilities, limits, and validation domains

Applications of DC-TMT must be interpreted in light of the dominant and irreducible uncertainty sources identified in Sections 3 and 4. While DC-TMT enables the simultaneous application of thermal and mechanical loading, it does not reproduce service conditions in a generic or thermodynamically equivalent sense. Instead, it imposes a specific electro–thermo–mechanical history characterised by rapid Joule heating, spatial temperature gradients, and limited gauge volumes. These characteristics fundamentally distinguish DC-TMT from furnace-based thermo-mechanical testing and condition the type of information that can be reliably extracted [208,209].

Consequently, the value of DC-TMT lies not in direct substitution for standardised bulk testing, but in its ability to probe deformation mechanisms, transformation kinetics, and relative property trends under tightly defined experimental constraints [53,154]. Numerous studies have employed DC-TMT to investigate coupled thermo-mechanical behaviour in aerospace alloys, additively manufactured materials, and nuclear materials [12,129,210–214]. In such cases, the technique has been used to examine thermal stability, oxidation-assisted damage, phase transformation behaviour, and high-temperature deformation response [154,215,216].

Accordingly, rather than cataloguing applications, this section highlights representative cases in which DC-TMT has demonstrated reproducible and interpretable insight, and clarifies the conditions under which such interpretation is valid.

## 5.1. Zero force thermomechanometry

### 5.1.1. Casting simulation

Establishing a robust and efficient casting process requires control over solidification mechanisms, thermal gradients, and defect formation mechanisms, and is therefore commonly supported by numerical casting simulations before process implementation [217]. Experimental validation of such simulations typically relies on large-scale physical simulators capable of reproducing liquid-phase behaviour, solidification shrinkage, interdendritic feeding, and mould–metal heat extraction. In contrast, DC-TMT operates exclusively in the solid state and therefore cannot address liquid-phase phenomena or solidification-controlled defect formation [12].

As a result, DC-TMT-based casting simulations are confined to post-solidification thermo-mechanical histories, such as cooling through the solidus under imposed strain. This distinction is especially relevant because conventional bulk testing averages over large material volumes and obscures the contribution of individual dendritic or microsegregated regions to later deformation behaviour [12]. While full-scale systems, such as the Gleeble, have been extensively used to simulate continuous casting and semi-solid processing, primarily in aluminium alloys [8] (discussed in Section 5.3.1), analogous ETMT studies on miniaturised specimens are scarce, focusing mainly on nickel-based superalloys [12].

### 5.1.2. Weld HAZ simulation

DC-TMT has been widely used to physically simulate weld heat-affected zones (HAZs) by imposing controlled thermal cycles and, where required, mechanical constraint in the solid state. Unlike full welding experiments, these simulations do not reproduce melt-pool dynamics or fusion-zone phenomena; however, they can replicate the rapid heating and cooling rates experienced by material volumes adjacent to the weld. This makes DC-TMT particularly suited to studying phase transformations, thermal expansion, and post-transformation mechanical response within HAZ sub-regions under well-defined thermal histories [218,219].

In structural steels, DC-TMT-based physical simulation has been used extensively to isolate individual HAZ regions that are only millimetres thick in real welds and therefore inaccessible to direct mechanical testing. Studies have shown that carefully prescribed thermal cycles can reproduce martensitic and bainitic transformation sequences, thermal strain evolution, and stress redistribution during simulated multi-pass welding, yielding responses that align with finite-element predictions when equivalent thermal histories are imposed [220]. These applications exploit DC-TMT's ability to decouple thermal history from geometric complexity.

In advanced high-strength steels, Midawi et al. [109] used a Gleeble-3500 to simulate the sub-critical HAZ (SCHAZ) of resistance-spot welds by applying temperature–time cycles below $Ac_1$ (≈350–650 °C). High-speed infrared imaging was used to verify spatial temperature uniformity during testing, and hardness changes associated with martensite tempering and carbide precipitation were quantified (Figure 17). Coupling the experimentally measured thermal histories with a Hollomon–Jaffe parameter enabled prediction of hardness and tensile response across the SCHAZ, demonstrating that softening or hardening depends on the base-metal microstructure and alloy chemistry rather than on the thermal cycle alone.

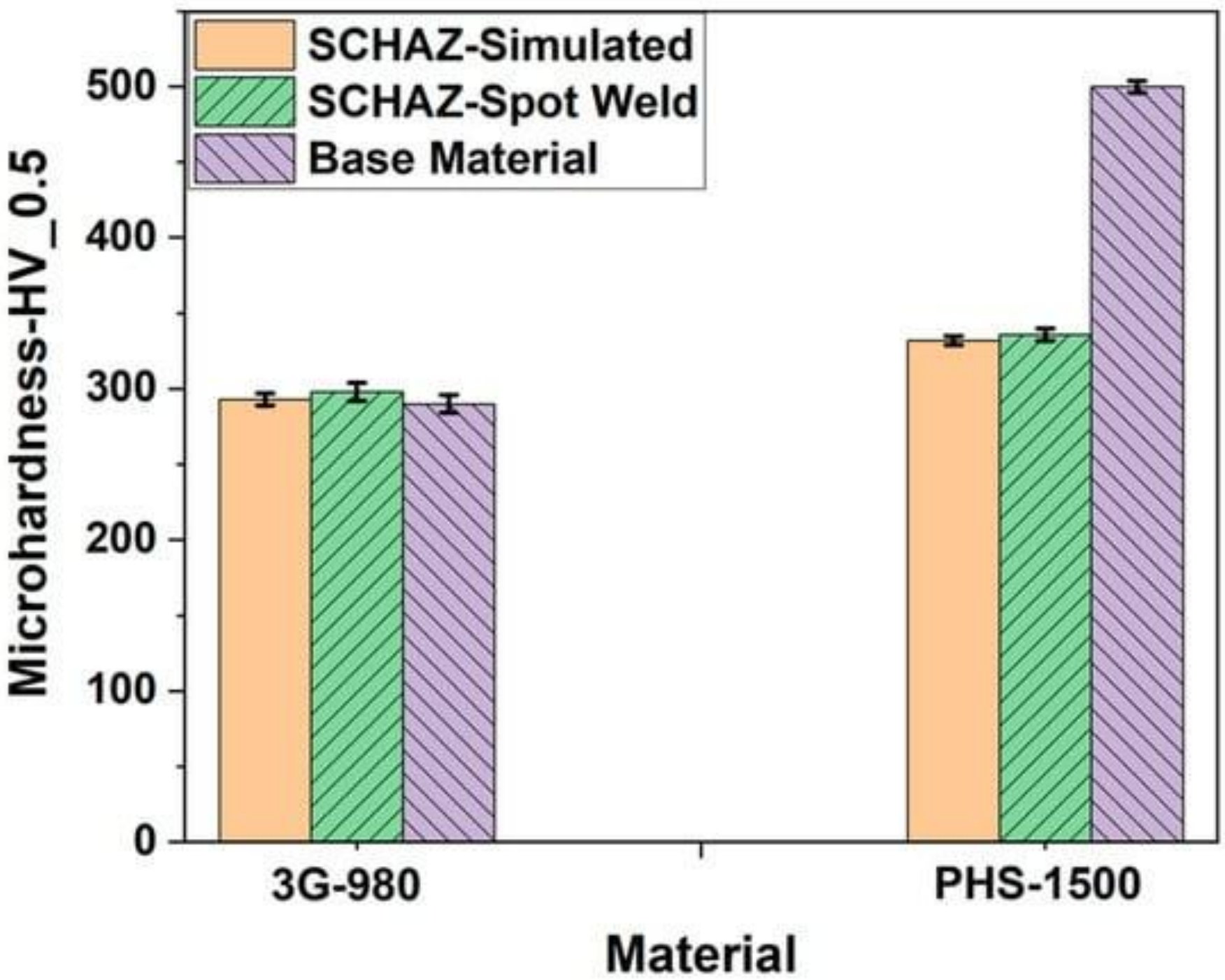

Figure 17: Comparison of Vickers microhardness ($HV_{0.5}$) measured in the sub-critical heat-affected zone (SCHAZ) of resistance-spot welds and in Gleeble-simulated SCHAZ specimens subjected to equivalent sub-$Ac_1$ thermal cycles. The agreement reflects comparable tempering and carbide precipitation kinetics under controlled thermal histories. Base-material hardness is shown for reference. Adapted from [109].

For heavy-wall pipeline steels, resistive heating has enabled targeted simulation of coarse-grained and intercritically reheated HAZ regions produced during multi-pass welding. Moeinifar et al. [221] imposed sequential peak temperatures of 1400 °C and 800 °C to simulate double-pass tandem submerged-arc welding of X80 steel, reproducing martensite–austenite constituent morphology observed in real welds. Their results showed that increasing heat input enlarged M/A constituents and reduced Charpy toughness, highlighting the sensitivity of fracture performance to thermal history within narrow HAZ regions. Fazeli et al. [222] extended this approach to multiple X80 grades using a Gleeble-3800, combining

Charpy and CTOD testing on thermally cycled specimens to link cooling time ($\Delta t_{800-500}$ = 1–100 s), transformation products, and fracture toughness. These studies demonstrate the value of DC-TMT for ranking HAZ performance under controlled thermal conditions.

Miniaturised ETMT-style testing has similarly been applied to stainless steels, where proof stress and hardness vary sharply across the weld region. Physical simulation studies have shown that HAZ properties often lie between those of the fusion zone and the parent material, and that DC-based heating enables systematic mapping of local hardening, recovery, and stress-relaxation behaviour driven by phase transformations [223,224]. Such approaches clarify metallurgical contributions to HAZ mechanical heterogeneity, although they do not capture macroscopic residual-stress redistribution.

Beyond ferrous systems, DC-TMT has been applied to nickel-based superalloys, where HAZ widths are extremely narrow and direct testing is impractical. Liu et al. [225] simulated TIG welding thermal cycles in IN617 between 1150 °C and 1350 °C, generating enlarged representative HAZ volumes for microstructural and mechanical characterisation. Their results linked carbide coarsening and lamellar grain-boundary formation to reductions in yield and tensile strength at high peak temperatures. Popoola et al. [226] extended this approach to Astroloy, a nickel-based superalloy, demonstrating that the application of compressive strain during thermal simulation promotes dissolution of liquated grain-boundary phases and improves resistance to HAZ cracking in powder-metallurgy nickel alloys.

DC-TMT thermo-mechanical simulation has also been used to study HAZ cracking susceptibility in aluminium alloys. In AA7017, controlled heating and cooling cycles were used to map the stress–temperature conditions for liquation cracking, showing that faster post-weld cooling reduces the time during which interdendritic liquid films experience tensile restraint and thereby lowers crack susceptibility [227]. In friction-welded Al–Fe joints, rapid post-weld thermal conditioning simulated via DC heating revealed thickening and phase transformation of Fe–Al intermetallic layers consistent with interdiffusion kinetics near 500 °C [228]. In a complementary modelling context, Heikebrügge et al. [229] incorporated DC-TMT-derived constitutive parameters for base metal, HAZ, and filler into finite-element simulations of post-weld deep rolling, demonstrating improved prediction of residual-stress modification near the weld toe when physically simulated material data were used.

Taken together, these studies demonstrate that DC-TMT offers a controlled and repeatable method for reproducing weld-relevant thermal histories in the solid state and for isolating individual HAZ sub-regions for microstructural and mechanical characterisation. Its strength lies in enabling quantitative ranking of HAZ response and informing constitutive and damage models under well-defined thermal cycles.

### 5.1.3. Phase transformation kinetics

Thermal analysis techniques such as differential thermal analysis (DTA) and differential scanning calorimetry (DSC) are routinely used to identify transformation temperatures associated with phase changes under controlled heating and cooling conditions. These methods provide robust equilibrium and near-equilibrium transformation data but offer limited insight into spatially resolved microstructural evolution or into transformation kinetics under rapid, non-isothermal thermal histories. In contrast, DC-TMT equipped with in situ resistivity measurement enables continuous monitoring of phase-fraction evolution during rapid heating and cooling in the solid state, under thermal conditions that are difficult to access using conventional calorimetric techniques [230,231].

In DC-TMT, changes in electrical resistance during thermal cycling have been used to identify phase dissolution, reprecipitation, and recovery processes, particularly under high heating rates. Comparative studies have shown that resistivity-based detection is more sensitive than DSC or DTA to transformation events during rapid heating (up to ~100 °C $s^{-1}$), where calorimetric signals are broadened or suppressed [207]. This approach has been widely applied to nickel-based superalloys to investigate in situ precipitation and dissolution behaviour of the γ′ phase during heating, isothermal holding, and cooling [207,232]. However, resistivity changes reflect the combined influence of temperature, phase fraction, defect density, and lattice strain, and therefore cannot be interpreted as direct measures of phase fraction without complementary structural information.

Recent developments increasingly combine DC-TMT with complementary in situ techniques, such as synchrotron or neutron diffraction, and digital image correlation, to decouple resistivity contributions arising from temperature, elastic strain, and phase evolution. These hybrid approaches are essential for constraining dominant uncertainties associated with temperature gradients and indirect phase sensing, and for rendering phase-transformation data obtained under DC-TMT conditions physically interpretable. For example, studies combining DC-TMT with diffraction and microscopy have shown how rapid thermal transients influence precipitate size, morphology, and spatial distribution, as well as lattice-parameter evolution of γ and γ′ phases under controlled thermo-mechanical histories [119,232,233]. Figure 18 illustrates an example in which lattice-parameter evolution measured during an isothermal hold is correlated with coarsening and redistribution of γ′ precipitates observed by microscopy.

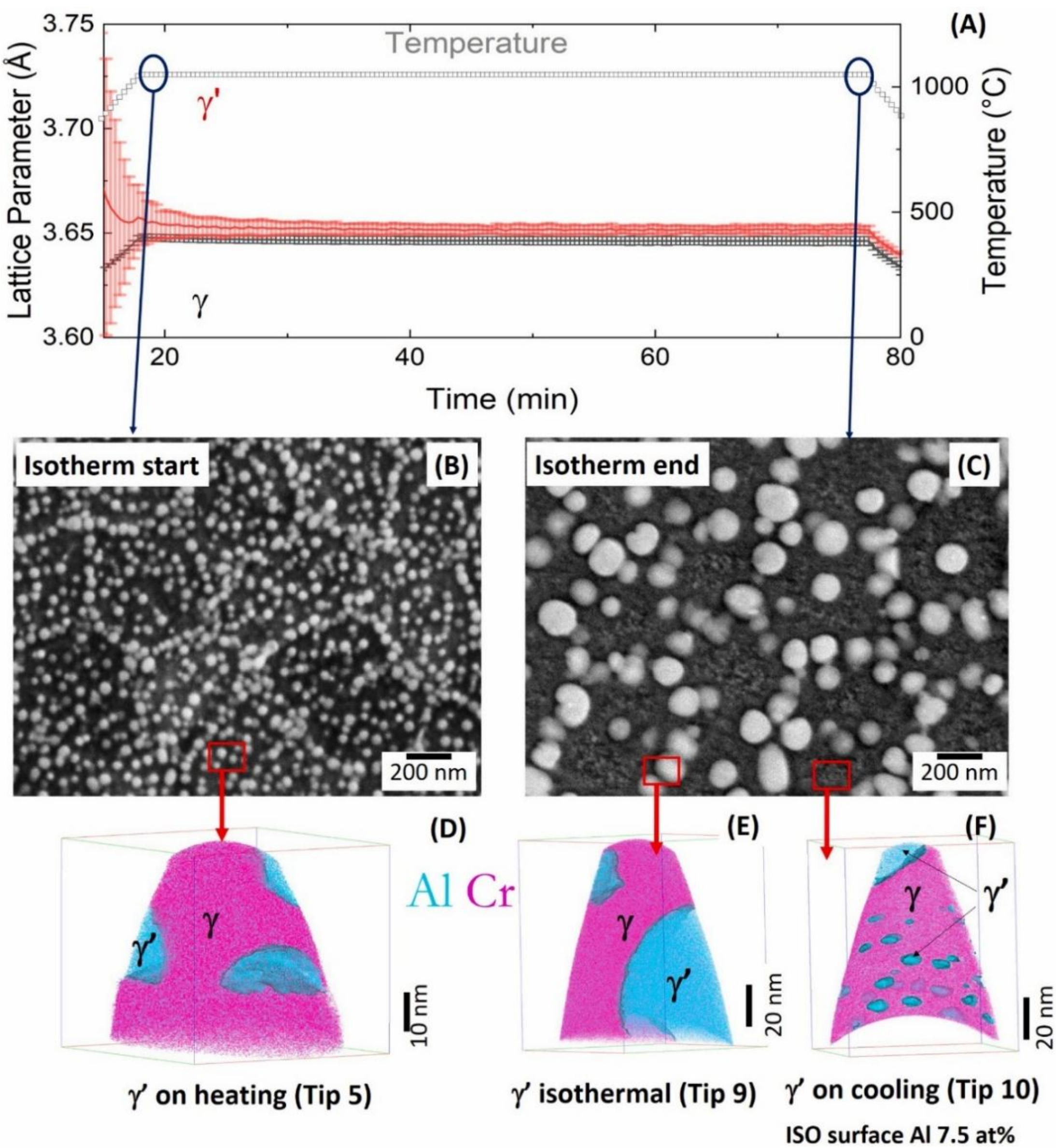

Figure 18: (a) Evolution of lattice parameters of the γ (black) and γ′ (red) phases during an isothermal hold measured in situ during DC-TMT; the imposed temperature history is shown for reference. (b, c) Representative microstructures at the start and end of the isothermal hold, illustrating the coarsening of γ′ precipitates. (d–f) Three-dimensional reconstructions highlighting changes in γ′ morphology

during heating, isothermal holding, and cooling. The combined data illustrate a correlation between lattice-parameter evolution and γ′ precipitate coarsening under controlled, non-equilibrium thermal histories. Adapted from [232], licensed under CC BY 4.0.

### 5.1.4. Environment-assisted cracking and oxidation-controlled embrittlement

Environmental-assisted cracking, driven by high-temperature oxidation, is a major limitation for the use of nickel-based superalloys in oxidising environments, such as gas turbines and power-generation systems [234]. Conventional test methods, including low-cycle fatigue and dwell-fatigue testing on standard-sized specimens, integrate multiple interacting phenomena, such as oxidation kinetics, stress redistribution, and microstructural evolution, making it challenging to isolate the mechanisms responsible for stress-assisted oxidation-induced embrittlement [235,236]. As a result, physically based interpretation of cracking mechanisms from bulk tests alone remains challenging.

DC-TMT, especially the ETMT, has therefore been employed as a mechanistic probe to accelerate oxidation-assisted damage and to isolate microstructural mechanisms associated with environmentally assisted cracking. For example, DC-TMT has been employed to investigate stress-assisted intergranular oxidation in nickel-based superalloys under controlled thermal and mechanical loading conditions. The high surface-to-volume ratio of miniature specimens increases oxygen ingress relative to the load-bearing section, enabling time- and temperature-dependent oxidation effects to be observed over experimentally accessible durations [11].

For the 720Li superalloy, DC-TMT studies have demonstrated that oxidation-assisted cracking is linked to internal intergranular oxidation along γ grain boundaries, particularly at incoherent γ/γ′ interfaces. Oxygen ingress promotes the formation of layered oxide structures involving NiO, $NiCr_2O_4$, $Cr_2O_3$, and $Al_2O_3$, which degrade grain-boundary cohesion and promote embrittlement under applied stress [237]. Figure 19 schematically illustrates the proposed oxidation morphologies observed under DC-TMT conditions, highlighting the differences between γ grain boundaries and γ/γ′ interfaces and their role in crack advance. In these experiments, failure is attributed to the combined effects of oxide-induced loss of cohesion and stress-assisted grain-boundary deformation, rather than to oxidation alone.

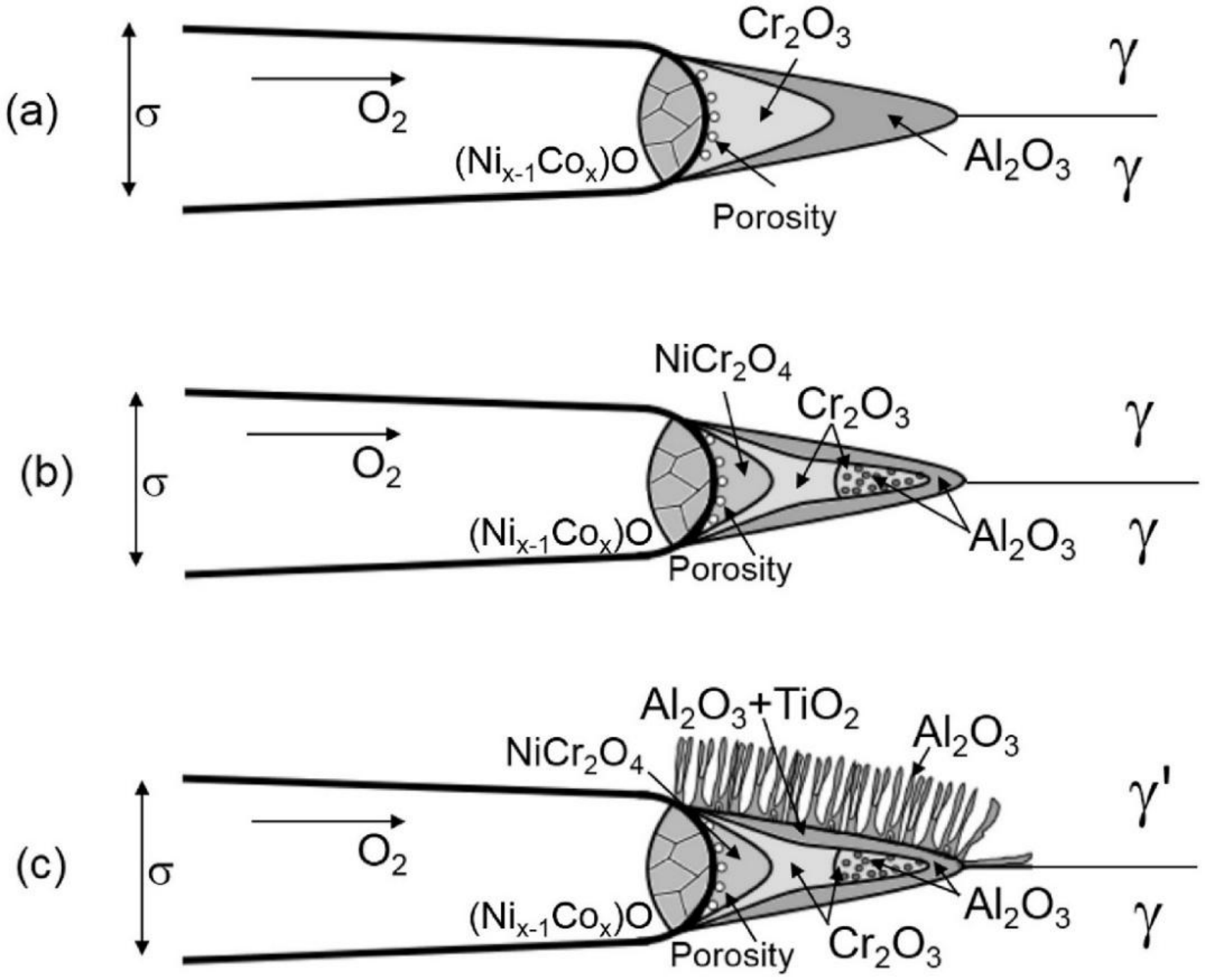

Figure 19: Schematic illustration of oxide morphologies observed during stress-assisted oxidation under DC-TMT conditions. (a) Layered crack-tip oxide structure illustrating the formation of NiO, $NiCr_2O_4$, $Cr_2O_3$, and $Al_2O_3$ under accelerated oxidation conditions. (b) and (c) depict oxidation morphologies reported for alloy 720Li at γ grain boundaries and at primary γ/γ′ interfaces, respectively. The schematic highlights interface-dependent oxidation mechanisms observed in miniature specimens subjected to combined thermal and mechanical loading. Adapted from [237], licensed under CC BY 4.0.

Taken together, these studies demonstrate that DC-TMT is effective for mechanistically interrogating oxidation-assisted embrittlement mechanisms in nickel-based superalloys, particularly in identifying susceptible interfaces and oxidation morphologies. However, the accelerated oxidation behaviour observed in miniature ETMT specimens must be interpreted with caution. The very high surface-to-volume ratio of typical DC-TMT coupons (on the order of millimetres in cross-section) means that even thin oxide scales represent a significant fraction of the load-bearing area, amplifying apparent section loss and stress uncertainty [161]. In addition, direct Joule heating produces rapid thermal transients and non-uniform temperature fields that can alter oxidation kinetics, oxide morphology, and scale adherence relative to furnace-heated specimens [162]. Consequently, oxidation proceeds more rapidly and often exhibits early transition to breakaway behaviour in DC-TMT, whereas oxidation in standard ASTM/ISO specimens is generally slower and contributes less directly to mechanical property degradation over comparable timescales [161,162]. Thus, the magnitude and kinetics of oxidation damage observed in miniature specimens are not directly transferable to bulk components, and DC-TMT results should be interpreted as upper-bound or accelerated representations of environmentally assisted cracking.

## 5.2. High-temperature mechanical testing under Joule heating

### 5.2.1. Monotonic deformation

DC-TMT has been extensively used for monotonic tensile (Figure 14b) and compressive testing of metallic alloys at elevated temperature, particularly where material availability, microstructural heterogeneity, or rapid thermal transients limit the applicability of conventional bulk tensile testing. Its use of miniature specimens enables targeted interrogation of local material response under controlled thermal histories. This has proven valuable for additively manufactured titanium alloys and nickel-based superalloys, where conventional specimen geometries are costly to produce and where spatial variability in microstructure motivates localised testing approaches [22].

Under monotonic loading, DC-TMT enables tensile and compressive deformation to be imposed over a wide range of temperatures and strain rates, while maintaining rapid heating and cooling through direct Joule heating. However, as established in Sections 3 and 4, interpretation of monotonic DC-TMT data is contingent on an accurate definition of the effective gauge volume, temperature field, and heating rate. In the absence of such control, apparent elastic stress, ductility, and strain-rate sensitivity may reflect thermal gradients or localisation effects rather than intrinsic material behaviour. Consequently, monotonic DC-TMT data are most robust when used for comparative assessment, trend identification, or calibration of physically informed constitutive models.

Beyond uniaxial loading, DC-TMT systems have been adapted for small-scale three-point bending tests by integrating displacement and force transducers [238]. Such configurations enable investigation of temperature-dependent flexural response in materials where tensile testing is impractical. For example, bending tests on NiTi shape-memory alloys under controlled thermal and mechanical loading have been used to correlate temperature-dependent deflection with thermoelastic martensitic transformation behaviour [239]. In these studies, the relationship between bending deflection and applied stress was interpreted using Euler–Bernoulli beam theory, providing qualitative insight into transformation-induced deformation.

### 5.2.2. Thermo-mechanical fatigue and cyclic deformation

DC-TMT has been applied to thermo-mechanical fatigue (TMF) studies where the objective is to probe fatigue damage mechanisms under combined cyclic thermal and mechanical loading using miniature specimens. Its primary advantage in this context is the ability to impose well-defined temperature–strain histories on small material volumes, enabling fatigue testing when material availability is limited or when localised microstructural response is of interest [59,240].

In nickel-based superalloys, DC-TMT-based TMF experiments have been used to investigate the effects of phase angle, thermal cycle, and strain amplitude on fatigue life and crack initiation behaviour. Studies on Nimonic 90 have shown that, under carefully matched thermal–mechanical loading mechanisms, miniature-specimen TMF lives and dominant failure modes can be comparable to those observed in larger-scale tests across different phase angles [59]. However, such an agreement depends critically on control of temperature gradients, gauge definition, and surface condition, and should not be interpreted as general equivalence between miniature DC-TMT and full-scale fatigue tests. Rather, these results demonstrate that DC-TMT can reproduce key fatigue damage mechanisms under constrained conditions.

DC-TMT has also been used to compare the thermo-mechanical fatigue response of coatings and surface-engineered systems for gas-turbine applications. In these studies, the technique enables systematic ranking of coating architectures based on their resistance to cyclic thermo-mechanical strain, facilitating identification of relative performance trends [241]. Similarly, in hardmetals such as WC–Co, DC-TMT-based fatigue testing has been employed to investigate the influence of binder content, carbide grain size, and carbon content on fatigue resistance. Clear trends in fatigue behaviour have been reported as a function of microstructural parameters, illustrating the sensitivity and repeatability of DC-TMT for discriminating material variants under controlled cyclic loading [30].

### 5.2.3. Creep and stress relaxation testing

DC-TMT has been applied to investigate time-dependent deformation under sustained load, including both creep and stress-relaxation behaviour, using miniature specimens subjected to controlled thermal histories. In creep testing, deformation is measured under constant applied stress, whereas stress-relaxation testing involves holding strain constant and monitoring the decay of stress as internal redistribution occurs. Although these responses are mathematically related for many viscoplastic materials, they probe different aspects of time-temperature-dependent deformation and together provide complementary constraints for constitutive modelling [242].

DC-TMT-based creep studies have shown that, under carefully controlled conditions, key features of creep behaviour can be captured, even in miniaturised specimens. Comparisons with conventional creep rigs indicate that steady-state creep rates and dominant deformation mechanisms can be broadly consistent when temperature fields, effective gauge length, and electrical artefacts are rigorously constrained [27,53,154]. However, agreement is conditional rather than general: rupture times, tertiary creep behaviour, and strain localisation are particularly sensitive to temperature uncertainty and specimen size. Consequently, DC-TMT creep data should not be interpreted as direct substitutes for standardised creep rupture tests, but as mechanistically informative measurements under accelerated and spatially confined conditions.

In nickel-based superalloys, DC-TMT has been used to investigate creep deformation mechanisms at approximately 1100 °C. Studies on single-crystal alloys have demonstrated that miniature-specimen testing can replicate transitions from precipitate shearing to dislocation climb under low-strain-rate and high-temperature conditions (Figure 20), which is consistent with established creep mechanism maps [27,206,243–245]. Nevertheless, rupture times obtained from DC-TMT are often underestimated relative to bulk tests, reflecting the combined effects of thermal gradients, representative volume limitations, and accelerated damage accumulation in small specimens. Ultrafast DC-TMT protocols, sometimes coupled with digital image correlation, have been used to extract apparent activation energies for creep from a single specimen, enabling rapid comparative assessment of alloy variants [53].

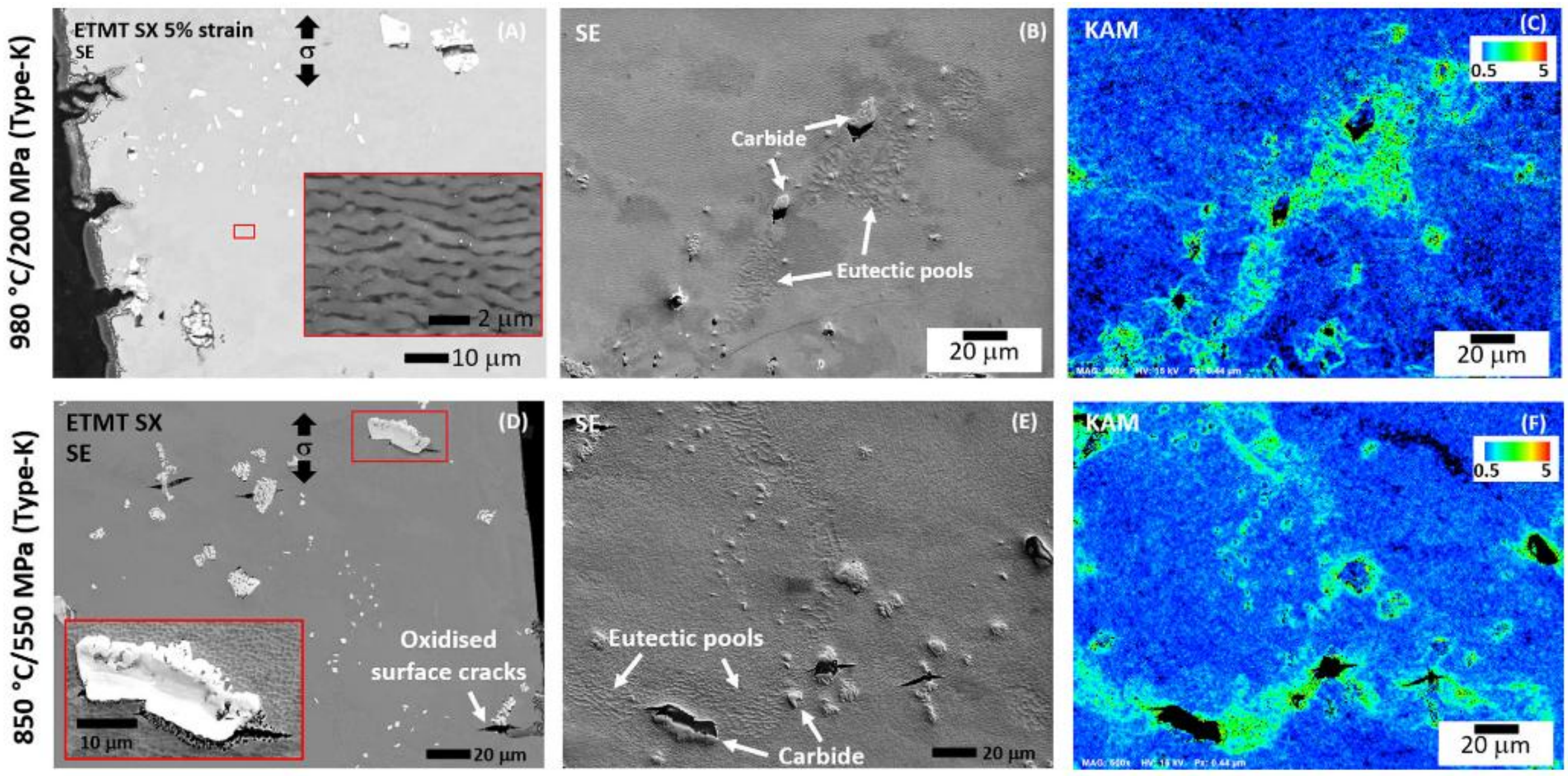

Figure 20: Microstructural features observed in a single-crystal Mar-M-247 specimen following DC-TMT creep testing. Specimens were tested at 980 °C and 200 MPa (a–c) and 850 °C and 550 MPa (d–f). The images illustrate deformation and damage features associated with creep under controlled thermal histories in miniature specimens, enabling identification of dominant creep mechanisms. Adapted from [27], licensed under CC BY 4.0.

DC-TMT has also been extended to other material classes where conventional creep testing is impractical or prohibitively time-consuming. In lead-free solders, rapid DC-TMT-based creep and thermo-mechanical fatigue experiments have been used to compare relative lifetime and deformation resistance of Sn–Ag–Cu and Pb–Sn systems under elevated temperature, providing an accelerated ranking [246]. In hardmetals, DC-TMT has enabled investigation of high-temperature creep mechanisms at 800–900 °C, revealing the roles of binder and carbide grain-boundary in long-term deformation [247]. High-temperature titanium alloys have similarly been studied, with DC-TMT results mapped onto classical creep-mechanism diagrams to distinguish regimes of dislocation creep, diffusion creep, and grain-boundary sliding [248].

It must be emphasised that the presence of electrical current during DC-TMT creep or stress-relaxation testing can influence deformation behaviour through non-uniformity in Joule heating and current-assisted effects. As discussed in Section 3.7, such effects represent conditional uncertainty sources that must be evaluated explicitly, particularly at low stresses and long dwell times. Failure to do so risks conflating intrinsic creep behaviour with test-specific artefacts.

## 5.3. Material-class-specific inference and limitations

### 5.3.1. Aluminium alloys

Aluminium alloys were among the first materials investigated using electro-thermal mechanical testing during the early development of the Gleeble system in the late 1950s [6]. Since then, DC-TMT has been widely applied to study high-temperature deformation, semi-solid behaviour, and thermally induced stress evolution across a broad range of aluminium alloy systems. Its principal value lies in enabling controlled investigation of temperature- and strain-rate-dependent deformation mechanisms under rapid, non-equilibrium thermal histories.

Early constitutive studies on alloys such as AA3003 and AA3102 established stress–temperature–strain-rate relationships within Zener–Hollomon frameworks using modified Gleeble-1500 configurations [249]. Interrupted deformation experiments on alloys, including 1050, 5182, and 7075, subsequently separated dynamic and static softening contributions during recovery and recrystallisation at 300–400 °C [250]. Related work using miniature ETMT specimens demonstrated that electrical resistance measurements are sensitive to microstructural evolution during thermal cycling in aluminium metal-matrix composites, enabling comparison of dimensional stability between reinforced MMCs and monolithic 2000- and 6000-series alloys [13,15].

Precipitation-hardenable aluminium alloys represent a critical application domain. Compression tests on alloys such as 2026 [251], 7150 [252], 7085 [253], and newly developed Al–Mg–Si–Cu systems (Figure 21) [254] consistently revealed peak-stress behaviour followed by dynamic softening governed by recovery, recrystallisation, and dynamic precipitation. Arrhenius-type constitutive descriptions yielded apparent activation energies of approximately 230–340 kJ/mol, reflecting the combined influence of dislocation annihilation and dynamic precipitation on hot-workability. Complementary studies on forging alloy 6005A identified temperature-dependent deformation regimes and mechanism transitions near 350 °C [255], while severe deformation using MAXStrain modules demonstrated the relationship between accumulated strain, grain refinement, and strength enhancement in alloys such as 1050, 2017, and 6060 [256].

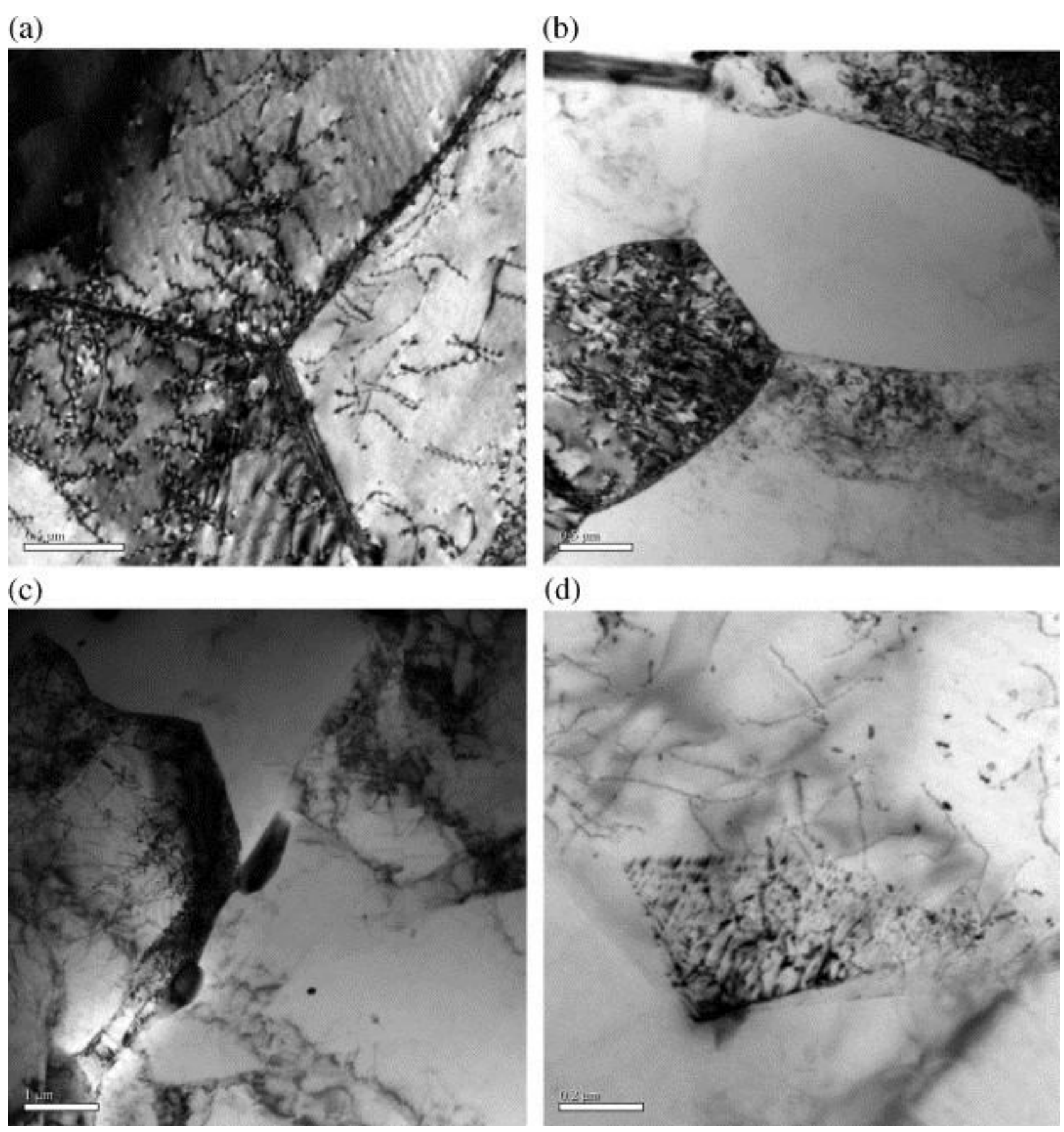

Figure 21: Transmission electron micrographs illustrating microstructural states observed in a novel Al–Mg–Si–Cu alloy following DC-TMT deformation under different temperature and strain-rate conditions: (a) 550 °C, 5 $s^{-1}$, showing high dislocation density near a high-angle grain boundary; (b) 450 °C, 5 $s^{-1}$, refined subgrain structures; (c) 450 °C, 0.05 $s^{-1}$, limited dynamic precipitation with partially coalesced particles along grain boundaries; and (d) 550 °C, 5 $s^{-1}$, fine precipitates within subgrains. The images illustrate the sensitivity of subgrain structure and precipitate morphology to imposed thermo-mechanical conditions in miniature specimens. Taken from [254].

DC-TMT has also enabled experimental access to deformation behaviour near the solid–liquid transition. For alloys such as AA6111, resistive heating reproduced parabolic temperature profiles characteristic of direct-chill casting, permitting tensile testing at liquid fractions below approximately 6% and stresses of 1–10 MPa [257]. Coupled with digital image correlation, these experiments quantified strain localisation relevant to hot-tearing susceptibility. Similar studies on AA3104, 5182, and 6111 confirmed the need for temperature- and strain-rate-sensitive constitutive descriptions when modelling casting deformation [258].

Beyond deformation studies, DC-TMT has been used to support quenching and thermo-mechanical simulations in high-strength aluminium alloys. Interrupted-quench experiments on thick 7xxx plates (AA7449 and AA7040) provided temperature-dependent tensile data representative of solutionised states, enabling thermo-viscoplastic modelling of residual-stress development without explicit precipitation modelling [259]. Related constrained-cooling experiments on AA2618 and other 7xxx alloys clarified the roles of recovery and localised plasticity in through-thickness stress gradients [259–261].

More recently, DC-TMT has been increasingly combined with advanced characterisation techniques. Out-of-phase thermo-mechanical fatigue tests in AA6061 demonstrated stable strain–temperature control once geometry and feedback parameters were carefully calibrated [262]. For quench-insensitive alloys such as AA6082, rapid heating and deformation combined with EBSD enabled isolation of dislocation storage and grain-size evolution while minimising dynamic precipitation effects [263]. Integration with synchrotron X-ray diffraction and SAXS has further enabled in situ investigation of precipitation kinetics in alloys such as AA7075 and 7021 under rapid Joule heating, revealing strain-rate-dependent vacancy generation and diffusion-controlled coarsening processes [264,265].

Overall, Al's high thermal conductivity and low resistivity flatten gradients yet make heating-rate and surface condition/emissivity decisive, especially when precipitation hardening is highly kinetic; thus, reported stress and precipitation trends should be read as mechanism-specific under rapid Joule heating rather than bulk-equivalent properties.

### 5.3.2. Steels

DC-TMT has been extensively applied to steels to investigate phase transformations, deformation behaviour, and microstructural stability under rapid thermal and thermo-mechanical loading. Its principal value lies in enabling controlled interrogation of temperature- and stress-dependent kinetics under non-equilibrium thermal histories.

In stainless, duplex, and lean-duplex steels, DC-TMT has been widely used to physically simulate weld heat-affected zones by imposing controlled peak temperatures and cooling rates. Studies on austenitic grades, such as AISI 304N, have demonstrated that repeated thermal cycles representative of multi-pass welding alter the δ-ferrite content, grain size, strength, and toughness in ways consistent with welding-induced microstructural evolution [266]. In situ DC-TMT–HEXRD clarified carbon redistribution during quenching-and-partitioning treatments [267], while grain-resolved 3DXRD measurements tracked stress evolution and load sharing in 316H stainless steel, providing constrained datasets for interrogation of crystal-plasticity models [268].

In duplex and lean-duplex steels, rapid heating and cooling were shown to suppress austenite reformation, promote $Cr_2N$ precipitation, and degrade corrosion resistance, with hardness and tensile behaviour correlating directly with phase balance and phase-boundary density [269–271]. Controlled DC-TMT experiments further enabled mapping of hot-ductility windows and brittleness limits in duplex steels, revealing the competing roles of dynamic recovery in ferrite and recrystallisation in austenite at elevated temperatures [272].

For structural and quenched-and-tempered steels, DC-TMT has clarified the coupling between cooling rate, phase constitution, and susceptibility to embrittlement. In Cr–Mo steels, resistive heating cycles demonstrated that fast cooling promotes martensitic transformation and hydrogen-assisted intergranular cracking, whereas slower cooling favours bainite formation and improved resistance to hydrogen damage [273]. In high-strength steels such as S690QL1, synchrotron-assisted DC-TMT revealed stress-assisted shifts in bainite start temperature and variant selection, with plastic accommodation occurring primarily in the austenite matrix [274–276].

Advanced high-strength steels, including TRIP and Q&P grades, have been a major focus of DC-TMT studies. Experiments demonstrated that martensitic transformation temperatures, bainite and ferrite kinetics, and retained-austenite stability are strongly stress dependent [277,278]. In Q&P steels, the role of silicon in suppressing carbide precipitation and stabilising austenite was clarified [277], while synchrotron-assisted DC-TMT provided quantitative links between alloy chemistry, transformation-induced plasticity, and work-hardening behaviour in TRIP steels (Figure 22) [278].

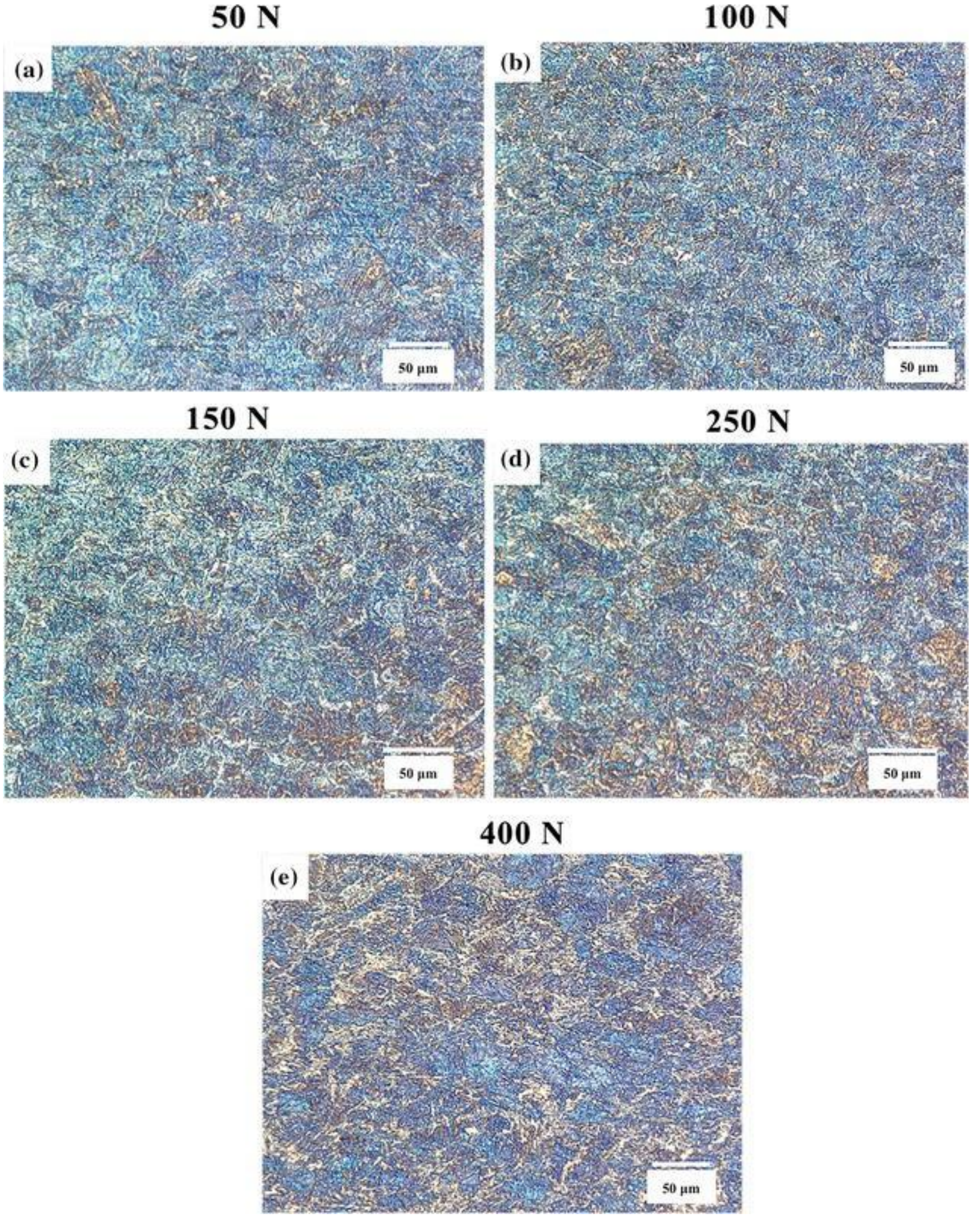

Figure 22: Microstructures of 3rd generation 1 GPa AhSS samples. ETMT was used to heat each sample to 1000°C at 10°C/s, with a dwell of 10s, followed by cooling at 40°C/s to 550°C where, over a period of 7s, an axial load was applied; (a) 50N, (b) 100N, (c) 150N, (d) 250N, (e) 400N, this load was held constant during further cooling to room temperature at 40°C/s. Micrograph (a) appears almost fully martensitic, as seen from the blue hue as a result of the LePera etchant, whilst ferrite, which has a tan colour, increasingly appears in microstructures (b) through (d). Taken from [278], licensed under CC BY 4.0.

DC-TMT has also been applied to low-temperature bainitic, microalloyed, and gear steels to resolve restoration and transformation kinetics under thermal-mechanical conditions. Studies have revealed supersaturation effects and unconventional partitioning mechanisms during isothermal bainite formation [279], as well as dynamic recrystallisation behaviour near the solidus in microalloyed steels. Additionally, correlations have been established between forging and heat-treatment routes and the resulting martensitic microstructures and fatigue performance in gear steels [280]. Hot-ductility mapping in Nb- and B-microalloyed steels further demonstrated the role of precipitation-controlled embrittlement, enabling the extraction of Zener–Hollomon relationships relevant to industrial processing [270]. Related synchrotron in situ diffraction study demonstrated how austempering temperature governs retained austenite stability and transformation kinetics in bainitic steel [281].

In hot-stamping steels such as 22MnB5, conductive hot-tensile testing using DC-TMT has been shown to reproduce fully austenitic deformation conditions and provide temperature-dependent stress data suitable for forming simulations, provided that specimen geometry and thermal boundary conditions are carefully controlled [282]. Subsequent studies employing digital image correlation demonstrated that water-cooled grips introduce significant axial temperature gradients, making calibration and thermal-field validation essential for reproducible thermo-mechanical data [283].

At a more fundamental level, DC-TMT has enabled in situ investigation of phase-specific stress partitioning and composite behaviour in multiphase steels. Synchrotron-assisted studies quantified load sharing between austenite and martensite during deformation and directly linked martensite strength to retained-austenite stability, transformation-induced plasticity, and work-hardening response (Figure 23) [284–286]. Complementary developments, such as laser-ultrasonic monitoring of recrystallisation kinetics and Joule-heated embrittlement testing, further illustrate the versatility of DC-TMT for probing dynamic restoration, liquid-metal embrittlement, and high-temperature fracture mechanisms under controlled conditions [287,288].

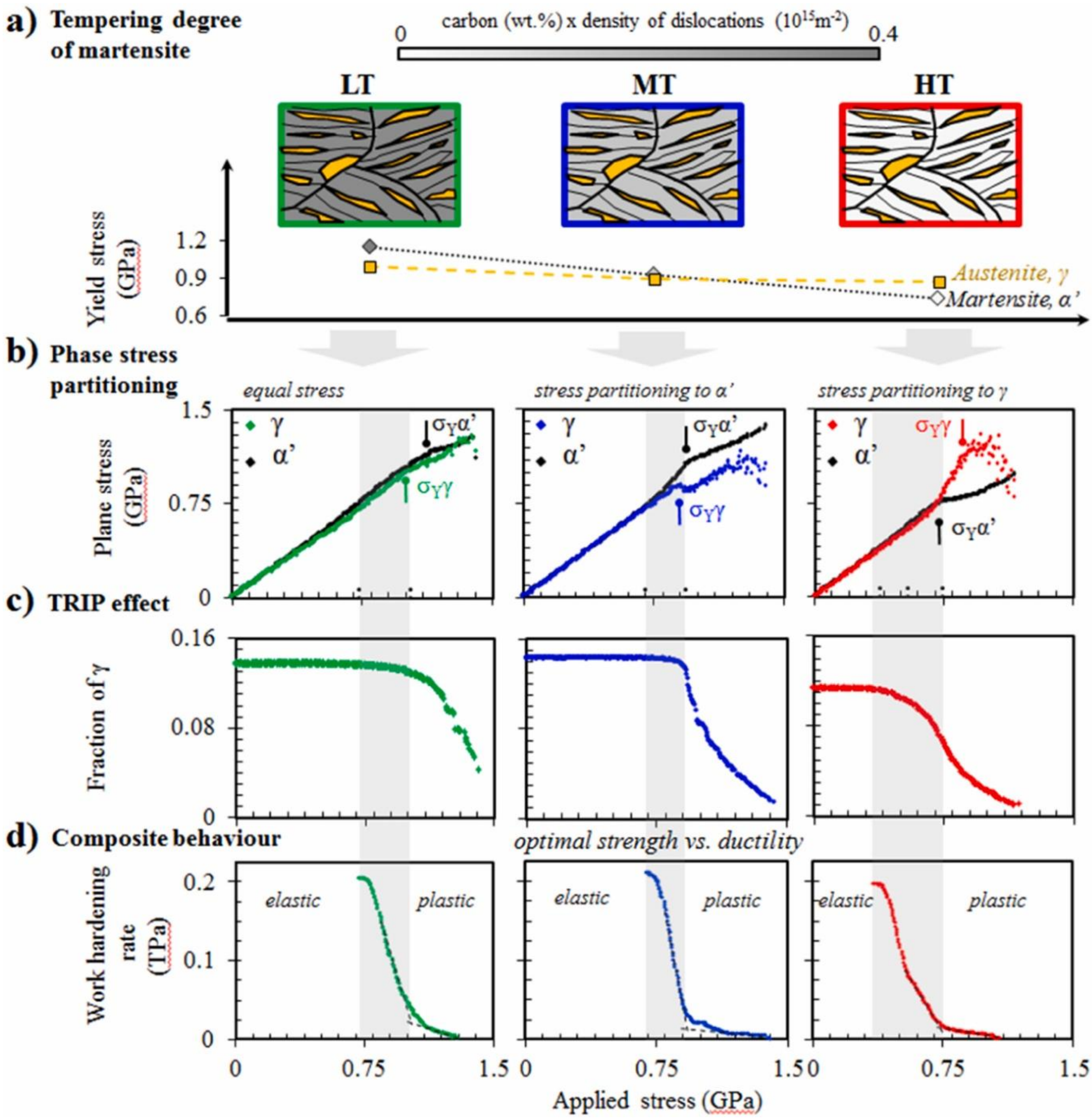

Figure 23: Phase-level stress partitioning and transformation-induced plasticity in multiphase steels studied using synchrotron-assisted DC-TMT. (a) Schematic microstructures with varying martensite tempering states. (b) Plane-stress evolution in austenite (γ) and martensite (α′) derived from diffraction. (c) Evolution of austenite fraction with applied stress, illustrating mechanically induced transformation. (d) Corresponding work-hardening response. The data illustrate correlations between phase strength contrast, stress partitioning, and TRIP behaviour under controlled thermo-mechanical loading. Adapted from [284], licensed under CC BY 4.0.

Overall, DC-TMT has proven to be a powerful mechanistic and comparative tool for steels, enabling the isolation of transformation kinetics, embrittlement mechanisms, and phase-level stress evolution under rapid thermal histories. However, the insights it provides are inherently conditioned by specimen size, temperature gradients, and heating mode, and results should be interpreted as accelerated or local representations, especially in multi-phase steels, which have strong transformation-kinetics dependence on heating rate and stress; hence, resistivity and microstructure evolve during tests.

### 5.3.3. Nickel-based superalloys

Nickel-based superalloys have been extensively studied using DC-TMT, as miniature specimens permit the controlled investigation of high-temperature deformation and microstructural evolution under rapid, non-equilibrium thermal histories. The technique has been applied to tensile, creep, and TMF testing, where material availability is limited or where localised response is of interest [59,207].

A major application of DC-TMT in superalloys has been the study of recrystallisation and γ′ phase evolution. Experiments on Ni single-crystal alloys such as CMSX-4 demonstrated that thermo-mechanical loading can trigger abnormal recrystallisation, with recrystallised depth and morphology strongly dependent on strain history, annealing temperature, and cooling rate [12,120,244,289,290]. Detailed studies revealed that slow cooling promotes solute drag, grain-boundary faceting, and thick recrystallised layers, whereas rapid cooling refines γ′ precipitates and suppresses boundary migration [55,107,237,245]. Complementary resistivity-based measurements have been used to track γ′ precipitation, dissolution, and coarsening kinetics under dynamic thermal–mechanical loading [25,119,180,232,291,292].

DC-TMT has also been employed to interrogate creep and TMF mechanisms in nickel-based superalloys. Miniature creep tests on alloys such as Nimonic 90 and CMSX-4 captured characteristic stress-exponent trends and activation-energy trends while revealing deformation features, including micro-twinning, dynamic recrystallisation, and serrated slips under specific loading mechanisms [293–296]. TMF studies conducted between approximately 400 and 850 °C demonstrated phase-angle-dependent fatigue behaviour and crack-initiation modes consistent with established mechanisms, provided that temperature control and gauge definition were carefully constrained [59,297].

Furthermore, DC-TMT has been utilised as a tool for comparative alloy assessment and environmental mechanism studies. Investigations of refractory-element additions (e.g., Ru,

Re) to Ni-based superalloys clarified their influence on rafting behaviour, creep resistance, and oxidation response under accelerated thermo-mechanical conditions [295]. Studies of environmentally assisted cracking and stress-accelerated oxidation have further demonstrated how grain-boundary chemistry, oxidation kinetics, and local strain localisation interact to degrade ductility and promote failure in miniature specimens [12,243,298]. These results provide mechanistic insight but must be interpreted with caution when extrapolating to service conditions, given the amplified oxidation kinetics and temperature gradients inherent to DC-TMT.

### 5.3.4. Titanium alloys

DC-TMT has been widely applied to titanium alloys because their mechanical response, phase stability, and damage tolerance are strongly sensitive to thermo-mechanical history. The technique enables controlled investigation of phase transformations, stress relaxation, and deformation mechanisms under rapid heating and constrained loading [299].

In metastable β-titanium alloys, such as Ti–5Al–5Mo–5V–3Cr (Ti-5553), DC-TMT has been employed to investigate β-decomposition, α precipitation, and the impact of grain-boundary α (GBA) on ductility and fatigue behaviour. Resistivity- and DIC-assisted experiments revealed that continuous GBA formation degrades toughness, while duplex ageing and slow-heating treatments refine α precipitation and improve mechanical response [300–302]. DC-TMT thermal cycles further clarified the role of athermal ω in controlling α plate refinement and precipitation morphology [301,303].

For α+β alloys, including Ti–6Al–4V and Ti–624x, DC-TMT tensile and stress-relaxation testing has been used to probe yield behaviour, creep strain, and stress redistribution up to approximately 750 °C under controlled thermal histories [77,304]. These studies demonstrated that residual stress evolution during post-forging and annealing significantly influences high-temperature response, and DC-TMT-derived data have been used as input for finite-element sensitivity analyses to assess heat-treatment sensitivity rather than to replace ASTM/ISO property datasets [304]. Stress-relaxation testing has also been applied to investigate cold dwell fatigue sensitivity, highlighting the role of alloy chemistry and β-phase content in controlling dwell-induced damage [305].

In advanced processing contexts, Rapid-heating DC-TMT–HE-XRD experiments revealed heating-rate-dependent shifts in β-transformation temperatures, highlighting the sensitivity of transformation behaviour to thermal history, which is relevant to welding and machining simulations [306]. DC-TMT has also enabled in situ investigation of rapid α↔β transformations, twinning variant selection, and texture evolution under high heating rates in Ti–6Al–4V [307,308]. In additively manufactured titanium alloys, miniature DC-TMT experiments have linked deformation to β-grain refinement via twinning, resolved transformation kinetics under extreme thermal transients, and supported a comparative assessment of constitutive models [193,309,310]. Related high-current electro-mechanical surface treatments, employing DC-TMT-style heating combined with mechanical stress, demonstrated near-surface microstructural refinement and associated improvements in hardness and fatigue resistance (Figure 24) [311].

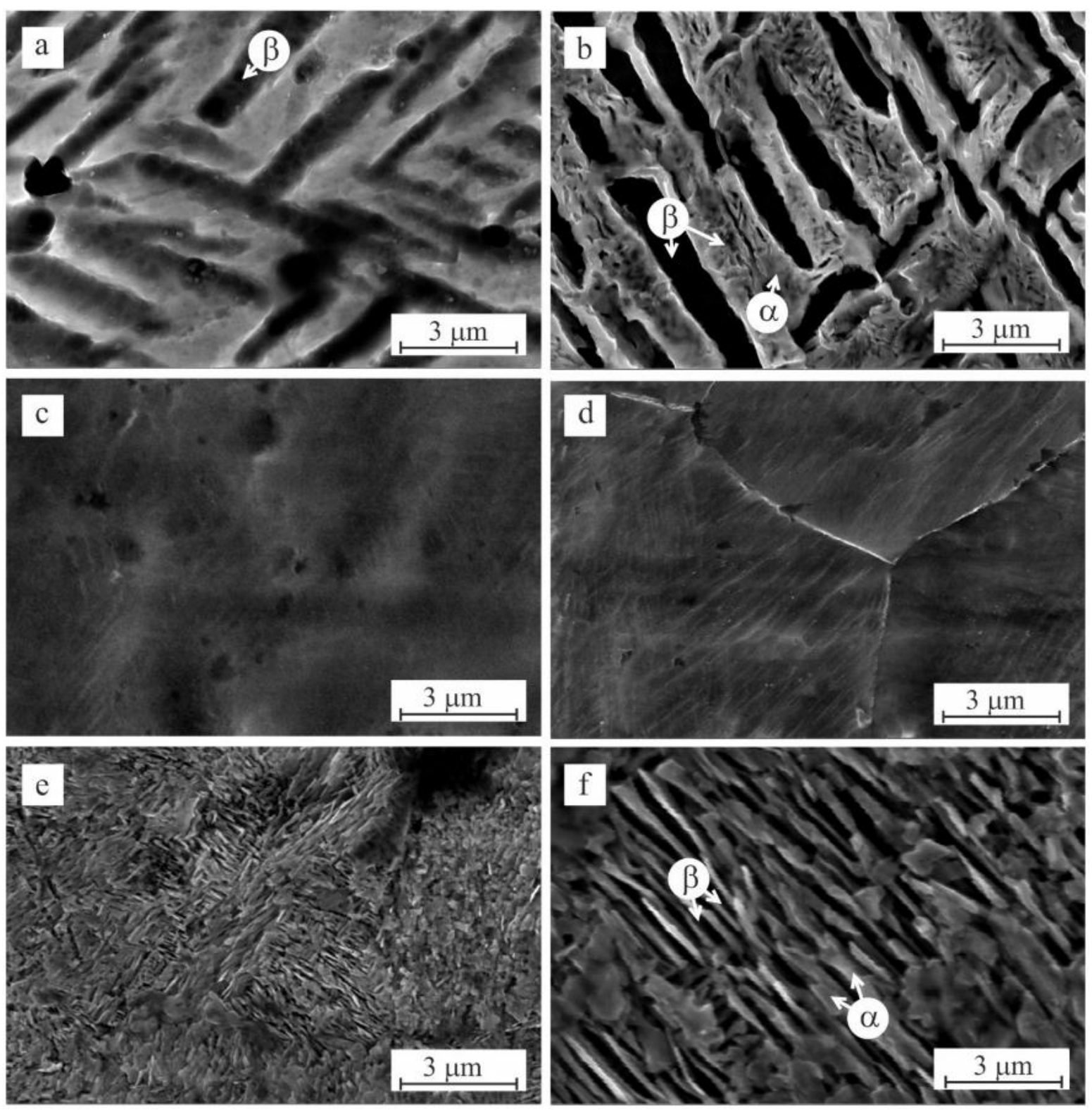

Figure 24: Microstructural evolution of an α+β titanium alloy following electro-mechanical treatment involving DC-based high-current heating combined with mechanical loading: (a) 300 A $mm^{-2}$; (b) 300 A $mm^{-2}$ followed by ageing at 600 °C for 14 h; (c and d) 600 A $mm^{-2}$; and (e and f) 600 A $mm^{-2}$ followed by ageing at 600 °C for 14 h. The images illustrate near-surface α/β refinement induced by accelerated thermo-mechanical treatment in miniature specimens. Adapted from [311] and panel (f) is reproduced with a corrected scale bar.

Moreover, γ-TiAl intermetallic alloys have been studied using DC-TMT and complementary thermophysical measurements to investigate their recovery, recrystallisation, and phase transformation behaviour at elevated temperatures. These studies linked deformation-induced microstructural evolution to changes in thermophysical properties such as heat capacity and viscosity, providing mechanistic insight into high-temperature processing behaviour [312–314].

Given the Ti's phase stability and precipitation are extremely sensitive to heating rate and prior thermal history, these results remain contingent on ultrafast heating-rate kinetics, texture/anisotropy in small gauges, and history-dependent precipitation; as such, reported transformation temperatures, dwell-related relaxation, and hot-workability windows should be treated as kinetic-mechanistic insights, not transferable property plateaus.

### 5.3.5. Hardmetals

Hardmetals, including WC–Co and related cemented carbide composites, were among the earliest material classes systematically investigated using electro-thermal mechanical testing with miniaturised specimens [9]. Their widespread application in cutting and forming tools demands reliable characterisation of high-temperature fatigue, creep, and thermo-mechanical response, and their inherently small characteristic length scales make them particularly amenable to DC-TMT-based investigation.

Early DC-TMT studies demonstrated that controlled rapid heating and cyclic mechanical loading enable reproducible thermo-mechanical fatigue (TMF) testing of hardmetals. By accessing heating rates up to approximately 200 °C $s^{-1}$, these studies revealed strong dependencies of TMF resistance on WC grain size, cobalt binder content, and carbon content. In particular, reductions in WC grain size were shown to increase fatigue life by more than an order of magnitude, while the cobalt mean free mechanism emerged as a dominant factor governing crack initiation and propagation [30,118,315,316].

Beyond fatigue, DC-TMT has been used to characterise tension–compression asymmetry, interphase cracking, binder cavitation, and WC grain-boundary sliding in cemented carbides, with compressive strength consistently exceeding tensile strength [118]. Phase-specific behaviour in Co–W–C systems has also been investigated, with resistivity measurements enabling the tracking of FCC–HCP transformations in the binder phase and correlating the phase state with high-temperature mechanical response [123].

The use of miniature specimens and high-throughput testing further enabled accelerated comparative assessment of WC–Co and TiC-based hardmetal grades. Cyclic thermo-mechanical screening experiments yielded comparative S–N behaviour and apparent activation energies for deformation, substantially reducing material consumption and testing time relative to conventional fatigue rigs [315,317,318]. However, creep and TMF rankings are strongly affected by surface oxidation in millimetre-scale sections and by phase-specific resistivity/thermoelastic contrasts that evolve near the FCC↔HCP Co transition (sub-500 °C). Thus collectively. Thus, these studies collectively demonstrate that DC-TMT is well-suited for comparative ranking of hardmetal grades and for probing localised deformation mechanisms, such as void formation, dislocation activity, and interphase cracking, under controlled thermal–mechanical loading.

### 5.3.6. Zirconium alloys

DC-TMT has been applied to zirconium alloys primarily because of their critical role as nuclear fuel cladding, where the coupled effects of texture evolution, hydride precipitation, and high-temperature phase transformations govern performance. The technique enables controlled investigation of these phenomena under rapid thermal and mechanical transients, especially in resolving localised transformation behaviour in miniature specimens.

A major focus has been the study of α→β→α phase transformations in alloys such as Zircaloy-2. DC-TMT experiments coupled with synchrotron X-ray diffraction demonstrated that recrystallisation before transformation weakens rolling textures, while variant selection during β→α cooling generates new texture components. In cases of incomplete transformation near the β-transus, a "texture memory" effect was observed, whereby the final α texture closely reproduces the initial state. These results clarify how peak temperature and thermal gradients during β-quenching influence texture inheritance and subsequent mechanical anisotropy [319,320]. The parabolic temperature profile intrinsic to DC heating further enables spatial correlation between maximum temperature and texture evolution within a single specimen (Figure 25).

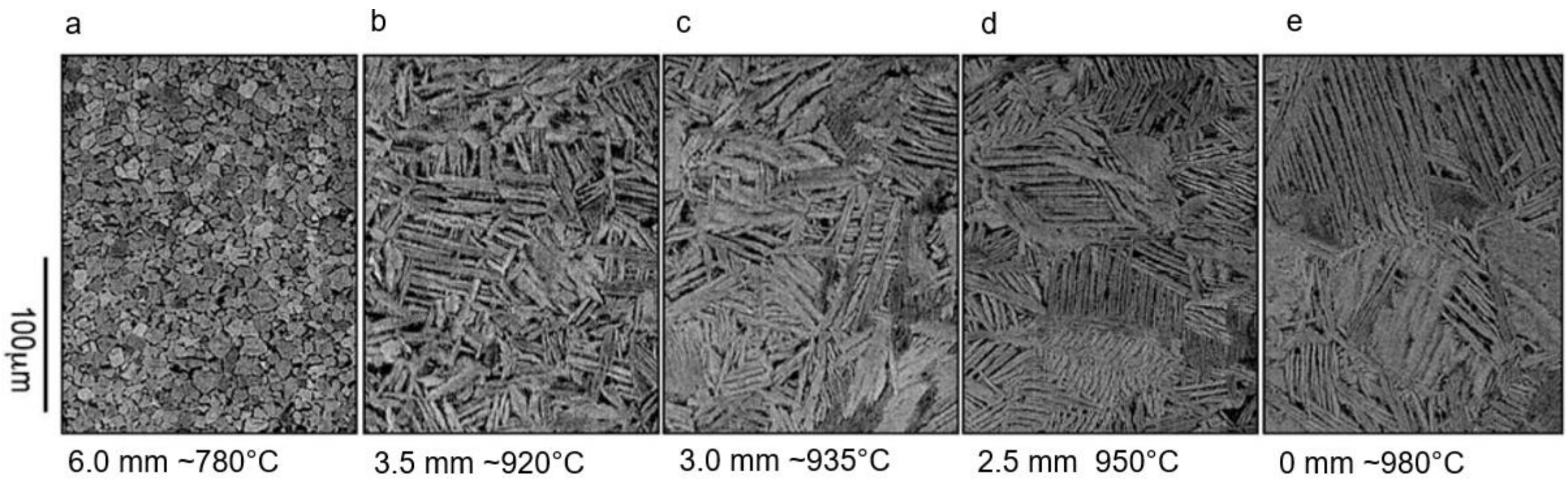

Figure 25: Backscatter electron micrographs illustrating texture evolution in a Zircaloy-2 specimen subjected to DC-TMT thermal cycling. Due to the parabolic temperature distribution generated by resistive heating, different positions along the specimen experience varying peak temperatures, allowing for a direct correlation between inherited texture, the maximum temperature reached, and transformation behaviour within a single test. Redrawn from [319].

DC-TMT has also been used extensively to study hydride precipitation and hydrogen embrittlement in zirconium alloys. Experiments on Zircaloy-4 integrated with synchrotron diffraction revealed anisotropic misfit strains associated with hydride formation, slow strain relaxation during isothermal holds, and the role of interstitial hydrogen in lattice dilatation. These observations provide mechanistic insight into how hydride orientation and stress–strain interactions contribute to embrittlement under service-relevant thermal–mechanical conditions [87,213].

In addition, DC-TMT has been applied to investigate creep, oxidation, and strain-hardening behaviour of zirconium alloys at elevated temperature, enabling a comparative assessment of microstructural stability and crack initiation under controlled loading mechanisms [321]. However, interpretation of such data must account for intrinsic limitations of DC-TMT, including non-uniform thermal gradients, constrained cooling rates, and the small representative volume of miniature specimens, which can amplify local effects relative to bulk cladding behaviour [322].

Considering that Zr alloys show strong texture effects and hydride-mediated strains coupled with using miniature tests specimens with limited RVE under gradients; these outcomes are inseparable from non-uniform temperature fields, small diffracting volumes, and hydride kinetics that are highly mechanism-dependent; consequently, reported texture inheritance, misfit strains, and creep/oxidation responses should be considered local, gradient-specific insights, not bulk cladding surrogates.

### 5.3.7. Shape memory alloys

DC-TMT has been widely applied to shape memory alloys (SMAs), for which functional behaviour is governed by stress-assisted martensitic transformations and their reversibility under thermal cycling. The technique is well-suited for SMA research because it enables the simultaneous control of temperature, stress, and electrical resistance, allowing for the direct interrogation of transformation mechanisms under coupled thermo-mechanical loading.

In NiTi-based alloys, DC-TMT combined with in situ synchrotron diffraction has resolved complex transformation sequences (B2 → R → B19/B19′) and quantified the influence of applied stress on variant selection, phase stability, and transformation hysteresis [239,323,324]. Repeated thermal–mechanical cycling experiments revealed texture evolution, dislocation accumulation, and progressive changes in transformation behaviour, providing insight into functional degradation mechanisms under accelerated testing conditions. Extensions of DC-TMT to bending configurations further enabled the measurement of transformation stresses and recoverable strains over hundreds of cycles, demonstrating that processed alloys, such as ECAP-treated NiTi, exhibit higher dislocation yield strength and improved cyclic stability compared with conventionally rolled material [238]. However, electroplasticity and modest magnetic fields generated by the DC can bias measured transformation stresses when thermal gradients are not fully quantified.

Beyond binary NiTi, DC-TMT has been applied to high-temperature SMAs alloyed with Hf, Zr, or Cu, where it enabled identification of shifts in transformation temperatures, negative thermal expansion behaviour, and rapid functional degradation associated with interface dislocation accumulation [325,326]. Ferromagnetic SMAs such as $Ni_2MnGa$ have also been investigated, with DC-TMT revealing coupling between martensitic transformations and magnetic ordering, thereby clarifying the role of Curie-temperature proximity in functional fatigue [176]. However, for ferromagnetic SMAs such as $Ni_2MnGa$ near ~110 °C Curie temperature, magneto-structural coupling is possible, but the DC-induced field is typically ~20–80 mT, so any magnetic contribution should be reported as conditional and material-specific.

At smaller length scales, modified DC-TMT configurations have been used to test SMA micro-wires, demonstrating how Joule heating combined with controlled actuation loads influences transformation temperatures, self-sensing capability, and actuation efficiency [327,328].

#### 5.3.8. Additively manufactured alloys

DC-TMT has been increasingly applied to additively manufactured (AM) alloys, where conventional testing is constrained by limited material availability, small feature sizes, and pronounced microstructural heterogeneity. The technique enables controlled thermal cycling and miniature mechanical testing under rapid heating conditions.

In metastable β-titanium alloys such as Ti–5Al–5Mo–5V–3Cr (Ti-5553), DC-TMT has been used to investigate post-build ageing strategies. Rapid heating produced fine intragranular α precipitation with discontinuous grain-boundary α, yielding bimodal microstructures that balanced strength and ductility (UTS ≈ 1345 MPa, elongation ≈ 11.6 %) [329,330]. Resistivity-based measurements enabled quantitative correlations between α-particle spacing, hardness, and yield strength (e.g., $Hv = 340 + 158/\sqrt{d}$), supporting comparative assessment of strength–ductility trade-offs under accelerated thermal histories [331]. Duplex and slow-heating schedules further refined α morphology, illustrating the sensitivity of AM microstructures to thermal mechanism selection.

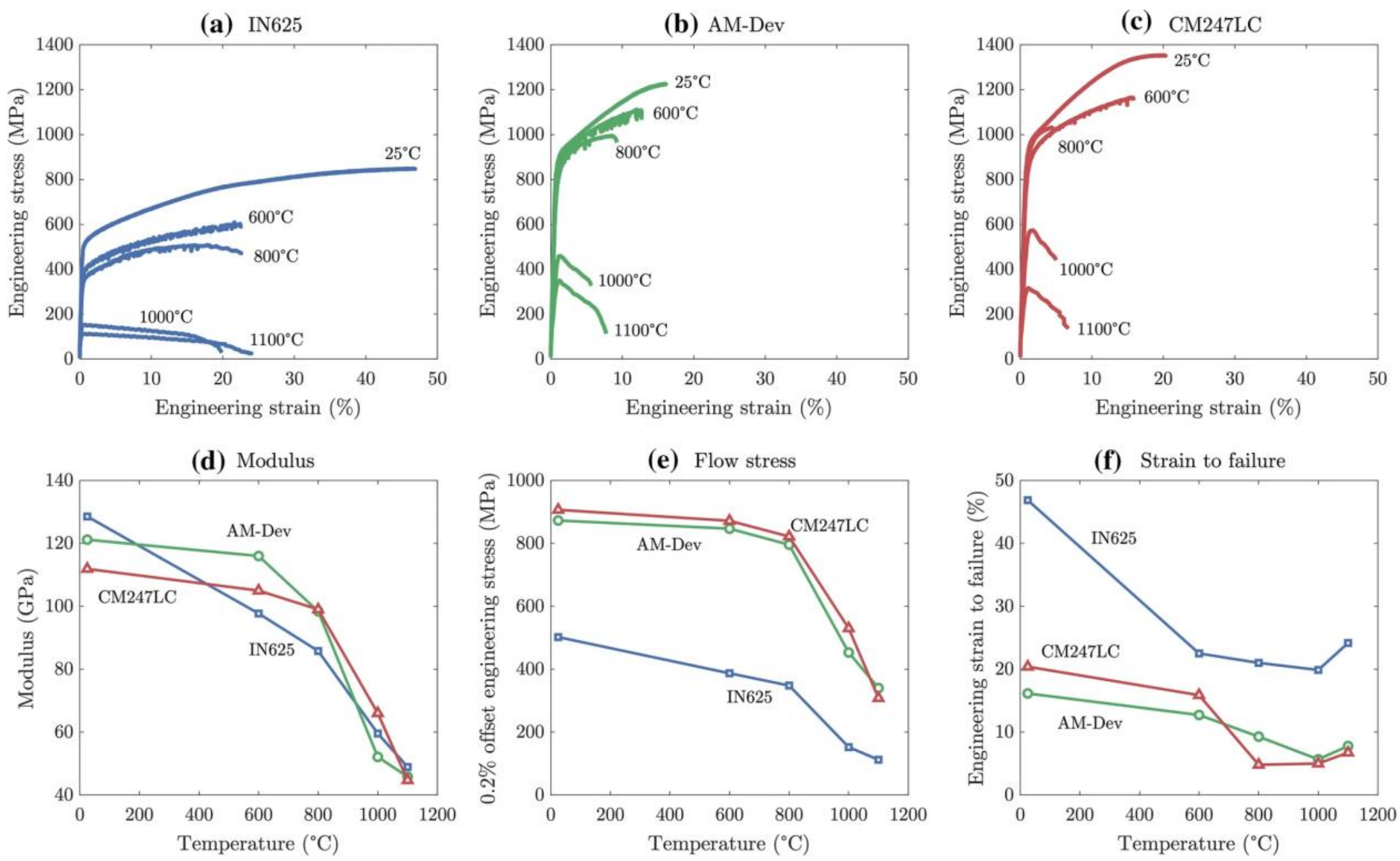

Figure 26: DC-TMT results showing the isothermal tensile response of (a) IN625, (b) AM-Dev, and (c) CM247LC at 25, 600, 800, 1000, and 1100 °C in the as-printed state. Panels (d–f) summarise the temperature dependence of elastic modulus, flow stress, and strain-to-failure, respectively. Adapted from [211], licensed under CC BY 4.0.

In AM nickel-based superalloys, DC-TMT has facilitated accelerated tensile and creep screening to examine γ′ strengthening, grain-boundary behaviour, and oxidation-assisted cracking. Studies have shown that increasing the (Nb + Ta)/Al ratio raises yield strength by increasing γ′ fault energies, but simultaneously degrades oxidation resistance and increases susceptibility to cracking. Sub-solvus heat treatments were necessary to restore ductility by limiting carbide coarsening and grain-boundary embrittlement, with DC-TMT enabling systematic comparison of these competing effects under constrained testing conditions [212,332]. Broader screening campaigns using DC-TMT-based creep relaxation and tensile ductility tests further highlighted the roles of solute segregation, ductility dip behaviour, and weak grain boundaries in both solid-state and solidification cracking, informing alloy design strategies for AM [211].

Beyond alloy chemistry, DC-TMT has been used to investigate specimen size, build orientation, and surface effects in AM Ti–6Al–4V. Reduced specimen dimensions were shown to increase strength through α-lath refinement and oxygen enrichment, but at the expense of ductility due to enhanced surface roughness and defect sensitivity. Orientation effects were equally significant, with vertically built specimens exhibiting inferior properties compared to horizontal orientations due to grain alignment and anisotropy [4].

These results highlight how AM-specific factors interact with the intrinsic size and temperature gradient sensitivities of DC-TMT. However, because AM coupons often exhibit columnar grains, segregation and strong orientation effects (e.g., Ti-6Al-4V vertical versus horizontal builds), and because miniature specimens have a limited RVE, use DC-TMT outputs for down-selection and model calibration rather than as production specification datasets.

### 5.3.9. New alloys development

DC-TMT has been increasingly utilised as a screening and mechanistic tool in the development of new alloys, where material availability is limited and rapid exploration of thermo-mechanical responses is required. By enabling miniature-specimen testing under controlled thermal and loading conditions, DC-TMT allows for a comparative assessment of deformation modes, phase stability, and strain-hardening behaviour across compositional or processing variants.

In high- and medium-entropy alloys such as CrCoNi, DC-TMT experiments have been used to probe deformation twinning, FCC→HCP transformations, and twin–lamella interactions under varying temperature and strain-rate conditions. These studies provided constrained datasets that supported discrimination between competing crystal-plasticity and EVP–FFT modelling frameworks, particularly with respect to texture evolution and intragranular lattice rotations [333,334]. Related work on FeNiCoAlTaB demonstrated precipitation hardening and hetero-deformation behaviour over a wide temperature range, clarifying the microstructural origins of strength and strain hardening under accelerated thermo-mechanical loading [335].

DC-TMT has also been applied in the exploratory development of alloys for extreme environments. In Nb-silicide systems processed by field-assisted sintering, DC-TMT-based compression creep tests enabled the extraction of apparent activation energies and stress exponents for comparative benchmarking [336]. In titanium-based developments such as Ti-407, DC-TMT revealed α/β interfacial sliding mechanisms absent in Ti–6Al–4V, suggesting an alternative mechanism to enhanced ductility at high strain rates [337]. For copper–steel metal–matrix composites intended for additively manufactured thrust chambers, DC-TMT experiments elucidated void coalescence and temperature-dependent hardening/softening under cyclic loading, informing processing-window selection [338]. In the dual-phase AlCoCrFeNi(CuTiZr) high-entropy alloy, DC-TMT compression testing at 800–1000 °C revealed persistent strain partitioning between FCC and BCC(B2) regions, with sustained work hardening up to 900 °C and premature softening at 1000 °C arising from localised dynamic recrystallisation triggered by excessive heterodeformation. Crucially, the tests demonstrated that retention of a Cr-rich spinodal phase within the BCC(B2) matrix stabilised high-temperature strength and delayed softening by maintaining phase strength contrast [339].

In steels and fusion materials, DC-TMT has supported rapid prototyping and comparative evaluation. Dual-phase steel development employed controlled thermal–mechanical cycles to guide ferrite refinement and martensite morphology optimisation [340]. In fusion-relevant alloys such as ODS EUROFER97 and V–4Cr–4Ti, DC-TMT combined with synchrotron diffraction (similar to Figure 27) enabled the measurement of lattice strain, dislocation density evolution, and anisotropic deformation behaviour at elevated temperatures, accelerating the down-selection of candidate compositions while retaining clear limits on extrapolation to reactor conditions [341,342].

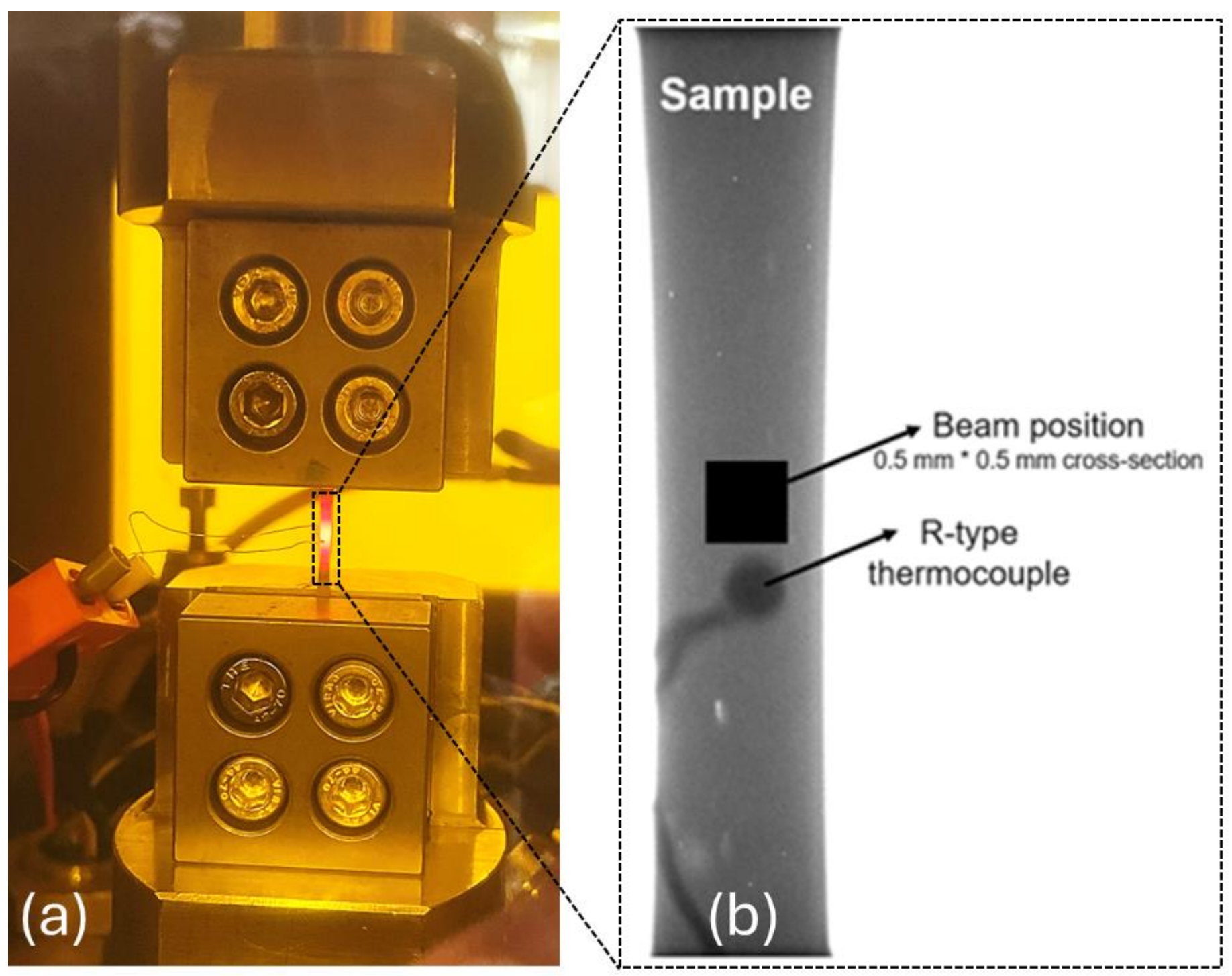

Figure 27: Experimental configuration for in situ synchrotron X-ray diffraction during DC-TMT testing. (a) Photograph of the DC-TMT tensile rig configured for diffraction measurements. (b) X-ray transmission image showing specimen geometry and beam position relative to the gauge section and thermocouple. The configuration enables simultaneous measurement of macroscopic stress and lattice-scale strain during thermo-mechanical loading [343].

Taken together, these studies illustrate that DC-TMT is most effective in new alloy development when used to identify deformation mechanisms, rank compositional variants, and generate physically constrained datasets for model interrogation. However, rankings must be framed by specimen-size effects, microstructural heterogeneity, and heating-rate-driven kinetics.

### 5.3.10. Metal–matrix composites and electrically conductive ceramics

In Al–Zn-based metallic-reinforced composites, DC hot-compression testing showed that Cu- and steel-particle reinforcements increased flow stress and stress exponents beyond those expected for self-diffusion-controlled deformation, identifying work hardening and dispersion strengthening as dominant mechanisms despite activation energies nominally consistent with dynamic recrystallisation [344]. Conversely, SiC-reinforced counterparts exhibited lower activation energies and smoother flow curves consistent with dynamic recovery–dominated behaviour [261].

Beyond metallic systems, DC-TMT–diffraction approaches have been extended to polygranular graphite and MAX phases. In graphite, combined resistivity and diffraction measurements during DC-TMT tensile testing linked porosity evolution to mechanical and transport properties at elevated temperature [345]. In MAX phases, in situ diffraction during DC-TMT loading revealed lattice-level origins of reversible hysteresis and non-linear elastic behaviour, demonstrating the technique's applicability to complex crystalline materials under coupled thermo-mechanical loading [346].

Overall, as shown throughout this section, coupling DC-TMT with X-ray diffraction is most effective for resolving mechanisms and discriminating models, particularly when spatially localised, time-resolved measurements are required. However, interpretation must account for intrinsic limitations, including temperature gradients, beam–gauge misalignment, and the small diffracting volume, which constrain direct extrapolation to bulk material behaviour. Additionally, neutron diffraction is a valuable tool for in situ analysis, particularly for steels and bulkier specimens [347–349]. Although neutrons offer lower spatial resolution than synchrotron X-rays, they penetrate larger volumes and are sensitive to light elements, making them suitable for studying lattice strain, phase transformations, and hydrogen behaviour under load.

Across these examples, DC-TMT is shown to be most effective when applied as a comparative and mechanistic tool within clearly defined thermo-mechanical and material-specific domains. Most importantly, across all material classes and applications reviewed in this section, the dominant uncertainty sources identified in Section 4 – most notably temperature-field definition, heating rate, and effective gauge volume – remain controlling, while secondary and conditional uncertainties modulate interpretation in material- and test-specific ways. Additionally, while the dominant uncertainty sources are common across materials, the effective RVE accessed by DC-TMT varies strongly with material class, microstructural length scale, and deformation mechanism, altering how local measurements relate to bulk behaviour. These challenges motivate the future research priorities outlined in the next section, where standardisation needs, hybrid measurement strategies, and irreducible constraints of DC-TMT are addressed explicitly.

# 6. DC-TMT standardisation challenges and future directions

Future progress in DC-TMT must be framed by the recognition that several uncertainty sources are intrinsic to Joule-heated miniature testing and cannot be eliminated through calibration or procedural refinement. Within these constraints, advancement can be structured around three tightly linked priorities: (1) standardisation and reporting practices that explicitly acknowledge irreducible uncertainty sources and domains of validity; (2) rapid specimen-screening strategies that deliberately exploit intrinsic temperature gradients, while resolving local temperature, strain, chemistry, and crystallography at high spatial resolution; and (3) the integration of spatially resolved experimental data with coupled thermo-electro-mechanical simulations to enable constrained inverse inference of material properties rather than unconstrained parameter fitting.

## 6.1. Standardisation and reporting requirements

ASTM or ISO has not formally adopted the DC-TMT technique as a standalone standardised testing method, which has constrained its uptake beyond academic and specialist research. Despite this, a substantial academic literature spanning over four decades demonstrates that DC-TMT can generate reproducible, mechanistically informative data under tightly defined thermo-mechanical conditions. Existing standards, such as ASTM E21 and ISO 6892-2, address conventional thermo-mechanical testing; however, they do not account for the coupled electrical, thermal-gradient, and size and RVE effects inherent to DC-TMT. Standardisation for DC-TMT should therefore focus on consistent test definition, reporting of temperature-field characterisation and spatial resolution limits, and uncertainty quantification, rather than enforcing equivalence with bulk properties.

Recently, ASTM E8/E8M-24 and ISO 6892-5/52909 were updated to guide specimen geometry, preparation, and testing protocols tailored for miniature specimens. The update was informed by an inter-laboratory comparison (ILC) [350], which demonstrated that miniature specimens can yield consistent elastic and strength measurements under controlled conditions. The ILC was performed on AM specimens, which, unlike wrought alloys, can exhibit spatially varying microstructures, including variations in grain size, phase distribution, and porosity, as well as strong crystallographic texture due to directional solidification and thermal gradients during layer-by-layer deposition. Still, testing of miniaturised AM specimens produced consistent results in elastic modulus, but with systematic discrepancies in ductility-related parameters attributable to size effects. Since creep deformation, particularly in the tertiary regime, is intrinsically coupled to ductility, similar discrepancies are unavoidable when creep behaviour is inferred from miniaturised specimens [351].

Looking ahead, future work should systematically compare heating methods in terms of their dominant uncertainty sources, reproducibility, and influence on inferred material response. While the present review focuses primarily on DC-resistive heating, induction heating also warrants explicit consideration because its uncertainty profile is materially different. In DC-resistive thermo-mechanical testing, the main uncertainties commonly arise from thermocouple-based temperature control, electrical measurement of current, voltage or resistance, specimen geometry, and axial temperature gradients caused by heat extraction through the grips. These gradients can in turn produce non-uniform strain, strain-rate and microstructural evolution, thereby biasing the apparent material response extracted from the test.

By contrast, in induction heating, the thermal field depends strongly on coil geometry, coupling efficiency, relative specimen position and orientation, and frequency-dependent electromagnetic penetration. Skin, proximity, end and edge effects can all contribute to non-uniform heat generation, making the relation between applied power and local specimen temperature more geometry-sensitive and more difficult to interpret directly [352–354]. These factors can produce stronger spatial non-uniformity of heating, greater sensitivity to specimen geometry and positioning, and more complex relationships between applied power and local specimen temperature. As a result, induction-heated tests may exhibit larger uncertainty in the actual thermal field and effective representative gauge volume, especially when lower frequencies, larger cross-sections, or magnetically active alloys are involved.

Accordingly, uncertainty quantification for thermo-mechanical testing should not treat the heating method as a secondary procedural detail. Instead, DC-resistive and induction heating should be evaluated as distinct experimental configurations, each with its own characteristic sources of thermal, electromagnetic, and mechanical variability. Unfortunately, at present, however, the literature appears to treat these issues largely within each heating method separately, and a systematic matched-condition comparison of uncertainty budgets between DC-resistive and induction-heated thermo-mechanical testing is still lacking. Such a comparison is needed and is currently being pursued in ongoing work.

## 6.2. Accelerated testing protocols and experimental design

Recent developments in DC-TMT have expanded its capabilities beyond traditional quasi-static strain rates and uniaxial loading. While conventional DC-TMT systems operate primarily at low to moderate strain rates, actuator optimisation enables reliable testing across the upper end of this range (≈1–100 $s^{-1}$). This allows for investigating dynamic phenomena such as recrystallisation and rate-sensitive behaviour in high-temperature alloys. In parallel, efforts to incorporate multi-axis loading into DC-TMT systems aim to approximate complex service stress states. Configurations involving torsion, biaxial tension, and bending have been explored using compact specimen geometries and specialised grips. These setups facilitate the study of anisotropic materials under realistic conditions, such as tension-torsion in turbine blade roots or biaxial membrane stresses in pressure vessels.

Enhancements in in situ measurement techniques can further improve the accuracy and resolution of DC-TMT experiments. Non-contact thermal imaging and full-field displacement measurements provide high spatial resolution data during testing, especially when temporally correlated [355]. Such techniques are essential for resolving dominant uncertainty sources, particularly spatial temperature gradients and effective gauge definition, rather than optional refinements, which can be further coupled with X-ray and neutron diffraction and imaging for studying lattice strain and phase transformations.

DC-TMT-based can also enable multi-temperature tests alloys for rapid screening, as a specimen can be resistance-heated with a thermal gradient along its length, enabling the extraction of static Time-Temperature-Transformation (TTT) [98] and dynamic creep behaviour at many temperatures from a single test when coupled with DIC and thermal imaging, allowing the creep strain rate to be measured at each temperature along the gauge [356,357]. This yielded an entire creep activation energy curve in a single experiment, with results consistent with those of conventional multi-test data [356]. This method cut required time and material by at least a factor of four while still producing accurate high-temperature creep properties. Such approaches are particularly attractive for alloy development, as a single miniature specimen can quickly provide a spectrum of high-temperature strength data.

Other innovative DC-TMT protocols that exploit the specimen thermal distribution enable substantial acceleration of high-temperature material screening, reducing testing times from months to days [53,154]. Stepped-temperature test (STT, Figure 28a) at a fixed load/strain, and stepped-load or stress relaxation tests at a fixed temperature enable rapid (Figure 28b) high-temperature screening, allowing for the mapping of time-dependent strength and creep properties from a single specimen in hours rather than weeks [154].

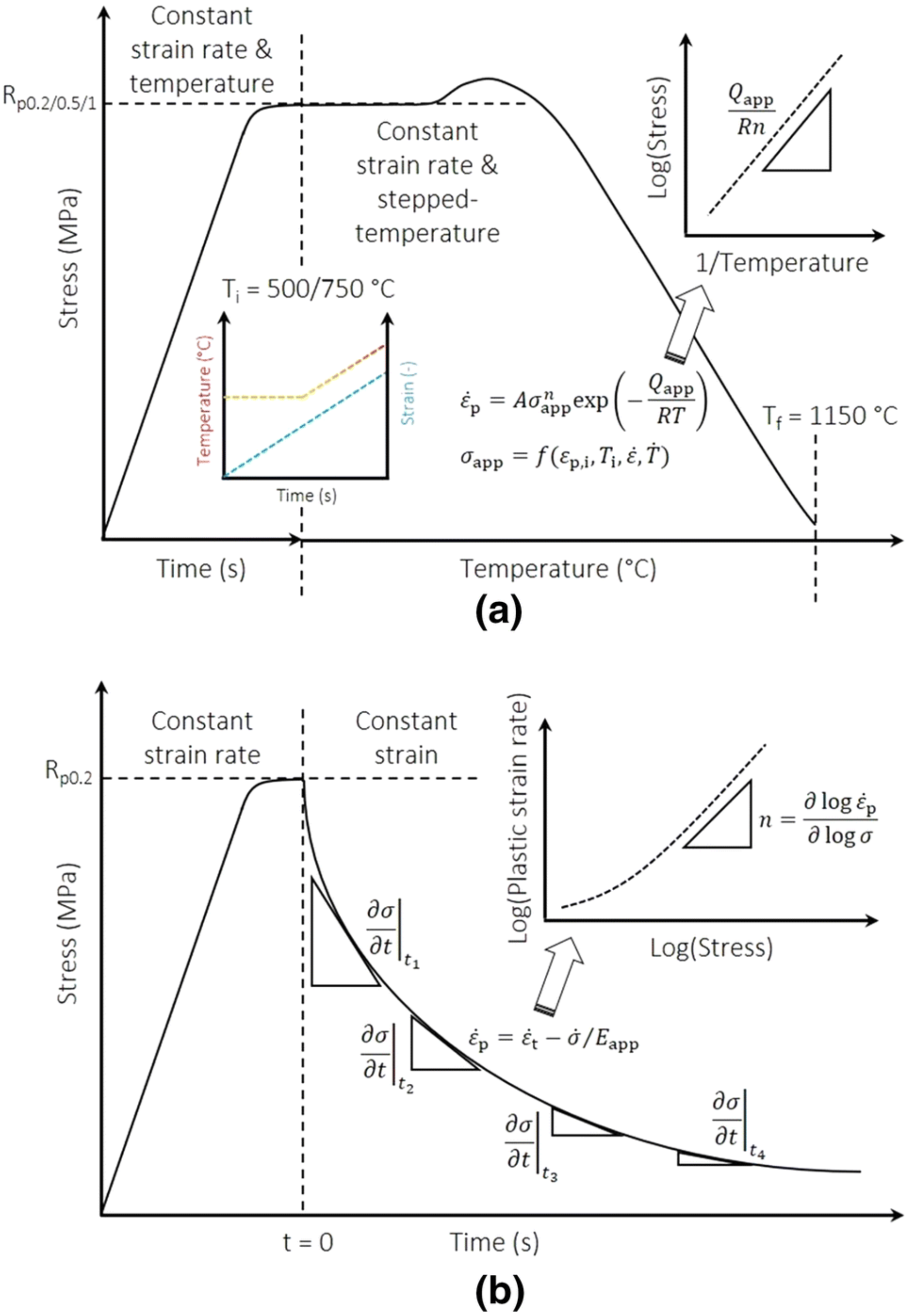

Figure 28. Schematic illustration of accelerated DC-TMT protocols: (a) stepped-temperature testing at fixed load or strain, and (b) stress-relaxation testing at fixed temperature, used for rapid screening of high-temperature deformation behaviour under controlled Joule heating. Adapted from [154], licensed under CC BY 4.

For stepped-load creep, two methods are becoming widely used for acceleration: (1) the Stepped Isostress Method (SSM), which increases stress in increments to assemble a master creep curve validated across materials [358,359]. It has been applied to metals such as IN718, achieving a ~64× reduction in test time while obtaining creep properties across multiple temperature "isotherms" [360]. (2) The Dynamic Negative Stepped Test (DNST), which further accelerates the process, involves sequential, decreasing stress steps at a constant temperature [361]. DNST dynamically adjusts each hold so that the minimum creep rate is reached at every step, and by descending the stress, it drives the material near tertiary creep in each step without failure. This maximises damage per step and gives alloys a rapid creep resistance ranking. For example, in a Ti–6Al–4V, DNST enabled quick identification of build directions with superior creep strength in AM material [361]. Such accelerated creep methods do not replace conventional creep tests but serve as rapid screening tools to identify promising alloy chemistries, heat treatments, or AM process parameters.

These approaches can be applied to conventional high-temperature alloys, as in engineering components, localised heat still forms temperature gradients that do not produce isolated microstructural zones, but create coupled fields where heat flow, stress, diffusion, and defect dynamics continuously interact. These coupled fields generate graded microstructures characterised by local phase variations and time–temperature dependent transformation, compositional shifts, and anisotropic defect structures, which can be studied explicitly using DC-TMT, rather than approximated as isolated isothermal zones.

## 6.3. Inference, inverse modelling, and identifiability

To improve data fidelity, real-time monitoring and computational analysis enable dynamic adjustment of testing parameters and more accurate stress–strain reconstruction, particularly during necking or phase transformations. Additionally, because DC-TMT utilises Joule heating, real tests develop axial/radial thermal gradients and non-uniform stresses that raw force–displacement (or crosshead strain) data cannot accurately represent. A practical inverse modelling workflow is:

(i) Measure the temperature field using spatially resolved thermal measurement techniques (e.g., infrared thermography, multi-wavelength pyrometry), and the deformation field using 2D or stereo DIC, with stereo DIC being preferred where out-of-plane deformation is non-negligible.

(ii) Reconstruct the full temperature map via coupled electro-thermal FE with realistic boundary conditions, grip temperatures, oxide scale effects, and control logic [31,103], then identify heat-transfer and contact/electrical parameters by inverse heat-transfer fitting [104,362,363].

(iii) With the calibrated model, recover local stresses/strains in the effective gauge length and back-correct nominal curves to obtain accurate deformation/creep data [31,104,105,138,147].

(iv) Throughout, measured materials properties (temperature- and rate-dependent elastic/plastic/creep laws) must be supplied or re-measured for the alloy tested, not borrowed from sparse handbooks, to avoid biasing gradients and stresses [31,364].

This approach enables the calibration of heat transfer coefficients and electrical contact conditions, ensuring that the simulated temperature field matches the measured distributions. Once validated, the FE model can be extended to predict local strain and stress distributions, allowing for the correction of nominal force–displacement curves into accurate constitutive data.

Thereafter, the forward FE reproduces the measured thermal fields in the effective gauge length and macroscopic force–displacement relations. Inverse identification can then proceed by minimising DIC/IRT/force misfits to extract constitutive parameters under the entire gradient temperature fields [138,147]. Parameter non-uniqueness is mitigated by multi-signal fits (force, local strain and temperature) and by physical bounds from independent tests (e.g., dilatometry for α/β or γ′ solvi, high-T modulus, conductivity) [104,105,364]. At higher current densities, include electro-assisted terms to capture the effect of direct current, especially when current densities influence creep rates [365]; e.g., modified Norton creep with current-dependent activation energy/stress exponent has been demonstrated in metallurgical systems (solder joints) [366].

For anisotropic or precipitation-strengthened alloys (Ti, Ni-base, steels), macro laws (von Mises with isotropic hardening) miss key physics (twinning, γ′ shearing, dwell). A thermo-mechanical crystal-plasticity (CP) framework, calibrated against the FE-recovered local fields, can capture orientation-dependent slip, twinning/phase activity, as well as temperature/rate coupling. Crystallographic information, captured via EBSD and elemental composition measured using energy-dispersive X-ray spectroscopy maps of the microstructure before and after testing, constrain slip/twin systems. CP parameters are then inversely identified using the same multi-signal misfit, but now at the grain-ensemble level, enabling consistent mapping from measured macro responses to micro-mechanisms. This tier incorporates evolving materials properties (e.g., temperature-dependent critical resolved shear stress and dynamic recovery) and links observed microstructural change to constitutive updates; thus, avoiding the steady-state assumptions that fail for IN718 or AM alloys under transient DC-TMT cycles [35,42,138].

However, common pitfalls arise at four levels: properties, modelling, identification, and scope, and each has a straightforward remedy. First, material properties must be correct over the tested thermal–mechanical window. Thermal conductivity ($k$), specific heat ($cp$), and linear thermal expansion ($\alpha$), as well as elastic moduli and rate sensitivities, should be measured for the alloy under study or drawn from vetted, time-temperature-dependent datasets spanning the exact ranges of Time–Temperature–Transformation (TTT) regime and strain rate ($\dot{\varepsilon}$). Using handbook constants outside their validity skews the reconstructed temperature field and the recovered stress state [364]. Thus, careful interpolation and clear citation of sources mitigate this bias. This combination – accurate $k$, $cp$, $\alpha$, multi-signal fits, realistic boundaries, and mechanism-aware constitutive forms – prevent common failure modes and yield reproducible DC-TMT property extractions.

Second, in addition to what was mentioned in Section 4.3, when modelling: geometric and microstructural simplifications, idealised assumptions of homogeneity and rate-independence, and grip compliance alter stress localisation [42]. Unrealistic, rigid, or frictionless grips should be replaced by measured stiffness, temperature, and friction.

FE element selection and contact/friction laws can affect load mechanisms [35,138], while coarse meshes miss steep thermal/mechanical gradients, and over-refined meshes invite instability/cost [35]. 3D analysis is often required to avoid 2D plane-strain misrepresentations. Moreover, Differences between finite element packages/software are non-negligible, as a NIST-style round-robin showed statistically significant spread across solvers, even for simple benchmarks [367].

Third, inverse identification must be well posed, as inverse modelling approaches depend heavily on optimisation routines that minimise the difference between measured and simulated outputs; however, they are sensitive to initial parameter guesses and vulnerable to local minima convergence that may not reflect actual material behaviour. Thus, it fits multiple, independent signals, i.e., global force–displacement, local strain fields from DIC, and time-resolved temperature, so the parameter set is constrained by orthogonal information rather than a single curve. Regularisation and physically informed bounds reduce non-uniqueness, while reporting confidence intervals (rather than a single best value) makes uncertainty explicit. When electric current flows during deformation, include current-dependent terms in the constitutive law so that fitted parameters reflect true material response rather than unmodeled electro-assistance. Crucially, guard against "validation by tuning": fitting parameters to match one curve without external checks harms generality [368].

Fourth, maintain model scope discipline. Empirical deformation laws such as Johnson–Cook or Cowper–Symonds should not be extrapolated beyond their calibrated TTT–$\dot{\varepsilon}$ domain. For anisotropic, textured, or precipitation-strengthened alloys, isotropic von Mises models often fail to capture slip asymmetry, twinning, or precipitate shearing. In such cases, anisotropic hardening or thermo-mechanical crystal plasticity with temperature- and rate-dependent critical resolved shear stress (CRSS) should be adopted.

Succinctly put, start macro (calibrate electro-thermal FE to measured temperatures; correct gradients and extract true deformation/creep), then bridge to inverse identification, and finally resolve micro-mechanisms via thermo-mechanical crystal plasticity constrained by EBSD/EDX, and always anchored by measured materials properties and transparent uncertainty reporting.

## 7. Conclusion

This review has examined direct current thermo-mechanical testing (DC-TMT) as an experimental methodology defined as much by its intrinsic constraints as by its capabilities. Rather than treating DC-TMT as a surrogate for conventional bulk thermo-mechanical testing, the analysis has shown that DC-TMT constitutes a distinct experimental paradigm, in which resistive heating, specimen geometry, and coupled thermal–mechanical loading fundamentally alter the relationship between measured response and inferred material properties. The principal contribution of this work is the explicit identification and hierarchical organisation of the uncertainty sources that govern the interpretation of DC-TMT data, and the clarification of when these uncertainties can be mitigated and when they are irreducible.

A central conclusion is that several defining features of DC-TMT, such as spatial temperature gradients, evolving gauge volumes, sensitivity to specimen geometry, and coupling between electrical current and deformation, are intrinsic to the technique and must be accounted for explicitly, rather than treated as experimental deficiencies. Consequently, DC-TMT data should not be assumed to be directly equivalent to bulk, isothermal furnace-based measurements, particularly for rate-sensitive deformation, phase-transformation kinetics, long-duration creep, or environmentally assisted damage. Discrepancies between DC-TMT and conventional data are therefore not necessarily experimental artefacts, but often reflect genuine differences in thermo-mechanical history and representative volume element.

At the same time, the review demonstrates that DC-TMT provides unique and physically meaningful insight when used within clearly defined domains of validity. Across materials, DC-TMT reveals precipitation kinetics and semi-solid behaviour in aluminium, stress-assisted transformations in steels, and creep-mechanism transitions in nickel-based superalloys. It clarifies α/β transformation pathways in titanium alloys, martensitic sequences and magneto-structural coupling in shape memory alloys, and texture memory and hydride strains in zirconium. These insights, however, are inseparable from dominant uncertainties – such as thermal gradients, heating-rate effects and limited representative volume – so results must be framed as mechanism-specific rather than bulk-equivalent properties.

The analysis further highlights that reliable interpretation of DC-TMT data increasingly depends on the integration of spatially resolved diagnostics and constrained modelling. Full-field strain measurements, spatially resolved temperature mapping, and in situ diffraction or scattering provide essential information for resolving heterogeneity and localisation that are otherwise hidden in macroscopic signals. When coupling with electro-thermal–mechanical finite element models and carefully constrained inverse identification enables back-correction of gradients and more reliable constitutive identification. With disciplined experimental design and transparent reporting, DC-TMT can deliver reproducible, mechanistically informative data that accelerate alloy development and support physics-based modelling. However, these approaches do not remove intrinsic uncertainty, but instead make it explicit and quantifiable.

Finally, this review emphasises that progress in DC-TMT practice is driven less by the elimination of uncertainty than by its transparent characterisation and consistent management. Standardisation efforts must focus on transparent reporting of test geometry, thermal boundary conditions, measurement fidelity, and domains of validity, rather than on enforcing artificial equivalence with bulk testing standards. Equally important is recognising which questions are fundamentally ill-posed for Joule-heated miniature testing, and which are uniquely accessible.

# Appendixes

## A. Governing Heat-Transfer Equation for DC-Heated Specimens

In direct-current thermo-mechanical testing (DC-TMT), the temperature field within the specimen evolves due to the coupled action of Joule heating, heat conduction, and surface heat losses. The governing equation for transient heat transfer in a deforming, electrically conductive specimen may be written as (with temperature-dependent properties):

$$\rho_m(T)\, c_p(T)\, \frac{\partial T(\mathbf{x},t)}{\partial t} = \nabla \cdot \big(k(T)\, \nabla T(\mathbf{x},t)\big) + \underbrace{\mathbf{J}(\mathbf{x},t) \cdot \mathbf{E}(\mathbf{x},t)}_{\text{Joule heating}} - \; q_{\text{loss}}(\mathbf{x},t), \quad 12$$

where $T(\mathbf{x},t)$ is the temperature, $\rho_m$ is the mass density, $c_p$ is the specific heat capacity, and $k$ is the thermal conductivity. The term $\mathbf{J} \cdot \mathbf{E}$ represents volumetric Joule heating, with $\mathbf{J}$ the current density and $\mathbf{E}$ the electric field. Under Ohmic conduction, this term can be expressed as

$$\mathbf{J} \cdot \mathbf{E} = \rho_e(T)\, |\mathbf{J}|^2 = \sigma(T)\ \mid \nabla\phi \mid^2, \quad 13$$

where $\rho_e$ and $\sigma$ are the electrical resistivity and conductivity, respectively, and $\phi$ is the electric potential. The loss term $q_{\text{loss}}$ accounts for heat exchange with the surroundings and may include convection and radiation,

$$q_{\text{loss}} = h\, \frac{A_s}{V}\, (T - T_\infty) \; + \; \varepsilon\, \sigma_{SB}\, \frac{A_s}{V}\, (T^4 - T_\infty^4), \quad 14$$

where $h$is the convective heat-transfer coefficient, $\varepsilon$the effective surface emissivity, $\sigma_{SB}$the Stefan–Boltzmann constant, $T_\infty$the ambient temperature, $A_s$the exposed surface area, and $V$the specimen volume. With equation 14 capturing convection (first term) and radiation (second term), where $A_s/V$ is the surface-area-to-volume ratio. In DC-TMT, this term becomes significant at high temperatures (typically above ~700–800 °C), especially for small-gauge specimens with high surface-area-to-volume ratios.

For the fully coupled electrical part, written explicitly (rather than using $\mathbf{J} \cdot \mathbf{E}$), we can add the quasi-static conduction equation:

$$\nabla \cdot (\sigma(T)\nabla\phi) = 0, \mathbf{E} = -\nabla\phi, \mathbf{J} = \sigma(T)\mathbf{E}, \quad 15$$

and then substitute into the heat equation to obtain

$$\rho_m c_p \frac{\partial T}{\partial t} = \nabla \cdot (k\nabla T) + \sigma(T) \mid \nabla\phi \mid^2 - q_{\text{loss}}, \quad 16$$

where $\rho_m$ is mass density; $\rho_e$ is electrical resistivity.

Although the governing equation is written in local (differential) form, specimen geometry enters the problem through both the electrical and thermal terms. For a flat specimen of length $L$, width $W$, and thickness $t$, geometry influences the temperature field in several ways. First, the current density scales inversely with cross-sectional area ($A = Wt$) for a given applied current, such that

$$|\mathbf{J}| \sim \frac{I}{Wt}. \quad 17$$

Consequently, thinner or narrower specimens experience higher current densities and thus higher volumetric Joule heating for the same applied current. Second, heat redistribution along the gauge length is governed by axial conduction over the characteristic length scale $L$, while transverse temperature gradients are constrained by $W$and $t$. Third, surface heat losses scale with the surface-area-to-volume ratio,

$$\frac{A_s}{V} \sim \frac{2(Wt + Lt + LW)}{LWt}, \quad 18$$

which increases sharply as the thickness decreases. As a result, thin specimens exhibit stronger coupling between internal Joule heating and external heat losses, leading to increased sensitivity to boundary conditions and environmental effects.

## B. Initial resistivity measurement at room temperature

A standard four-point probe configuration is typically employed to obtain accurate room-temperature electrical resistivity measurements for metallic and composite specimens. The setup consists of a DC power supply (constant current or voltage), a digital multimetre (DMM) for voltage measurements, a calibrated standard resistor for current verification, and a probe assembly with precisely spaced contact points (Figure 29). Regular calibration of the DMM and the standard resistor is essential to ensure accuracy.

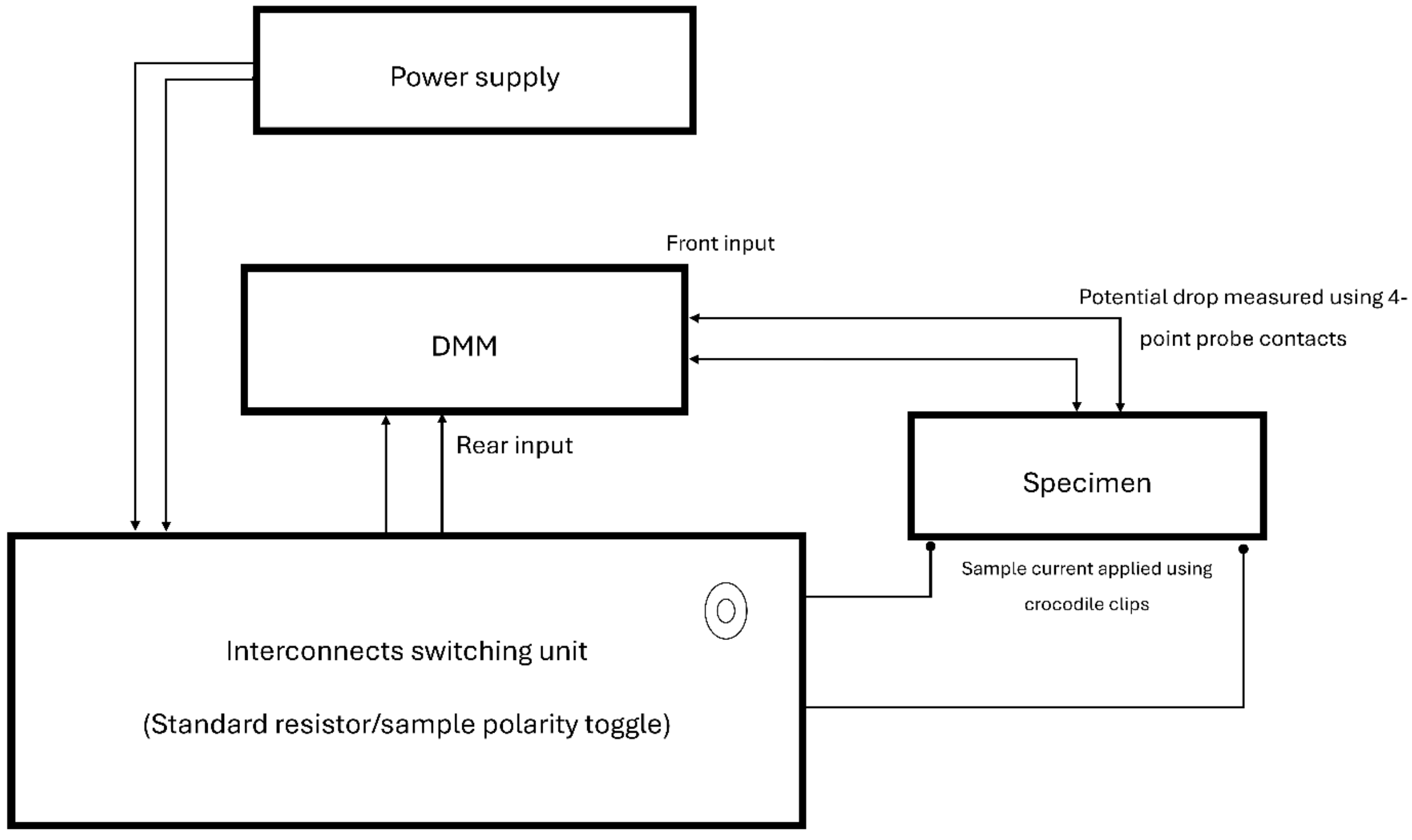

Figure 29: Schematic of the resistivity measurement system.

Specimen dimensions, including width $w$, thickness $t$, and the spacing $L$ between the inner voltage-sensing contacts is measured precisely using a micrometre or optical methods. These values define the cross-sectional area and current path length, both of which are critical for accurate resistivity calculations.

Current is applied through the outer probe contacts while the voltage drop is measured across the inner contacts, ensuring that voltage sensing remains independent of contact resistance. Forward and reverse polarity measurements are taken and averaged to minimise thermal EMF or parasitic voltages induced during the testing process. The applied current is verified via the standard resistor using Ohm's law ($I = V_{SR}/R$), ensuring deviations from the target current are detected and corrected before specimen measurements.

The resistivity is then calculated using

$$\rho = \frac{V_{Tp}\, w\, t}{I\, L}, \tag{19}$$

where $V_{Tp}$ is the averaged potential drop across the specimen, $I$ is the applied current, $L$ is probe spacing, and $w$ and $t$ define the cross-sectional area. Forward/reverse measurements, thermal stability, and accurate dimension measurements are critical to minimising uncertainties in resistivity determination.

## C. Magnetic field estimation

For DC-TMT specimens with a flat rectangular cross-section, the self-induced magnetic field is geometry-dependent and non-uniform across the specimen width. Therefore, the first cylindrical-conductor expression in equations (20) is not strictly applicable. For a rectangular DC-TMT specimen with a width $w$ and thickness $t$ carrying a uniform DC ($I$), the magnetic field immediately outside the broad face near the centre can be estimated, as

$$B = \frac{\mu_0 I}{2\pi r} \approx \mu_0 I \left[\frac{1}{\pi w} \tan^{-1}\left(\frac{w}{2t}\right) + \frac{1}{4\pi t} \ln\left(1 + \frac{4t^2}{w^2}\right)\right], \\ \mu_0 = 4\pi \times 10^{-7}\,\mathrm{Hm^{-1}}, \tag{20}$$

where $t$ is the specimen thickness and $\mu_0$ is the magnetic constant or permeability of free space. Take $I$ = 100 A and $t$ = 1 mm, the estimated centre-surface field is:

- 1 mm width, $B \approx 34.6$ mT
- 2 mm width, $B \approx 22.6$ mT
- 3 mm width, $B \approx 16.8$ mT

Thus, for a typical flat DC-TMT specimen, the self-induced magnetic field is of the order of a few 20 mT. Electromagnetic pinch (magnetic pressure) at the surface:

$$p \sim \frac{B^2}{2\mu_0} \Rightarrow p(34.6\ \mathrm{mT}) \approx 4.8 \times 10^2\ \mathrm{Pa} \approx 0.0048\ \mathrm{bar}, \tag{21}$$

which remains mechanically negligible compared with the stresses typically applied during ETMT. This pressure estimate should likewise be regarded as approximate, since the actual electromagnetic self-stress distribution depends on specimen geometry.

## Acknowledgement

This work was informed by discussions initiated at the ETMT interest group meeting and by subsequent exchanges with colleagues in the field. The authors thank Ms Maria Lodeiro and Dr Gavin Sutton (National Physical Laboratory), Dr Neil D'Souza (Rolls-Royce), Dr Christina Reinhard (University of Manchester and Diamond Light Source), Dr Michael King (University of Strathclyde), and Mr Andrew Pearce (Instron) for their valuable comments, input, and technical discussions. The authors also thank Professor Mark Gee and Professor Tony Fry (National Physical Laboratory) for careful proofreading and constructive feedback on the manuscript. AK, BR and RW acknowledge support from the National Measurement System (NMS) programme of the UK Department for Science, Innovation and Technology (DSIT), and YTT acknowledges support from the Royal Society (RGS\R2\252493)..

## CRediT authorship contribution statement

**AK:** Conceptualization, Methodology, Software, Validation, Formal analysis, Investigation, Writing - Original Draft, Visualization. **SAK:** Investigation, Writing - Original Draft, Visualization. **OFO:** Writing - Original Draft, Visualization. **RW:** Writing - Review & Editing, Visualization. **JUA:** Writing - Original Draft, Visualization. **YTT:** Validation, Investigation, Writing - Review & Editing. **BR:** Data Curation, Writing - Original Draft.